\documentclass[a4paper,11pt]{article}
\pdfoutput=1 % if your are submitting a pdflatex (i.e. if you have
\usepackage{jheppub} % for details on the use of the package, please
\usepackage[T1]{fontenc} % if needed

\usepackage{scalerel}
\usepackage{color}
\usepackage{latexsym}
\usepackage{mathrsfs}
\usepackage{amssymb}
\usepackage{amsfonts, epsf,epsfig,color}
\usepackage{graphicx}
\usepackage{tensor}
\usepackage{manfnt}
\usepackage{slashed}
\usepackage{simpler-wick} %For Wick contractions
\usepackage[all]{xy}
\usepackage[]{tensor}
\include{macro}

\newcommand{\R}{\mathbb{R}}

\newcommand{\cH}{\mathcal{H}}

\newcommand{\Z}{\mathbb{Z}}

\newcommand{\cO}{\mathcal{O}}

\newcommand{\cV}{\mathcal{V}}

\newcommand{\be}{\begin{equation}\label}
\newcommand{\ee}{\end{equation}}
\newcommand{\bea}{\begin{eqnarray}\label}
\newcommand{\eea}{\end{eqnarray}}

\DeclareSymbolFont{bbold}{U}{bbold}{m}{n}
\DeclareSymbolFontAlphabet{\mathbbold}{bbold}

\title{\textbf{Operator Deformations and T-duality via Parallel Transport}}
\author{H. Mahmood \& R. A. Reid-Edwards}

\affiliation{Department of Applied Mathematics and Theoretical Physics \\ University of Cambridge, Cambridge, CB3 0WA, UK}

\emailAdd{hm516@cam.ac.uk}
\emailAdd{rar31@cam.ac.uk}

\abstract{
We consider deformations of CFTs from the perspective of parallel transport in moduli space. In particular, we show how the deformations of individual operators can be computed and we also explore how these ideas can be extended to more general QFTs lacking conformal symmetry. We explore how to write one theory in terms of operators defined in a nearby theory, related to the first by a parallel transport in theory space. Using this construction, we describe how T-duality can be realised and how this provides a different perspective from the usual Buscher construction.}

\begin{document} 
\maketitle

\flushbottom

\section{Introduction}

This paper builds on work presented in \cite{Mahmood:2020mtq}, where an alternative approach to understanding T-duality was presented. Much recent progress in understanding T-duality has come from string theory on flat target spaces where the worldsheet theory is free or, for more general backgrounds, has been inspired by advances in supergravity. The aim in \cite{Mahmood:2020mtq} was to provide a framework that put both the interacting worldsheet quantum field theory and the role of T-duality as a broken gauge symmetry \cite{Dine:1988nrl} at the conceptual heart of the formalism. A key idea, due to \cite{Evans:1995su}, is that the CFT of a free boson on a circle of radius $R$ may written in terms of the CFT at the self-dual radius $R=\sqrt{\alpha'}$. This then allows one to see how the enhanced symmetry at the self-radius is broken $SU(2)_L\times SU(2)_R\rightarrow U(1)_L\times U(1)_R$ and how a residual $\Z_2$ symmetry is preserved at any radius. This $\Z_2$ is T-duality and the set-up generalises straightforwardly to higher dimensions. 

Exact solutions are hard to find and much of the progress in applications of the duality have centred on toy models - simple examples that are not CFTs, but may play a role in exact string solutions\footnote{See, for example,  \cite{Kachru:2002sk,Chaemjumrus:2019ipx}.}. As such, we would like to understand how the framework of \cite{Evans:1995su,Mahmood:2020mtq} can be applied to more general nonlinear sigma models in a genuine departure from the CFT analysis. The key step in these constructions is to write the theory at general radius in terms of the states defined at the self-dual radius. For torus backgrounds, there is a straightforward way to to this, by using universal coordinates \cite{Kugo:1992md, Ranganathan:1992nb} (reviewed and generalised in section \ref{S: toroidal backgrounds}). However, for more general backgrounds, that may have fluxes or curvature, it was shown in \cite{Mahmood:2020mtq} how the universal coordinate approach works well as an adiabatic limit at large radius, but is not sufficient for a more general analysis. One of our aims in this paper is to describe how the use of a connection on the space of backgrounds can be used to define a parallel transport that allows one to write the operators of a theory in one background in terms of the operators in another background (at least perturbatively). 

Much attention has been given to changes in the worldsheet stress tensor $T(z)$ under a change of background. A secondary aim is to develop a formalism that correctly treats operators, such as $\partial X^{\mu}(z)$, which are not target space scalars. We will initially consider toroidal backgrounds which are exact CFTs, where we will show that our formalism reproduces the known deformation results obtained in \cite{Evans:1995su, Evans:1989xq} to all orders. In later sections we will then extend the formalism to more complicated backgrounds which are not CFTs.

A rough outline of the framework is as follows. Details will be given in the text. We start with a family of sigma models, characterised by a metric, $B$-field and any other fields in the target space. We assume that, near a theory of interest, this data defines a space of backgrounds ${\cal M}$. Over ${\cal M}$ there is a bundle ${\cal E}\rightarrow {\cal M}$ whose fibres are operators of the worldsheet theory defined at the point in ${\cal M}$. We will really only be interested in a subset of these operators relevant to the deformation we are considering. 

The starting point is a string theory at a point of enhanced symmetry $p_0 \in {\cal M}$, where this symmetry acts as an automorphism on the operator algebra
\begin{equation}
    {\cal A}_0\rightarrow U^{-1}{\cal A}_0U,
\end{equation}
for any operator ${\cal A}_0$ in the fibre at $p_0$, where $U$ is an element of the symmetry group. We then define a connection $\Gamma$ on ${\cal E}$ and use it to parallel transport to a background of interest $p\in {\cal M}$. This allows us to describe the theory at $p$ in terms of the operators of the theory at $p_0$. The advantage of this is that, if we know the action of the symmetry group on operators at $p_0$, we can use this to deduce the action on operators at $p$. In general, the enhanced symmetry is spontaneously broken in the new background, but a subgroup may remain. One of the benefits of this approach is that, even if the background we start at is a CFT, we can deform the theory `off-shell' to a background which is not a CFT, so it can be used for a large class of backgrounds, including a number of toy models. When we come to the Nilfold and $H$-flux in section \ref{S: Nonlinear Sigma Models and Off-shell string theory}, we will see that this is useful for investigating the T-duality between these backgrounds.

Although we find the conceptual framework is compelling, there are several drawbacks to this approach: the main issue is that the calculations are hard. This is, in part, reflective of the fact that the worldsheet quantum field theory is at the heart of this formalism and calculations in interacting quantum field theories are hard. The challenges of doing explicit calculations are a reflection of this fact. There is a sense in which this is not the whole story. Even the free theory calculations in this formalism can be involved and there are examples of dualities between toy models that may be done straightforwardly in the Buscher construction \cite{Buscher:1987sk, Buscher:1987qj}, but are technically challenging in the formalism presented here. We also avoid issues of topology change and degenerations in the background and we generally lift to a covering space to perform the parallel transport. %This is a common shortcoming of most approaches to generalising the Buscher construction and we shall not discuss issues of topology change here, although it would be interesting to investigate this further.

What \emph{is} to be gained from this formalism is a different perspective on T-duality, one that is potentially applicable to a wider class of examples than the Buscher construction. For example, the challenges with non-isometric duality are particularly clear from this perspective and no longer seem insurmountable.

The outline of this paper is as follows. In section \ref{S: Deformations and connections}, we will describe an approach to deforming operators in a given theory space of sufficient generality for our purposes. We will review the discussions of \cite{Ranganathan:1993vj} and explain how the notion of parallel transport is used to deform operators to some target point in the moduli space of backgrounds. In section \ref{S: toroidal backgrounds}, we will apply this formalism to the case of a CFT with constant metric and $B$-field. We will briefly review the universal coordinates approach of \cite{Kugo:1992md, Evans:1995su}, and then we will re-derive these results using our approach. We will discuss our prescription for recovering the known deformations to all orders and we will discuss different choices of connection. In particular, we will explore the relationship between the $\hat{\Gamma}$ connection \cite{Ranganathan:1993vj}, universal coordinates \cite{Kugo:1992md} and doubled geometry \cite{Hull:2004in, Hull:2009sg}. In section \ref{S: Nonlinear Sigma Models and Off-shell string theory}, we will move away from CFTs and discuss how our formalism can be applied to backgrounds with non-trivial coordinate dependence. In particular, we will look at the Nilfold and $T^3$ with $H$-flux, and we will explore how the worldsheet interactions change the deformations of the $\partial X_\mu$ operators. These backgrounds were discussed in this context in \cite{Mahmood:2020mtq}, where an adiabatic approximation, which neglects such interactions, was assumed. One of the results of this section will be to compute leading order corrections to this approximation. In section \ref{S: T-duality}, we will discuss T-duality and how to approach it in our formalism. Finally, in section \ref{S: Discussion}, we briefly discuss the results of this paper and some potential further directions.

\section{Deformations and Connections}\label{S: Deformations and connections}

We start by reviewing the general idea of deforming both CFTs and QFTs more generally. The CFT case was covered in detail in \cite{Ranganathan:1993vj}, where a process for deforming a CFT using the marginal operator and surface states was given, and in \cite{Sonoda:1991mv, Sonoda:1992hd} the more general procedure was outlined for QFTs by looking at deformations of correlation functions.

\subsection{General deformations and connections}

We shall consider a family of sigma models $\{E_{\mu\nu}\}$, where $E_{\mu\nu}=g_{\mu\nu}+B_{\mu\nu}$ specifies a target space metric $g_{\mu\nu}$ and antisymmetric $B$-field $B_{\mu\nu}$, parameterised in some convenient way. Let ${\cal E}$ be a bundle of operators ${\cal A}$ over this space, where an element $A\in{\cal A}$ may be written schematically in terms of a local worldsheet operator as $A=\int\Phi$. Then, the integrated correlation functions will change as we move about this space, not simply because of any direct metric dependence of the field, but also due to the change in the measure $e^{S[\Phi]}$ used to define the correlation function. If 
\begin{equation}
    \big\langle \Phi_1...\Phi_n \big\rangle=\int\limits{\cal D}\Phi e^{-S[\Phi]}\;\Phi_1...\Phi_n,
\end{equation}
where $\Phi_i$ is a local operator inserted at the point $z_i$ on a worldsheet $\Sigma$ embedded in the target space\footnote{To compute the correlation functions, the locations of the operators would ultimately be integrated over and it is probably more natural to think in terms of integrated operators $\int_{\Sigma}\Phi$ when talking about objects in the fibres of ${\cal E}$. We will also not explicitly include the ghost contributions to such correlation functions.}, then to leading order,
\begin{equation}\label{def}
    \delta\big\langle \Phi_1...\Phi_n\big\rangle=\sum_{i=1}^n\int\limits{\cal D}\Phi e^{-S[\Phi]}\;\Phi_1...\delta_{E}\Phi_i...\Phi_n-\int\limits{\cal D}\Phi e^{-S[\Phi]}\delta S[\Phi]\;\Phi_1...\Phi_n+...,
\end{equation}
where $\delta_{E}\Phi$ is a `classical' change in $\Phi$ -  a change in the background field that preserves $S[\Phi]$. 

If we want to understand how an individual operator, say $\Phi_i$, changes we need a more subtle tool. The above expressions suggest that this would be given schematically, to first order, by
\begin{equation}
    \delta\Phi_i=-{\cal O}_i[X](\Phi_i)+\delta_{E}\Phi_i+...,
\end{equation}
where the effect of the change in the action on the contribution $\Phi_i$ makes to the correlation function is given by the insertion of a non-local operator ${\cal O}_i[X]$ that has the same functional form as $\delta S[X]$,
\begin{equation}
    {\cal O}_i[X]=\int\limits_{\Sigma_i^\epsilon}:\delta \widehat{{\cal L}}:,
\end{equation}
where $\delta\widehat{\cal L}$ is the integrand\footnote{${\cal O}_i$ is not local and the deformation need not be of Lagrangian type, but for simplicity we shall assume it does take the form $\delta S=\int\limits_{\Sigma}\delta {\cal L}$.} of $\delta S[X]$, lifted to an operator expression and $\Sigma_i^\epsilon\subset \Sigma$ has holes of size $\epsilon>0$ cut out around the locations of the $\Phi_{j\neq i}$ fields. We recover the variation of the correlation function when all fields $\Phi$ are allowed to change. Here, we will mainly be interested in cases where the starting point is a free theory. In this case, we can use Wick's theorem to evaluate the correlation function, and the action of ${\cal O}$ that affects the field $\Phi_i$ directly is given by the contraction
\begin{equation}\label{delta}
    \delta_{\cal O}\Phi_i(z_i)=\wick{\c1{{\cal O}}_i[X]\c1{\Phi_i}(z_i)}.
\end{equation}
Sequential applications of ${\cal O}$ will be discussed in detail in section \ref{S: toroidal backgrounds} and Appendix \ref{A: Sen}.

This construction extends simply to cases where the sigma model is specified by other target space fields. We will flesh out what the terms in (\ref{delta}) mean in the following sections. %Our main focus is on the deformation generated by ${\cal O}$, but, by way of an example, we shall briefly say a little about $\delta_{E}$. 
If $\Phi$ is a target space scalar, then this is the whole story. However, if $\Phi$ is a vector or higher tensor then there may be an additional contribution, which preserves the action and therefore is not included in the transformation generated by ${\cal O}$. For example, in a flat torus background, $\partial X^{\mu}=e^{\mu}{}_a\partial X^a$ has an explicit dependence on the background metric through the vielbein $e^{\mu}{}_a$. We now imagine changing the radius $R$ of one of the circles. With all of the $R$ dependence in $e^{\mu}{}_a$, we think of $\partial X^a$ as a universal field for such backgrounds and the change in $\partial X^{\mu}$ may be written as 
\begin{equation}
    \partial X^{\mu}\rightarrow \partial X^{\mu}+\delta R (\partial_Re^{\mu}{}_a)e^a{}_{\nu}\partial X^{\nu}+... .
\end{equation}
Such contributions are absent in the previously studied fields such as the worldsheet stress tensor, but play an important role in recovering the correct transformation properties of non-scalar fields. We denote such contributions to the variation by $\delta_{E}$.\footnote{Schematically, $\delta_{E}$ may be thought of as being of the form
$
    \delta_{E}\sim\delta g_{\mu\nu}\frac{\delta}{\delta g_{\mu\nu}}.
$}
It will be helpful to think of $\delta_{E}$ as the part of the deformation that leaves the action invariant and ${\cal O}$ as that part which changes the action.

\subsection{Connections on theory space}

The deformation (\ref{def}) can be cast in a more concrete geometric setting. Let us assume the existence of a space of backgrounds ${\cal M}$ (at least locally), each point of which corresponds to a nonlinear sigma model\footnote{Ultimately, this is a story about a bundle operator algebras over theory space and no target space interpretation is necessary; however, our wish to discuss T-duality means we will focus on cases where a target space interpretation is available.}. The data that defines the sigma model - its coupling constants, in the form a a metric, $B$-field, etc. - define a point $p\in{\cal M}$. We can define a fibre bundle with base ${\cal M}$ and fibre given by the Hilbert space of states ${\cal H}_p$ of the theory at $p$ \cite{Ranganathan:1993vj}. Here, we choose to work with the bundle ${\cal E}\rightarrow {\cal M}$ with fibres given by the operators\footnote{We have in mind here some generating set of operators $\{{\cal O}_\alpha\}$ that are, at the least, rich enough to construct the deformation operators in the cotangent bundle $T^*{\cal M}$.} of the theory.

The fibres of the tangent space $T{\cal M}$ are spanned by the beta-functions of the theory. Similarly, and in a sense made precise in \cite{Sonoda:1993dh,Sonoda:1991mv}, the fibres of the dual space $T^*{\cal M}$ are spanned by the deformation operators ${\cal O}_\alpha$. These operators are conjugate to local coordinates $\mathfrak{m}^\alpha(p)$ in the neighbourhood of a point $p\in{\cal M}$. We shall be interested in studying connections on ${\cal E}$ and the associated deformations of the sigma models given by parallel transport in ${\cal E}$.

To make the above discussion precise, we can think of the correlation function $\big\langle \Phi_1(z_1)...\Phi_n(z_n) \big\rangle_p$ at $p\in {\cal M}$ as a formal Taylor expansion about a reference point $p=p_0$,
\begin{align}
    \big\langle \Phi_1(z_1)...\Phi_n(z_n) \big\rangle_p&=\big\langle \Phi_1(z_1)...\Phi_n(z_n) \big\rangle_{p_0}
    +\delta\mathfrak{m}^\alpha\frac{\partial}{\partial \mathfrak{m}^\alpha}\big\langle \Phi_1(z_1)...\Phi_n(z_n) \big\rangle_{p_0}+....
\end{align}
Given the above discussion, the naive guess would be to associate the derivative with an insertion of the (conjugate) deformation operator
\begin{equation}
    {\cal O}_{\alpha}=\int_\Sigma{d}^2z\, O_\alpha(z, \Bar{z}).    
\end{equation}
It has been proposed that the derivative should be more properly understood to mean\footnote{The variational formula (\ref{corrDeriv}) was proposed in \cite{Sonoda:1991mv}, but has not been derived from first principles as far as we are aware. A similar formula was suggested in \cite{Wilson:1969zs}.}
\begin{align}
    -\frac{\partial}{\partial \mathfrak{m}^\alpha}\big\langle \Phi_{a_1}(z_1)&...\Phi_{a_n}(z_n)\big\rangle_{p}=\lim_{\epsilon\rightarrow 0}\left[
    \int\limits_{\Sigma - \bigcup_i {\cal D}_i^\epsilon}{d}^2z\big\langle  O_\alpha(z)\Phi_{a_1}(z_1)...\Phi_{a_n}(z_n)\big\rangle_{p}\right.\nonumber\\
    &\left.+\sum_{i=1}^n\Bigg(\Gamma_{\alpha,a_i}{}^b(p)-\int\limits_{{\cal D}_i^1 - {\cal D}_i^\epsilon}{d}^2z C_{\alpha,a_i}{}^b(z,p)\Bigg)\big\langle \Phi_{a_1}(z_1)...\Phi_b(z_i).
    ..\Phi_{a_n}(z_n)\big\rangle_{p}
    \right] \label{corrDeriv},
\end{align}
where ${\cal D}_i^\epsilon$ is a disk of radius $\epsilon$ around $z_i$, and ${\cal D}_i^1$ is the unit disk \cite{Sonoda:1991mv, Sonoda:1992hd, Ranganathan:1993vj}. The insertion of the conjugate operator ${\cal O}_\alpha$ is expected from (\ref{def}). What of the other terms? The divergences caused by the short-distance singularities are removed by extracting the OPE contribution on a finite disc. The covariance of the expression under field transformations $\Phi_a\rightarrow {\cal S}_a{}^b\Phi_b$ on the space of backgrounds is ensured  by introducing the finite counter-terms $\Gamma_\alpha(p)$. The $\Gamma_\alpha(p)$ are not unique. There are different ways in which one may remove divergences in (\ref{corrDeriv}) and different prescriptions correspond to different choices of connection in ${\cal E}$. These are independent of $z$ and transform as connections on the space of backgrounds,
\begin{equation}
   \Phi_a\rightarrow {\cal S}_a{}^b(p)\Phi_b, \qquad  \Gamma_{\alpha a}{}^b(p)\rightarrow {\cal S}_a{}^c(p)\Big(\Gamma_{\alpha c}{}^d(p)+\delta_c^d\partial_\alpha\Big)({\cal S}^{-1})_d{}^b(p). 
\end{equation}
% Defining the covariant derivative $D_\alpha=\partial_\alpha +\Gamma_\alpha$, this can more usefully be written as
%\begin{align}
%    -D_{\alpha}\big\langle \Phi_{a_1}(z_1)...\Phi_{a_n}(z_n)\big\rangle_{p}&=\lim_{\epsilon\rightarrow 0}\left[
%    \int\limits_{\Sigma - \bigcup_i {\cal D}_i^\epsilon}{d}^2z\big\langle O_\alpha(z)\Phi_{a_1}(z_1)...\Phi_{a_n}(z_n)\big\rangle_{p}\right.\nonumber\\
%    &\left.-\sum_{i=1}^n \;\int\limits_{{\cal D}_i^1 - {\cal D}_i^\epsilon}{d}^2z\, C_{\alpha,a_i}{}^b(z,p)\big\langle \Phi_{a_1}(z_1)...\Phi_b(z_i)...\Phi_{a_n}(z_n)\big\rangle_{p}
 %   \right].
%\end{align}
The choice of connection is, in part, given by a choice of how we deal with the divergences. One may also combine this transformation with an automorphism of the operator algebra (a symmetry of the theory at a given point).  Defining the covariant derivative $D_\alpha=\partial_\alpha +\Gamma_\alpha$, this can more usefully be written as
\begin{eqnarray}
    D_{\alpha}\big\langle \Phi_{a_1}(z_1)...\Phi_{a_n}(z_n)\big\rangle_{p}&=&\lim_{\epsilon\rightarrow 0}\left[-
    \int\limits_{\Sigma - \bigcup_i {\cal D}_i^\epsilon}{d}^2z\big\langle O_\alpha(z)\Phi_{a_1}(z_1)...\Phi_{a_n}(z_n)\big\rangle_{p}\right.\nonumber\\
    &+&\left.\sum_{i=1}^n\Omega_{\alpha,a_i}{}^b(p)\big\langle \Phi_{a_1}(z_1)...\Phi_b(z_i)...\Phi_{a_n}(z_n)\big\rangle_{p}
    \right]\label{ODef},
\end{eqnarray}
where
\begin{equation}\label{Omega}
    \Omega_{\alpha,a_i}{}^b(p)=\int\limits_{{\cal D}_i^1 - {\cal D}_i^\epsilon}{d}^2z\; C_{\alpha,a_i}{}^b(z,p)+\omega_{\alpha,a_i}{}^b(p).
\end{equation}
$\omega$ generates a symmetry around each puncture. For CFTs, one can relate one value of $\epsilon$ to another by a dilation and so there is no need to take the $\epsilon\rightarrow 0$ limit. Indeed, different choices of $\epsilon$ may be compensated by including an appropriate conformal transformation in $\Omega$.

Renormalization flow defines a trajectory on ${\cal M}$. The relationship between this construction and renormalization group flow was explored in \cite{Sonoda:1991mv,Sonoda:1992hd}\footnote{In fact, the requirement that the connection was compatible (in a way made precise in \cite{Sonoda:1991mv,Sonoda:1992hd}) with renormalization group flow was one of the main considerations in constructing it.} (see also \cite{Sonoda:1993dh}). In particular, compatibility of the renormalization flow of the operator ${\cal O}$ with the variational formula (\ref{corrDeriv}) places constraints on the coefficients $C_{\alpha,a_i}{}^b(z,p)$, which are detailed in \cite{Sonoda:1991mv,Sonoda:1992hd}. This relationship with renormalization is central. The deformation operator ${\cal O}$ relates one theory to a theory of a similar kind, with a different value of the background fields - the theory is qualitatively the same, but quantitatively different. This quality of self-similarity, familiar from renormalization, is the feature that defines a particular path between two points in ${\cal M}$.

\subsubsection{Choices of connection}

In \cite{Ranganathan:1993vj}, three connections were highlighted, denoted by $\hat{\Gamma}, c$ and $\bar{c}$. Here we will focus mainly on the $\hat{\Gamma}$ connection and the $c$ connection (which is the connection originally used in \cite{Sen:1990hh} to computing the deformation of the stress tensor for string theory on a circle and is the natural choice for a general QFT). Each of these connections is described by the pair $({\cal D}, \Omega)$, where $\Omega$ is as in (\ref{Omega}), and ${\cal D} = \bigcup_i{\cal D}_i$ is the region around the punctures that we remove to regulate the divergences. We briefly describe these connections here. 

The $\hat{\Gamma}$ connection is defined by the pair $(\bigcup_i{\cal D}_i^1, 0)$, i.e. we remove a disk of radius 1 around each puncture $z_i$. The radius 1 is arbitrary and is conformally equivalent to any other radius. In this case, there are no divergences to subtract since we are not shrinking the discs to zero, so $\Omega=0$. Note that this is only possible for a CFT since, due to the conformal symmetry, any calculation will be independent of the radii of the disks, so these radii may be freely chosen. It was pointed out in \cite{Ranganathan:1993vj} that, using uniformising coordinates on $\Sigma$, this connection preserves the metric\footnote{The metric is given by $\mathcal{G}_{\alpha\beta}(p)=\left\langle \Phi_{\alpha}\Phi_{\beta}\right\rangle_p$, where $\{\Phi_{\alpha}\}$ is a basis describing the theory and $p\in {\cal M}$ and $\Sigma$ is a sphere.} $\mathscr{G}_{\alpha\beta}$ on the space of CFTs. Indeed, as observed in \cite{Sen:1993mh}, string field theory seems to favour this connection.

The connection $c$ is defined to subtract away any divergences as we approach the punctures. Therefore, the disk around each puncture has radius $\epsilon\rightarrow 0$, and for a given puncture $z_i$, $\Omega_{\alpha,a_i}{}^b$ is the coefficient of $\Phi_b$ in the OPE between the deformation operator $O_\alpha$ and the operator at $z_i$ which diverges as $z\rightarrow z_i$. This divergent part is computed in an annulus $\epsilon<|z-z_i|<1$ (1 chosen here to match up with $\hat{\Gamma}$), and therefore we can also think of the $c$ connection as follows: we always integrate up to the disk of radius 1 around a given puncture. Then, operators in the OPE which diverge as we take $z\rightarrow z_i$ are not integrated any further, and operators which do not diverge are integrated fully within the annulus. Practically, the way we would use this connection is that we would work with disks of radius $\epsilon$ removed around each puncture, and at the end of the calculation we try to take the limit $\epsilon \rightarrow 0$. If this limit exists, we take it, and if not, we set $\epsilon = 1$. This is done for each operator appearing in the OPE. 

The $\bar{c}$ connection is a kind of `minimal subtraction' variation on the $c$ connection. We get this by noticing that not all terms in the divergent part of the OPE between the deformation operator and an operator insertion need be divergent in the limit $\epsilon\rightarrow 0$. We therefore define $\Omega$ such that we subtract only the parts of the OPE which diverge in the limit. In \cite{Ranganathan:1993vj}, it is explained that this is essentially just the diagonal part of the $c$ connection, which is itself the upper triangular part of the $\hat{\Gamma}$ connection. We will not be particularly interested in the $\bar{c}$ connection in this paper. The focus will be on the deformation of the sigma model and, in general, a specific choice of connection (regularisation of divergences) will not be made, but we will state explicitly if we ever do so.

\subsubsection{Parallel transport}\label{SS: parallel transport}

Given a suitable path in ${\cal M}$ and a covariant derivative of a correlation function in terms of some deformation operator $\cal O$, one can parallel transport operators from one theory to another. In other words, we can describe an operator at one point in ${\cal M}$ in terms of a basis of operators at another. We recall the salient points of the discussion in \cite{Ranganathan:1993vj} for completeness. 

Suppose we have a path in moduli space  ${\mathfrak{m}^\alpha}(s')$, where $s'\in[0,s]$ parameterises the path and the ${\mathfrak{m}^\alpha}$ can be thought of as local coordinates in ${\cal M}$. We are interested in expressing operators $\Phi(s)$ in terms of operators $\Phi(0)$ via parallel transport. This is defined by the vanishing of the covariant derivative $\frac{D}{Ds'}\Phi(s')$, which can be written in terms of the connection as
\begin{equation}
    \frac{D}{Ds'}\Phi(s') = \frac{\partial}{\partial s}\Phi(s') + \frac{{d}\mathfrak{m}^{\alpha}(s')}{{d} s'}\Gamma_{\alpha}(s')\Phi(s') = 0,
\end{equation}
where the connection $\Gamma$ can be written in terms of $( {\cal D}, \Omega)$. \cite{Ranganathan:1993vj} showed that the solution to this is given by 
\begin{align}
    \Phi(s) &= \Phi(0)P\exp\left( -\int\limits_0^s ds'\frac{d{\mathfrak{m}^\alpha}(s')}{ds'}\Gamma_\alpha(s') \right) \notag \\
    &= \Phi(0)\left(  1 - s\frac{d{\mathfrak{m}^\alpha}}{ds}\Gamma_\alpha(0) - \frac{s^2}{2}\left(  \frac{d^2{\mathfrak{m}^\alpha}}{ds^2}\Gamma_\alpha(0) + \frac{d{\mathfrak{m}^\alpha}}{ds}\frac{d\mathfrak{m}^\beta}{ds}\Big(\partial_\alpha\Gamma_\beta- \Gamma_\alpha\Gamma_\beta\Big)(0)      \right)+...\right), \label{parallel}
\end{align}
where $P$ denotes path ordering and $\mathfrak{m}^{\alpha}, \Gamma_{\alpha}$ and their derivatives are all evaluated at $s=0$ on the right hand side. For our purposes, ${\mathfrak{m}^\alpha}$ will essentially be the background tensor $E_{\mu\nu}$ and our connection will be defined by the ${\cal O} \text{ and } {\delta_{E}}$ operators and a choice of regularisation prescription. Note that this path ordered exponential is invariant under reparameterisations of $s'$, so that the parallel transport of $\Phi(0)$ is in fact invariant of the parameterisation we choose, as we would expect. In the calculations that follow, we will see this explicitly, in particular for the circle deformation, where we can either choose to work with $\delta g$, the full metric variation, or $\delta R$, the deformation of the radius of the circle, and both choices give the same results. 

Alternatively, suppose we locally have a section of ${\cal E}$, given by a choice of some $\Phi(\mathfrak{m})$ in some subset of ${\cal M}$. Take some path through moduli space $\mathfrak{m}(s')$, $s'\in [0,s]$, and a connection $\Gamma_\alpha(s')$, as before. Unless $\Phi(\mathfrak{m})$ was generated by parallel transport using this connection, it is not covariantly constant along this path. However, it is possible to define a tensor along this path, $\tilde{\Phi}(s')$, such that $\tilde{\Phi}$ is covariantly constant, given by \cite{Ranganathan:1993vj}
\begin{equation}\label{parallel2}
    \tilde{\Phi}(s) = P\exp\left( -\int\limits_0^sds' \frac{d{\mathfrak{m}^\alpha}(s')}{ds'}D_\alpha \right)\Phi(s).
\end{equation}
It can be easily checked that $D\Tilde{\Phi}/Ds=0$. It is also easily seen that, in the case where $D\Phi/Ds=0$, i.e. $\Phi$ is already covariantly constant along the path $\mathfrak{m}(s')$, the above equation reduces to $\tilde{\Phi}(s) = \Phi(s)$, as we would expect. \eqref{parallel2} can be used to relate parallel transport by different connections. Given two distinct connections $\Gamma_1$, $\Gamma_2$, use $\Gamma_1$ to compute the deformation of some operator $\Phi_0$ along some path between points in ${\cal M}$ defined by parameters $s'=0$ and $s'=s$. Call this parallel transport $\Phi_1(s')$, so that $\Phi_1(0) = \Phi_0$. This deformation is defined by parallel transport with respect to $\Gamma_1$, but it is \textit{not} defined by parallel transport with respect to $\Gamma_2$. However, using \eqref{parallel2}, we can define a new deformation of $\Phi_0$, $\Phi_2(s')$, which is covariantly constant wrt $\Gamma_2$, and in this way we can relate deformations defined by different connections. If we write $D_2 = \partial + \Gamma_2 = D_1 + \Gamma_2 - \Gamma_1$, then $\eqref{parallel2}$ becomes
\begin{equation}\label{parallel3}
    \Phi_2(s) =P\exp\left( -\int\limits_0^sds' \frac{d{\mathfrak{m}^\alpha}(s')}{ds'}\Big(\Gamma_2 - \Gamma_1\Big)_\alpha \right)\Phi_1(s),
\end{equation}
so the difference between the two is given by the difference between connections. In general, the difference between connections will include differences in regularisation and any local automorphisms. We will not take these to be physically significant and so take $\Phi_1(p)\sim\Phi_2(p)$.

A point that should be made clear here is that there is a difference between the choice of connection and the choice of deformation operator ${\cal O}$. Given two points $p_1,p_2\in{\cal M}$, the first choice to be made is the path between the points. This essentially defines the deformation operator ${\cal O}$. Then, given the path, we need to define a connection on that path, and, as discussed above, there are many different choices of connection one can make. 

Section \ref{S: Nonlinear Sigma Models and Off-shell string theory} will discuss the specific case of a flat background\footnote{We will have in mind a $T^3$, but we will work with $\R^3$ or $T^2\times \R$ and impose identifications on the coordinates after the deformation.} with constant $H$-flux. There, the gauge transformations of the $B$-field play an important role. It is important to bear in mind that these gauge transformations are different to the redundancies discussed here that arise from different connection choices. If we have two $B$-fields that differ by a large gauge transformation, $B_1\sim B_2$, these correspond to \textit{different} points $p_1\neq p_2\in{\cal M}$ (unless we impose appropriate identifications in ${\cal M}$). Therefore, this choice of $B$-field gauge is made before any connection considerations. Another example is the physical equivalence of two sigma models under a diffeomorphism that is not isometric, which would be described by different points in ${\cal M}$ (unless an appropriate identification is imposed).

Thus, there are two types of symmetry here. On the one hand, there are symmetries which preserve a given sigma model, i.e. which do not change the point in ${\cal M}$. \eqref{parallel3} is an example of this, since $\Phi_{1,2}$ must be physically equivalent, so the relation between them can be described as a symmetry. Since this symmetry is induced by a change of connection on the same path $\mathfrak{m}(s')$, the point in ${\cal M}$, $\mathfrak{m}(s)$, is unchanged. On the other hand, there are symmetries between \textit{distinct} points on ${\cal M}$. Physically, sigma models related by such symmetries are the same, but this is not apparent on the worldsheet.

We shall be interested in the case where the starting point for the parallel transport is a free theory, with the destination the theory with non-trivial interaction. As such, the OPE coefficients which appear in \eqref{Omega}, and play a key role in the connection\footnote{The connection coefficients are determined by the OPE coefficients, the regularisation procedure used and any additional symmetries that are included. Specific examples are derived in \cite{Ranganathan:1993vj}.} $\Gamma$, are those of the free theory.

\subsection{Example: stress tensor deformation}

We start with a worldsheet embedding into $S^1$. In \cite{Sen:1990hh}, the deformation of the stress tensor of this free boson CFT to first order under an infinitesimal deformation of the background was derived\footnote{See also \cite{Evans:1989xq, Evans:1991qf} for further discussion on the deformation of the stress tensor.}. Here, we briefly recap this derivation as it will be important for much of our later discussions. We start by considering the addition of the term to the action
\begin{equation}
   {\cal O}= \lambda\int\limits_{\Sigma} d^2z \,O(z,\bar{z}), 
\end{equation}
where $O$ is some $(1,1)$ primary field and $\lambda$ is a constant. The correlation function in the deformed background (\ref{def}), denoted by a prime, is
\begin{align}
    \big\langle\Phi_1(z_1,\bar{z}_1)...\Phi_N(z_N,\bar{z}_N)\big\rangle' &= \big\langle\Phi_1(z_1,\bar{z}_1)...\Phi_N(z_N,\bar{z}_N)\big\rangle \notag \\&
    + \lambda\int\limits_{\Sigma^\epsilon}d^2z \big\langle O(z,\bar{z})\Phi_1(z_1,\bar{z}_1)...\Phi_N(z_N,\bar{z}_N)\big\rangle+...,
\end{align}
where we have defined the shorthand
\begin{equation}\label{sigma'}
    \Sigma^\epsilon = \Sigma - \bigcup_i{\cal D}^\epsilon_i.
\end{equation}
Note that we are not taking the limit $\epsilon\rightarrow 0$ here since we are dealing with a CFT. Of course, we may still choose to take the limit if we wish, but we have the freedom to leave $\epsilon$ as it is for now. However, if we insert the stress tensor at some point $z=w$, we will always choose our connection such that the radius of the disk around $z=w$ goes to zero, and we will assume that this is understood throughout the paper. Thus, given the correlator\footnote{We have used $$
    (w-z)^{-2} = \sum_{n\geq 0}n(w-z_i)^{-n-1}(z-z_i)^{n-1}.
$$}
\begin{equation}
    \Big\langle T(w)\prod_{i=1}^N\Phi_i(z_i,\bar{z}_i)\Big\rangle = \sum_{i=1}^N\sum_{n\geq -1}(w-z_i)^{-n-2}\Big\langle\prod_{j\neq i}\Phi_j(z_j,\bar{z}_j)L_n\Phi_i(z_i,\bar{z}_i)\Big\rangle, 
\end{equation}
we insert the marginal operator $\cal O$ and compute the OPEs of $T(w)$ with $\cal O$ and with the $\Phi_i$ (see \cite{Sen:1990hh} for the full details). We find, to leading order in $\lambda$, that
\begin{align}
    \Big\langle T(w)\prod_{i=1}^N\Phi_i(z_i,\bar{z}_i)\Big\rangle' &=\sum_{i=1}^N\sum_{n\geq -1}(w-z_i)^{-n-2}\Big\langle\prod_{j\neq i}\Phi_j(z_j,\bar{z}_j)L_n\Phi_i(z_i,\bar{z}_i)\Big\rangle' \notag\\
    &- \lambda \oint\limits_{C^\epsilon} d\bar{z} \frac{1}{w-z}\Big\langle O(z,\bar{z})\Phi_1(z_1,\bar{z}_1)...\Phi_N(z_N,\bar{z}_N)\Big\rangle,
\end{align}
where $C^\epsilon$ denotes the collection of anticlockwise contours of radius $\epsilon$ around each of the points (with $\epsilon\rightarrow 0$ for the contour around $w$). Noting then that the RHS contains no singularities at $z=w$ and therefore that the integral around $w$ vanishes in the $\epsilon\rightarrow 0$ limit, we are finally left with 
\begin{align}
    \Big\langle T(w)\prod_{i=1}^N\Phi_i(z_i,\bar{z}_i)\Big\rangle' =  \sum_{i=1}^N\sum_{n\geq -1}(w-z_i)^{-n-2}\bigg\langle&\prod_{j\neq i}\Phi_j(z_j,\bar{z}_j)\Big(L_n\Phi_i(z_i,\bar{z}_i) \notag \\
    &- \lambda\oint\limits_{C_i^\epsilon} d\bar{z}(z-z_i)^{n+1}O(z,\bar{z})\Phi_i(z_i,\bar{z}_i)\Big)\bigg\rangle', 
\end{align}
where $C_i^\epsilon$ is the contour around the point $z_i$ of radius $\epsilon$, from which we compare coefficients of $(w-z_i)^{-n-2}$ and read off the shift in the Virasoro modes as\footnote{Naively, it looks as though this will lead to non-trivial commutation relations between $L_n$ and $\bar{L}_n$ in the deformed CFT. It was shown in \cite{Campbell:1990dz} that this is not the case.}
\begin{equation}
    L_n \rightarrow L_n - \lambda\oint\limits_{C_0^\epsilon} d\bar{z}z^{n+1}O(z,\bar{z}),
\end{equation}
where $C_0^\epsilon$ is the contour of radius $\epsilon$ around $z=0$. This example provides a concrete illustration of how it is possible to isolate the change in a single operator from the change of a correlation function containing that operator. Beyond first order, the prescription must be more carefully defined since we have multiple ${\cal O}$ insertions. We must ensure that we define these insertions in such a way that we are able to deal with potential singularities when different ${\cal O}$ insertions coincide. The details of our prescription for this are given in Appendix \ref{A: Sen}, and we show that this does indeed give the expected deformation for $\partial X$.

We note that the natural objects with which to describe the deformation in ${\cal M}$ are, as anticipated, non-local objects - the modes - rather than local fields on $\Sigma$.

\section{Exploring the Space of Toroidal Backgrounds}\label{S: toroidal backgrounds}

We begin by discussing the space of CFTs on toroidal backgrounds and parallel transport around this space generated by marginal deformations. The space of such backgrounds, modulo symmetries, is
\begin{equation}\label{M}
    {\cal M}=O(d)\times O(d)\backslash O(d,d)/ O(d,d;\Z),    
\end{equation}
and provides a helpful toy example. For such cases, the worldsheet theories are free and such theories admit a useful characterisation in terms of \textit{universal coordinates} \cite{Kugo:1992md}. Universal coordinates provide a convenient way to efficiently describe one theory in the basis of another without the need to explicitly evaluate the effect of marginal deformations on fields. This appears to be a feature of free theories and does not appear to generalise in any simple way\footnote{There is a sense in which one can still work with universal coordinates in the adiabatic limit \cite{Mahmood:2020mtq}.}.

We start by looking at operator deformations in toroidal backgrounds. We will give the basic setup first and then explore the circle case in detail, before looking at the general dimensional case. 

On the cylinder, for a given metric $g$ and $B$-field $B$, we have the mode expansions
\begin{align}
&X^\mu(\sigma,\tau)=x^\mu+\omega^\mu\sigma+\tau g^{\mu\nu}(p_\nu-B_{\nu\rho}\omega^\rho)+\frac{i}{\sqrt{2}}\sum_{n\neq 0}\frac{1}{n}\left(\alpha_n^\mu e^{-in(\tau-\sigma)}+\bar{\alpha}_n^\mu e^{-in(\tau+\sigma)}\right),\label{XP}\\
&\Pi_\mu(\sigma, \tau) \equiv 2\pi P_\mu(\sigma,\tau) = p_\mu + \frac{1}{\sqrt{2}}\sum_{n\neq 0}\left(E^T_{\mu\nu}\alpha_n^\nu e^{-in(\tau-\sigma)}+E_{\mu\nu}\bar{\alpha}_n^\nu e^{-in(\tau+\sigma)}\right),
\end{align}
where $P_\mu = \frac{1}{2\pi}(g_{\mu\nu} \Dot{X}^\nu+B_{\mu\nu}X'^\nu)$ is the conjugate momentum ($\Dot{X}^\mu=\partial_\tau X^\mu, X'^\mu=\partial_\sigma X^\mu$) and $E=g+B$ uniquely specifies $g,B$. Using these, we have that 
\begin{equation}
    \partial X_\mu(\sigma^-) = \frac{1}{2}\Big(\Pi_\mu(\sigma,\tau) - E_{\mu\nu}X'^{\nu}(\sigma,\tau)\Big), \qquad \bar{\partial}X_\mu(\sigma^+) = \frac{1}{2}\Big(\Pi_\mu(\sigma,\tau)+E_{\mu\nu}^TX'^\nu(\sigma,\tau)\Big),
\end{equation}
where on the cylinder we define $\partial = \frac{1}{2}(\partial_\tau - \partial_\sigma)$ and $\bar{\partial} = \frac{1}{2}(\partial_\tau + \partial_\sigma)$. However, we will be computing correlators on the plane, and therefore we will need to change to Euclidean conventions, i.e. $\tau\rightarrow -i\tau$. This changes the mode expansions to 
\begin{align}
&X^\mu(z,\bar{z})=x^\mu-\frac{i}{\sqrt{2}}\Big(\alpha_0^\mu\log(z)+\bar{\alpha}_0^\mu\log(\bar{z})\Big)+\frac{i}{\sqrt{2}}\sum_{n\neq 0}\frac{1}{n}\left(\alpha_n^\mu z^{-n}+\bar{\alpha}_n^\mu\bar{z}^{-n}\right),\\
&\Pi_\mu(z,\bar{z}) \equiv 2\pi P_\mu(z,\Bar{z}) = p_\mu + \frac{1}{\sqrt{2}}\sum_{n\neq 0}\left(E^T_{\mu\nu}\alpha_n^\nu z^{-n}+E_{\mu\nu}\bar{\alpha}_n^\nu \bar{z}^{-n}\right),
\end{align}
where $z=e^{\tau-i\sigma}$ and
\begin{equation}
    g_{\mu\nu}\alpha^\nu_0=\frac{1}{\sqrt{2}}\Big(p_\mu-E_{\mu\nu}\omega^\nu\Big),      \qquad  g_{\mu\nu}\bar{\alpha}^\nu_0=\frac{1}{\sqrt{2}}\Big(p_\mu+E^T_{\mu\nu}\omega^\nu\Big).
\end{equation}
Then, we have
\begin{equation}
    z\partial X_\mu(z) = -\frac{i}{2}\left(\Pi_\mu(z,\bar{z}) - E_{\mu\nu}X'^\nu(z,\bar{z})\right), \qquad \bar{z}\bar{\partial}X_\mu(\bar{z}) = -\frac{i}{2}\left(\Pi_\mu(z,\bar{z})+E^T_{\mu\nu}X'^\nu(z,\bar{z})\right),  
\end{equation}
where now $\partial = \partial_z$, $\bar{\partial} = \partial_{\bar{z}}$. Note the extra $z,\bar{z}$ factors. These play an important role in relating $\partial X_\mu$ at different backgrounds. 

For fixed $\tau$ (which we usually set to zero), we think of $X(\sigma),\Pi(\sigma)$ as being \textit{universal} \cite{Kugo:1992md, Mahmood:2020mtq}, i.e. they are independent of the background $E$.\footnote{Universality only holds for fixed $\tau$ because it is a consequence of the equal-time commutation relations being background independent.} It is the $\alpha_n(E)$ which are background dependent. On the plane, fixed $\tau$ corresponds to fixed radius and we usually set $|z|=1$. 

\subsection{Deformations using universal coordinates}

In what follows, we will mostly be interested in deriving the transformation of the operator $\partial X_\mu$, since this is the simplest operator to look at and it is used to construct the stress tensor. Before describing the operator approach, we will review the derivation of the $\partial X_\mu$ transformation using universal coordinates derived in \cite{Evans:1995su}, and then show that we recover this using the parallel transport method.

We will first look at the case of a circle target space of radius $R$ being deformed to a circle of radius $R+\delta R$. Using the expressions above, we have
\begin{equation}\label{univ}
    z\partial X(R)(z) = -\frac{i}{2}\Big(\Pi(z,\Bar{z}) - R^2X'(z,\Bar{z})\Big), 
\qquad
    \bar{z}\bar{\partial} X(R)(\bar{z}) = -\frac{i}{2}\Big(\Pi(z,\bar{z}) + R^2X'(z,\bar{z})\Big).
\end{equation}
Note that these relations hold for all $z$ and are not a consequence of universal coordinates. If we now restrict to $|z|=1$, so that $\Pi, X'$ are universal, we can use \eqref{univ} to write $\partial X(R+\delta R)$ in terms of $\partial X(R), \bar{\partial}X(R)$ by rearranging the equations. Doing so, we obtain\footnote{We have removed the explicit $z,\Bar{z}$ dependence to keep these equations uncluttered.}
\begin{align}
    &\partial X(R+\delta R)= \partial X(R) + \frac{\delta R}{R}\left(\partial X(R) - \frac{\bar{z}}{z}\bar{\partial}X(R)\right)  +\frac{1}{2}\left(\frac{\delta R}{R}\right)^2 \left(\partial X(R) - \frac{\bar{z}}{z}\bar{\partial}X(R)\right),  \label{dXcircle} \\
    &\bar{\partial} X(R+\delta R) = \bar{\partial} X(R) + \frac{\delta R}{R}\left(\bar{\partial} X(R) - \frac{z}{\bar{z}}\partial X(R)\right)  +\frac{1}{2}\left(\frac{\delta R}{R}\right)^2 \left(\bar{\partial} X(R) - \frac{z}{\bar{z}}\partial X(R)\right).
\end{align}
From this, we can also find the mode transformations using the mode expansion
\begin{equation}\label{dXexpansion}
    \partial X(R)(z) = -\frac{iR^2}{\sqrt{2}}\sum_n\alpha_nz^{-n-1}.
\end{equation}
Doing so, we obtain
\begin{equation}
    \alpha_n(R+\delta R) = \alpha_n(R) - \frac{\delta R}{R}(\alpha_n(R)+\bar{\alpha}_{-n}(R)) + \frac{3}{2}\left( \frac{\delta R}{R}\right)^2(\alpha_n(R) + \bar{\alpha}_{-n})+...,
\end{equation}
\begin{equation}
    \bar{\alpha}_n(R+\delta R) = \bar{\alpha}_n(R) - \frac{\delta R}{R}(\bar{\alpha}_n(R)+\alpha_{-n}(R)) + \frac{3}{2}\left( \frac{\delta R}{R}\right)^2(\bar{\alpha}_n(R) + \alpha_{-n})+...,
\end{equation}
where the $...$ indicates higher order terms arising from the expansion of the inverse metric $(R+\delta R)^{-2}$. In this relatively simple $d=1$ case, we can easily obtain the finite transformation as
\begin{equation}\label{UniversalA}
    \alpha_n(R+\delta R) = \alpha_n(R) - \frac{\lambda}{2(1+\lambda)}(\alpha_n(R)+\bar{\alpha}_{-n}(R)),
\end{equation}
where $\lambda = g^{-1}\delta g= (2R\delta R+ \delta R^2)/R^2$. Note that, in this finite case, $\delta R$ need not be a small deformation of $R$. We can also do the same for the higher dimensional case, i.e. for the toroidal target space. The calculation is very similar, so we will just state the results \cite{Evans:1995su}. We have
\begin{align}
    \partial X_\mu(E')(z) &= \frac{1}{2}g^{\nu\rho}\Big((E'_{\mu\nu}+E^T_{\mu\nu})\partial X_\rho(E)(z)+(-E'_{\mu\nu}+E_{\mu\nu})\frac{\bar{z}}{z}\bar{\partial} X_\rho(E)(\Bar{z})\Big), \label{dXtorus}\\
    \bar{\partial} X_\mu(E')(\Bar{z}) &= \frac{1}{2}g^{\nu\rho}\Big((-E'^T_{\mu\nu}+E^T_{\mu\nu})\frac{z}{\bar{z}}\partial X_\rho(E)(z)+(E'^T_{\mu\nu}+E_{\mu\nu})\bar{\partial} X_\rho(E)(\Bar{z})\Big), \label{dbarXtorus}
\end{align}
and the mode transformations are
\begin{equation}
2g'_{\mu\nu}\alpha^\nu_n(E')=\big(E_{\mu\nu}^T+E'_{\mu\nu}\big)\alpha^\nu_n(E)+\big(E_{\mu\nu}-E'_{\mu\nu}\big)\bar{\alpha}^\nu_{-n}(E),
\end{equation}
\begin{equation}
2g'_{\mu\nu}\bar{\alpha}^\nu_n(E')=\big(E_{\mu\nu}^T-{E'}_{\mu\nu}^T\big)\alpha^\nu_{-n}(E)+\big(E_{\mu\nu}+{E'}_{\mu\nu}^T\big)\bar{\alpha}^\nu_{n}(E).
\end{equation}
We could then rearrange this to get $\alpha_n(E')$ on its own by expanding $g'$ perturbatively around $g$. These results may be recovered using the connection formalism and the $\cal O$ and $\delta_{E}$ operators written down previously, which we now do.

\subsection{$d=1$: Circle deformations by parallel transport}

To compute the deformation, we first need to write down the action and in particular the marginal operator.  

\subsubsection{The action \& deformation}

The action for the flat torus is\footnote{We use conventions where $d^2z = dz\wedge d\Bar{z}/2\pi i$ and $\alpha'$ has been absorbed into $X$ and the background, so that both the coordinates and the background tensors are dimensionless.}
\begin{equation}
    S = \int_{\Sigma}E_{\mu\nu}\partial X^\mu\Bar{\partial}X^\nu,
\end{equation}
and so under a shift $E_{\mu\nu}\rightarrow (E+\delta E)_{\mu\nu}$, the marginal operator is
\begin{equation}
    \delta E_{\mu\nu}\int_{\Sigma}\partial X^\mu\Bar{\partial}X^\nu.
\end{equation}
For the circle of radius $R \rightarrow R+\delta R$, the marginal operator required is thus
\begin{equation}
    {\cal O} = \frac{2\delta R}{R^3}\int\limits_{\Sigma} d^2z\partial X(z)\bar{\partial}X(\bar{z}).
\end{equation}
Unlike the stress tensor deformation, this is not the full story and this is where the spacetime tensor structure of the operator of interest is important. The full deformation of the operator is given by the sum of two parts: a part coming from the marginal operator ($\cal O$) and a part coming from the deformation of the background, which for the circle is given by
\begin{equation}
    {\delta_{E}} = \delta R \frac{\partial}{\partial R}.
\end{equation}
The full deformation operator is given by the path ordered exponential
\begin{equation}
    P\exp\left(\int\limits_0^s ds' \frac{d\mathfrak{m}}{ds'}\frac{1}{R^4}\int\limits_{\Sigma} d^2z \partial X(z)\bar{\partial}X(\Bar{z})+\int\limits_0^s ds' \frac{d\mathfrak{m}}{ds'}\frac{1}{2R} \frac{\partial}{\partial R}\right),
\end{equation}
where $\mathfrak{m}$ is our parameterisation of moduli space. As discussed, we are free to choose this parameterisation, but the one which we will use here is $\mathfrak{m} = r^2$ for a circle of radius r, i.e. the metric. Thus, if our path is between the radii $R$ and $R'=R+\delta R$, parameterised by $s'\in [0,R'-R]$, so that $s'=r-R$, where $r$ is the radius at $s'$, we have
\begin{equation}
    \int\limits_0^s ds_1 ... \int\limits_0^{s_{n-1}}ds_{n} \frac{d\mathfrak{m}}{ds_1}...\frac{d\mathfrak{m}}{ds_{n}} = \frac{1}{n!}(R'^2-R^2)^n,
\end{equation}
so that our deformation operator is 
\begin{equation}\label{circleOP}
    P\exp\left((2R\delta R+\delta R^2)\left(\frac{1}{R^4}\int\limits_{\Sigma} d^2z \partial X(z)\bar{\partial}X(\Bar{z})+\frac{1}{2R} \frac{\partial}{\partial R}\right)\right).
\end{equation}
We now compute the transformation of $\partial X$ to first order in $\delta R$. The higher order case will be discussed properly in section \ref{SSS: Higher orders}; there are subtleties about how we define multiple $\cal O$ insertions due to potential divergences when different $\cal O$ coincide. 

\subsubsection{$\partial X$ deformation using parallel transport in ${\cal M}$}

We now show how \eqref{dXcircle} can be recovered from the parallel transport construction outlined in the previous section. We are interested in the correlator $\big\langle\partial X(w)\Phi(0)\big\rangle$, where $\Phi$ is a generic operator (inserted at $z=0$ for convenience). We will use this correlation function to deduce the deformation of $\partial X(w)$. From the discussion that led to \eqref{ODef}, we have, to first order, 
\begin{equation}
    \big\langle\partial X(w)\Phi(0)\big\rangle' = \big\langle\partial X(w)\Phi(0)\big\rangle + \big\langle {\cal O}\partial X(w)\Phi(0)\big\rangle+\big\langle {\delta_{E}}\partial X(w)\Phi(0)\big\rangle.
\end{equation}
First consider the action of $\cal O$. We have
\begin{align}
    \big\langle\partial &X(w)\Phi(0)\big\rangle' \notag \\
    &= \big\langle\partial X(w)\Phi(0)\big\rangle + \frac{2\delta R}{R^3} \int\limits_{\Sigma^\epsilon} d^2z \big\langle\partial X(z)\bar{\partial}X(\bar{z})\partial X(w)\Phi(0)\big\rangle +\big\langle {\delta_{E}}\partial X(w)\Phi(0)\big\rangle\notag\\
    &= -\frac{iR^2}{\sqrt{2}}\sum_{i= 1}^N\sum_{n\geq 0}(w-z_i)^{-n-1}\left(\big\langle\alpha_n\Phi(0)\big\rangle+\frac{2\delta R}{R^3}\big\langle\alpha_n\Phi(0)\int\limits_{\Sigma^\epsilon} d^2z\partial X(z)\bar{\partial}X(\bar{z})\big\rangle\right) \notag\\
    &-\frac{2\delta R}{R} \int\limits_{\Sigma^\epsilon} d^2z \frac{1}{(w-z)^2}\big\langle\bar{\partial}X(\bar{z})\Phi(0)\big\rangle +\big\langle {\delta_{E}}\partial X(w)\Phi(0)\big\rangle \notag\\
    &=-\frac{iR^2}{\sqrt{2}}\sum_{i,n}(w-z_i)^{-n-1}\big\langle\alpha_n\Phi(0)\big\rangle' \notag \\
    &+ \frac{\delta R}{R}\int\limits_{\Sigma^\epsilon} d^2z \partial_z\left(\frac{1}{z-w}\right)\big\langle\bar{\partial}X(\bar{z})\Phi(0)\big\rangle +\big\langle ({\delta_{E}}\partial X(w))\Phi(0)\big\rangle, \quad \label{dXCalc2}
\end{align}
where the action of $\delta_{E}$ on $\Phi$ has been absorbed into the first term in \eqref{dXCalc2}, and we recall the definition of $\Sigma^\epsilon$ from \eqref{sigma'}. Comparing this first term with the LHS, we see that it is simply the zeroth order term, so we will focus on the second term. We can write it as\footnote{We define the contour integral so that the $2\pi i$ factor is absorbed into the integral, so that  $$\oint\limits_{|z|=\text{const.}}\frac{dz}{z}=-\oint\limits_{|z|=\text{const.}}\frac{d\bar{z}}{\bar{z}}=1.$$}
\begin{equation}
    \frac{\delta R}{R}\int\limits_{\Sigma^\epsilon} d^2z \partial_z\left(\frac{1}{z-w}\right)\big\langle\bar{\partial}X(\bar{z})\Phi(0)\big\rangle = -\frac{\delta R}{R}\oint\limits_{C^\epsilon} \frac{d\bar{z}}{z-w}\big\langle\bar{\partial}X(\bar{z})\Phi(0)\big\rangle,\label{firstOrderdXIntegrals} 
\end{equation}
where the contour $C^\epsilon$ simply consists of circles around $z=0,w$ of radius $\epsilon$. Around $w$ there is no contribution since there are no negative powers of $\bar{z}-\bar{w}$. 

Around $z=0$, we use 
\begin{equation}
    (w-z)^{-1} = \sum_{n\geq 0}(w-z_i)^{-n-1}(z-z_i)^n
\end{equation}
with $z_i=0$ to expand the integrand around zero to get
\begin{equation}
    \frac{\delta R}{R}\sum_{i,n}w^{-n-1}\oint\limits_{C_0^\epsilon} d\bar{z} z^n\left\langle\bar{\partial}X(\bar{z})\Phi(0)\right\rangle.
\end{equation}
Comparing this to what we have on the left hand side, as in  \cite{Sen:1990hh}, gives
\begin{align}
   \big\langle\partial X(w)\Phi(0)\big\rangle' = &\sum_{n\geq 0}w^{-n-1}\left(\frac{-i}{\sqrt{2}}(R^2+2R\delta R)\right)\big\langle\alpha_n'\Phi(0)\big\rangle \notag
   \\
   =&\sum_{n\geq 0}w^{-n-1}\Big\langle\left(\frac{-iR^2}{\sqrt{2}}\right)\alpha_n\Phi(0)+\frac{\delta R}{R}\oint\limits_{C_0^\epsilon} d\bar{z} z^{n}\bar{\partial}X(\bar{z})\Phi(0)\Big\rangle +\big\langle ({\delta_{E}}\partial X(w))\Phi(0)\big\rangle\notag \\
   =&\sum_{n\geq 0}w^{-n-1}\left\langle\Bigg[\left(\frac{-iR^2}{\sqrt{2}}\right)\alpha_n+\frac{\delta R}{R}\oint\limits_{C_0^\epsilon} d\bar{z} z^{n}\bar{\partial}X(\bar{z})+\frac{\delta R}{R} \partial X(w)\Bigg]\Phi(0)\right\rangle,\label{firstOrderdX}
\end{align}
where ${\delta_{E}}\partial X(R) = R^{-1}\delta R\,\partial X(R)$ has been used. This can be seen by noting that ${\delta_{E}}$ is a derivative with respect to the einbein which acts trivially on spacetime scalars. In one dimension, the einbein is simply $R$. We can rearrange \eqref{firstOrderdX} to read off
\begin{equation}
    \delta \alpha_n = -\frac{\delta R}{R}\left(\alpha_n+\bar{\alpha}_{-n}\epsilon^{2n}\right).\label{alpha}
\end{equation}
Note that at this point we would usually make a specific choice of connection, for example by taking the limit $\epsilon\rightarrow 0$ and dropping divergent terms, or setting $\epsilon=1$. In this case, in order to make contact with the results of \cite{Evans:1995su}, we set $\epsilon=1$ (the $\hat{\Gamma}$ connection), which recovers their result. This may also be written as a first order shift of $\partial X$ as 
\begin{equation}
    \partial X(R+\delta R)(z) = \partial X(R)(z)+\frac{\delta R}{R}\left(\partial X(R)(z) - \frac{\bar{z}}{z}\bar{\partial}X(R)(\Bar{z})\right),
\end{equation}
where this expression is taken to hold on the contour $|z|=1$\footnote{Note that this expression for $\partial X(R+\delta R)$ in general requires that $|z|=\epsilon$ to get the correct $\Bar{\partial}X(R)$ term. As discussed earlier, the choice of $\epsilon=1$ for the $\hat{\Gamma}$ connection is arbitrary for a CFT and we choose $|z|=1$ to match up with this choice. }. Note that the results of \cite{Evans:1995su} relied only on the existence of universal coordinates, which suggests a relationship between universal coordinates and the $\hat{\Gamma}$ connection. Indeed, we find that this is the case and we show how this works shortly. First, we extend the analysis of \cite{Sen:1990hh} and recover the result of \cite{Evans:1995su} to all orders in $\delta R$.

\subsubsection{Higher orders}\label{SSS: Higher orders}

For $\partial X$, we can simply obtain the deformation to all orders in $\delta R$ by replacing $2R\delta R \rightarrow \delta g = 2R\delta R + \delta R^2$. If we then work to first order in $\delta g$ instead of $\delta R$, we will obtain the same results as above, but with $2R\delta R$ replaced by $\delta g$:
\begin{equation}
    \delta \partial X(z) = \frac{1}{2}\delta g\left(\partial X(z) - \frac{\Bar{z}}{z}\Bar{\partial}X(\Bar{z})\right),
\end{equation}
for a background deformation $g\rightarrow g+\delta g$. This is in fact the full deformation, since we already know that the finite transformation is given by \eqref{dXtorus}, and indeed we see that this is only first order in the metric deformation. 

However, the story for the mode transformation is different because this transformation does \textit{not} truncate at first order in $\delta g$ (or second order in $\delta R$), but has corrections to all orders, as may be seen by expanding the right hand side of (\ref{UniversalA}) in powers of $\lambda$. This is essentially because the mode deformation involves the inverse metric, the deformation of which involves an infinite expansion in $\delta g$. Therefore, we need a way of computing $\delta_{\cal O}$ and $\delta_{E}$ transformations to all orders.

${\delta_{E}}$ has a straightforward action as a local transformation, so there is no difficulty in computing higher order variations $(\delta_{E})^n$ simply as repeated applications of $\delta_E$. Higher order ${\cal O}$ insertions require more care. In order to generalise the path integral derivation of the first order stress tensor deformation in section \ref{S: Deformations and connections} to higher orders, a prescription is needed to specify how to treat the otherwise ambiguous insertions of higher powers of ${\cal O}$. The prescription for computing higher orders is given by
\begin{equation}
  {\cal O}^n\partial X(w)=  \int\limits_{\Sigma_n}d^2z_n...\int\limits_{\Sigma_1}d^2z_1O(z_n,\Bar{z}_n)...O(z_1,\Bar{z}_1) \partial X(w),
\end{equation}
where
\begin{equation}\label{domain}
    \Sigma_i = \{ z_i \in \mathbb{C} | \quad |z_i|\geq \epsilon, |z_i-w|>0, |z_i-z_j|\geq \epsilon \quad \forall j>i  \}, 
\end{equation}
where the order of the integral signs denotes the order in which the integrations are done, i.e. we remove discs around all punctures corresponding to $O(z_j,\bar{z}_j)$ insertions which have not yet been integrated out. For example, in the above integral we would compute the $z_1$ integral first, and therefore we would need to remove discs around $z_2, ..., z_n$. For the $z_2$ integral, since $z_1$ has already been integrated out, we now only need to remove discs around $z_3, ..., z_n$, and so on.

The evaluation of the resulting deformation is potentially  complicated. However, we claim\footnote{Further details and a justification of this prescription are given in Appendix \ref{A: Sen}.} that the only contraction that gives a finite contribution at order $\delta g^n$ is the one where the contractions are taken by order of integration, i.e. schematically, 
\begin{equation}
    \int\limits_{\Sigma_n}...\int\limits_{\Sigma_1} [O_n[O_{n-1}[...[O_2[O_1,\partial X]]...]],
\end{equation}
i.e. we first contract $\partial X$ with $O_1$, then contract the result with $O_2$, and so on until $O_n$ has been contracted. Here, we use commutator notation to avoid the confusing notation of multiple Wick contractions. Since a contraction of $O_i$ with $\partial X$ or $\bar{\partial}X$ leads to a $\bar{\partial}X$ or $\partial X$ respectively, this prescription is unambiguous. For other deformation operators, such as those we shall consider in section \ref{S: Nonlinear Sigma Models and Off-shell string theory}, matters are more complicated. This prescription is verified explicitly to second order in Appendix \ref{A: Sen}. In terms of the modes, this gives a justification for being able to apply $\cal O$ sequentially to $\alpha_n$, which is how we will compute the higher order corrections to $\delta\alpha_n$. With this justification, we now come to the calculation itself. 

Working with $\delta g$ instead of $\delta R$, we have seen that 
\begin{equation}
    \delta_{\cal O}(R^2\alpha_n) = \frac{i}{\sqrt{2}}\delta g g^{-1} \oint\limits_{C_0^\epsilon} d\Bar{z}z^n \Bar{\partial}X(\Bar{z}) =-\frac{1}{2}\epsilon^{2n}(2R\delta R+\delta R^2)\bar{\alpha}_{-n} = -\frac{1}{2}\epsilon^{2n}\delta g \bar{\alpha}_{-n},
\end{equation}
which we can simply divide by $g=R^2$, since the metric is unaffected by $\cal O$, to get
\begin{equation}
    \delta_{\cal O}\alpha_n = -\frac{1}{2}\epsilon^{2n}\lambda \bar{\alpha}_{-n}.
\end{equation} 
where we introduce $\lambda:=g^{-1}\delta g$. We note that ${\delta_{E}}\, g=\delta g$ and ${\delta_{E}}\delta g=0$, and so ${\delta_{E}}\lambda=-\lambda^2$. Therefore, acting with ${\delta_{E}}$ on the mode gives
\begin{equation}
    \delta_{E}(R^2\alpha_n) = \left(R\delta R + \frac{1}{2}\delta R^2\right)\alpha_n = \frac{1}{2}\delta g\alpha_n,
\end{equation}
which we rearrange to get
\begin{equation}
    \delta_{E}\alpha_n = -\frac{1}{2}\lambda\alpha_n.
\end{equation}
Thus,
\begin{equation}
    (\delta_{\cal O}+\delta_{E})\alpha_n = -\frac{1}{2}\lambda\Big(\alpha_n+\epsilon^{2n}\bar{\alpha}_{-n}\Big).
\end{equation}
This fist order result, coupled with the prescription described above, allows us to systematise the calculation of higher order terms. We can thus iterate these results to obtain the transformation to all orders.  We will work with $\delta g$ here because it is simpler than working with $\delta R$, but it is easy to switch between the two (the two are related by a reparameterisation of the path  on ${\cal M}$ connecting the initial and final backgrounds). The advantage of working with $\delta g$ is that we can say ${\cal O},{\delta_{E}}$ are first order in $\delta g$, and the $n$th order operator insertion is simply $\frac{1}{n!}({\cal O}+\delta_{E})^n$, as we can see from the path ordered exponential \eqref{circleOP}.

It is easy to show that\footnote{This may be shown via induction:
\begin{equation}
    \frac{1}{(m+1)!}(\delta_{\cal O} + \delta_{E})^{m+1}\alpha_n = \frac{1}{m+1}\frac{(-1)^m}{2}\lambda^m(\delta_{\cal O}+\delta_{E})\left(\alpha_n+\epsilon^{2n}\bar{\alpha}_{-n}\right).
\end{equation}
Then, using ${\delta_{E}}(g^{-1}\delta g)^m = m(g^{-1}\delta g)^{m-1}{\delta_{E}}(g^{-1}\delta g)$, we get to the desired result, i.e. 
\begin{equation}
    \frac{1}{(m+1)!}(\delta_{\cal O} + \delta_{E})^{m+1}\alpha_n =(-1)^{(m+1)}\frac{1}{2}(g^{-1}\delta g)^{m+1}(\alpha_n + \epsilon^{2n} \bar{\alpha_{-n}}),
\end{equation}
as required. $\square$}
\begin{equation}
    \frac{1}{m!}\Big(\delta_{\cal O}+\delta_{E}\Big)^m\alpha_n = (-1)^m\frac{1}{2}\lambda^m(\alpha_n + \epsilon^{2n} \bar{\alpha_{-n}}).
\end{equation}
Summing over $m$ recovers the all orders result (\ref{UniversalA}) found using the universal coordinate method. From \eqref{dXexpansion}, we can also recover the transformation of $\partial X$, evaluated on the contour $|z|=\epsilon$, to all orders 
\begin{equation}
    \partial X(R+\delta R)(z) = \partial X(R)(z)+\delta g g^{-1}\left(\partial X(R)(z) - \frac{\bar{z}}{z}\bar{\partial}X(R)(\Bar{z})\right),
\end{equation}
which indeed truncates at second order (since the transformation of $R^2\alpha_n$ truncates at second order) and agrees with \cite{Evans:1995su}.

\subsection{Interlude: connections and universal coordinates}

It is interesting to consider the relationship between universal coordinates and choice of connection. As we shall see, $X'$ and $\Pi$ only remain universal under parallel transport in ${\cal M}$ (given by (\ref{M})) with connection $\hat{\Gamma}$. For a fixed background, we have $X'(z,\Bar{z}) = \frac{i}{R^2}(z\partial X(z)-\Bar{z}\bar{\partial}X(\Bar{z}))$ and $\Pi(z,\Bar{z}) = i(z\partial X(z)+\Bar{z}\bar{\partial}X(\Bar{z}))$. In terms of modes,
\begin{equation}
    X'(z,\bar{z}) = \frac{1}{\sqrt{2}}\sum_n(\alpha_nz^{-n}-\bar{\alpha}_n\bar{z}^{-n}),
\qquad
     \Pi(z,\bar{z})  = \frac{R^2}{\sqrt{2}}\sum_n(\alpha_nz^{-n}+\bar{\alpha}_n\bar{z}^{-n}),
\end{equation}
and so using the results of the previous section we can compute how $X',\Pi$ change to first order. 

\subsubsection*{The $\hat{\Gamma}$ connection}

For this connection, where $\epsilon=1$,
\begin{equation}
    \delta X'(z,\bar{z}) = \frac{1}{\sqrt{2}}\sum_n(\delta\alpha_nz^{-n}-\delta\bar{\alpha}_n\bar{z}^{-n}) =0,
\end{equation}
on the contour $z\bar{z}=1$. The $R^2$-dependence makes the calculation for $\Pi$ only slightly more involved
\begin{equation}
    \Pi(z,\bar{z})+\delta \Pi(z,\bar{z}) = \frac{(R+\delta R)^2}{\sqrt{2}}\sum_n\Big((\alpha_n+\delta\alpha_n)z^{-n}+(\bar{\alpha}_n+\delta\bar{\alpha}_n)\bar{z}^{-n}\Big).
\end{equation}
Substituting the transformations for the modes in, we find that $\delta \Pi =0$. Thus, we see that $X',\Pi$ are indeed universal for the $\hat{\Gamma}$ connection. This is not the case for the $c,\bar{c}$ connections. 

\subsubsection*{The $c$ connection}

For the $c$ connection, we integrate up to $\epsilon=1$ and then, for those operator coefficients in the OPE for which the integral gives a finite result, we take the limit $\epsilon \rightarrow 0$. In this case at hand, the only OPE which may have potential singularities that need subtracting is the OPE between the marginal operator and $\partial X(w)$, which gives
\begin{equation}\label{c_conn}
    \oint\limits_{C_w^1 - C_w^\epsilon}\frac{d\bar{z}}{z-w}\Big( \bar{\partial}X(\bar{w}) + (\bar{z} - \bar{w})\bar{\partial}^2X(\bar{w})+... \Big),
\end{equation}
where $C_w^1$ and $C_w^{\epsilon}$ are circles around $z=w$ of radius $1$ and $\epsilon$ respectively. Evaluating (\ref{c_conn}) gives zero since there are no negative powers of $(\Bar{z}-\Bar{w})$, so there are in fact no divergences to subtract. This is also evident in the fact that \eqref{alpha} is finite in the limit $\epsilon\rightarrow0$, since we are only summing over $n\geq 0$. We see in this limit that all of the terms vanish apart from $n=0$. Extending the result to $n<0$, the transformation of the modes for the $c$ connection is given by
$$
    \delta \alpha_n = -\frac{\delta R}{R}\alpha_n,  \qquad
    \delta \bar{\alpha}_n = -\frac{\delta R}{R}\bar{\alpha}_n, \qquad n\neq 0,
$$
\begin{equation}
    \delta \alpha_0 =\delta \bar{\alpha}_0= -\frac{\delta R}{R}(\alpha_0 + \bar{\alpha}_0).
\end{equation}
In this case, after some brief calculation, we find that
\begin{equation}
    \delta X'(z,\bar{z}) = -\frac{\delta R}{R\sqrt{2}}\sum_{n\neq 0}(\alpha_nz^{-n} - \bar{\alpha}_n\bar{z}^{-n}),
\qquad
    \delta \Pi(z,\bar{z}) =  \frac{R\delta R}{\sqrt{2}}\sum_{n\neq 0}(\alpha_nz^{-n} + \bar{\alpha}_n\bar{z}^{-n}),
\end{equation}
i.e. they are not universal with respect to the $c$ connection. Note that the variation of the zero modes still cancel out in both cases. In fact, what we notice is that the transformation of $\alpha_0, \bar{\alpha}_0$ is independent of the $\epsilon$-dependent part of the connection. We should perhaps have expected this since we recall that $\alpha_0, \Bar{\alpha}_0$ commute with $L_0^+$, the generator of dilations. Since changes in the radii of the disks, $\epsilon$, correspond to dilations, we would expect $\alpha_0, \Bar{\alpha}_0$ to be independent of $\epsilon$. 

\subsubsection*{The $\bar{c}$ connection}
In this case, since there are no divergences to subtract, the $c$ and $\bar{c}$ connections are actually the same, so $X', \Pi$ are only universal for the $\hat{\Gamma}$ connection, as suggested by our calculations earlier. We shall see in section \ref{S: Nonlinear Sigma Models and Off-shell string theory} that, for more general non-CFT deformations, the $X^{\mu}$ will not be universal even for the $\hat{\Gamma}$ connection.

We now look at the higher dimensional analogue of the previous section, i.e. when we have a toroidal target space. In some ways things are clearer in this more general case since the vielbein structure is more explicit and thus it is easier to see the contrast between the $\cal O$ and $\delta_{E}$ transformations and why we need both of them to get the full transformation.

\subsection{$d>1$: torus deformations by parallel transport} \label{SS: general dimensions}

For $d>1$, the above discussion generalises straightforwardly. The only new feature is the possible presence of a constant $B$-field. Due to a subtlety in the ${\delta_{E}}$ transformation, it will be helpful to distinguish between the $B=0$ and $B\neq 0$ cases.

As with the $d=1$ case, we can choose a parameterisation of our path along which we deform. We have deformation operator
\begin{equation}
    P\exp\left(\int\limits_0^s ds' \frac{dg_{\mu\nu}(s')}{ds'}\left(\int\limits_{\Sigma} d^2z \partial X^\mu(z)\bar{\partial}X^\nu(\Bar{z}) +\frac{\partial}{\partial g_{\mu\nu}}\right)\right),
\end{equation}
where $g(s') = g+\frac{s'}{s}\delta g$, so that
\begin{equation}
    \int\limits_0^s ds_1...\int\limits_0^{s_{n-1}}ds_{n}\frac{dg_{\mu\nu}}{ds_1}...\frac{dg_{\rho\sigma}}{ds_{n}} = \frac{1}{n!}\delta g_{\mu\nu}...\delta g_{\rho\sigma},
\end{equation}
giving deformation operator
\begin{equation}
    P\exp\left(\delta g_{\mu\nu}\left(\int\limits_{\Sigma} d^2z \partial X^\mu(z)\bar{\partial}X^\nu(\Bar{z}) +\frac{\partial}{\partial g_{\mu\nu}}\right)\right).
\end{equation}

\subsubsection{Metric deformations}\label{metricdef}

We start with the simpler case where $B_{\mu\nu}=0$ and the only change is due to the metric. It is useful to introduce vielbeins $e\indices{_\mu^a}$ such that
\begin{equation}\label{ge}
    g_{\mu\nu} = \tensor{e}{_\mu^a}e_{a\nu}.
\end{equation}
The generalisation of the $\delta_{E}$ transformation to higher dimensions is given by
\begin{equation}\label{delta_E}
    {\delta_{E}} = \delta g_{\mu\nu}\frac{\partial}{\partial g_{\mu\nu}},
\end{equation}
and we identify ${\delta_{E}}\tensor{e}{_\mu^a} = \delta \tensor{e}{_\mu^a}$, where $\delta e$ is induced by $\delta g$ (this could also be taken as a definition of ${\delta_{E}}$ in this case). To compute this, we vary \eqref{ge} and rearrange to get to 
\begin{equation}\label{delta_e}
    \delta \tensor{e}{_\mu^a} = \delta g_{\mu\nu}e^{\nu a} - \tensor{e}{_\mu^b}\delta\tensor{e}{_b_\nu}e^{\nu a}.
\end{equation}
Now, if we define $\tensor{U}{_a^b} = \delta\tensor{e}{_a_\nu}e^{\nu b}$,
%then we see that the second term in the vielbein variation looks like a frame transformation except for the fact that $U$ is not antisymmetric, which is required to give a Lorentz transformation. Therefore, we will extract the symmetric part from it and take the anti-symmetric part as the "unphysical" part. Doing so, we find that
%\begin{align}
%   U_{(ab)} =& \delta %e_{(a|\mu}\tensor{e}{^\mu_{|b)}} %\notag\\
 %   =& \tensor{e}{_a^\mu}\delta g_{\mu\nu}\tensor{e}{^\nu_b} - V_{(ab)},
%\end{align}
%where $V_{(ab)} = \tensor{e}{_{(a|}^\mu}\delta e_{\mu|b)}$, i.e. we have 
%\begin{gather}
 %   U_s = \delta e^Te^{-T}+e^{-1}\delta e,\\ V_s = e^{-1}\delta e + \delta e^Te^{-T},
%\end{gather}
%and so $U_s = V_s$. Thus, rearranging the above equation gives
%\begin{equation}
   % U_{ab} = \frac{1}{2}\tensor{e}{_a^\mu}\delta g_{\mu\nu}\tensor{e}{^\nu_b} + \Lambda_{ab}, 
%\end{equation}
%where $\Lambda_{ab} = U_{[ab]}$, and
the variation of the vielbein can be written as
\begin{equation}\label{delta_E2}
    \delta \tensor{e}{_\mu^a} = \frac{1}{2}\delta g_{\mu\nu}e^{\nu a} - \tensor{e}{_\mu^b}\tensor{\Lambda}{_b^a},
\end{equation}
where $\Lambda_{ab} = U_{[ab]}$ is a local frame transformation. The transpose can similarly be written as
\begin{equation}
    \delta \tensor{e}{^a_\mu} = \frac{1}{2}e^{a\nu}\delta g_{\nu\mu} + \tensor{\Lambda}{^a_b}\tensor{e}{^b_\mu},
\end{equation}
where the sign change comes from the fact that $\Lambda$ is antisymmetric\footnote{We can also invert these to get
\begin{gather}
    \delta \tensor{e}{^\mu_a} = -\frac{1}{2} g^{\mu\nu}\delta g_{\nu\rho}\tensor{e}{^\rho_a} - \tensor{e}{^\mu_b}\tensor{\Lambda}{^b_a},    \qquad
    \delta \tensor{e}{_a^\mu} = -\frac{1}{2} \tensor{e}{_a^\rho}\delta g_{\rho\nu}g^{\nu\mu} + \tensor{\Lambda}{_a^b}\tensor{e}{_b^\mu}.\label{delta_e_inv}
\end{gather}}.
Thus, we have (the Lorentz transformation being compensated by a similar transformation of $\partial X_a$)
\begin{equation}
    \delta_{E}\partial X_{\mu}(w)=\frac{1}{2}\delta g_{\mu\nu}\partial X^{\nu}(w).    
\end{equation}

For the $\cal O$ insertion, the deformation operator is
\begin{equation}
    {\cal O}[X]=\delta g_{\mu\nu}\int_{\Sigma}{d}^2z\partial X^{\mu}(z)\bar{\partial}X^{\nu}(\Bar{z}),    
\end{equation}
and, following a calculation similar to the $d=1$ case, we find that
\begin{equation}
    \delta (g_{\mu\nu}\alpha_n^\nu) =  -\frac{1}{2}\delta g_{\mu\nu}\Big(\alpha_n^{\nu}+\epsilon^{2n}\bar{\alpha}^{\nu}_{-n}\Big).
\end{equation}
If we are on a contour of constant $|z|=\epsilon$, we can also deduce the transformation for $\partial X_\mu(z)$ from this, which is
\begin{equation}
    \delta \partial X_\mu(z) = \frac{1}{2}\delta g_{\mu\nu}\left(\partial X^\nu(z) - \frac{\Bar{z}}{z}\Bar{\partial}X^\nu(\Bar{z})\right),
\end{equation}
agreeing with \eqref{dXtorus}.

\subsubsection{Deformations with a constant $B$-field}

The doubled formalism \cite{Tseytlin:1990nb,Hull:2004in} provides an efficient way to generalise \eqref{delta_E2} in the presence of non-vanishing constant $B$-field. This will shed light on how to understand this in the `undoubled' case, which we explain in detail in \ref{SSS: why do we need doubled?}.

The embedding coordinate in the original ($X^{\mu}$) and dual ($\widetilde{X}_\mu$) descriptions are related as \cite{Hull:2009sg, Hull:2004in, Kugo:1992md}
\begin{equation} \label{SDC_1}
    \partial \tilde{X}_\mu = -E^T_{\mu\nu}\partial X^\nu, 
\qquad
    \bar{\partial} \tilde{X}_\mu = E_{\mu\nu}\bar{\partial}X^\nu.
\end{equation}
The doubled formalism combines the two in an $O(d,d)$-covariant doubled coordinate $\mathbbold{X}^I = (X^\mu, \tilde{X}_\mu)$. An $O(d,d)$-invariant metric on the doubled space $L_{IJ}$ is given by \cite{Hull:2009sg}
\begin{equation}
    {d} s^2=\frac{1}{2}L_{IJ}{d}\mathbbold{X}^I {d}\mathbbold{X}^J={d} X^{\mu}{d} \widetilde{X}_{\mu},
\end{equation}
where $\mu = 1, ..., d$. This is the metric that is used to raise/lower indices. Additionally, for a given constant background $E_{\mu\nu}=g_{\mu\nu}+B_{\mu\nu}$, we have the doubled metric
\begin{equation}
    \cH_{IJ} = \begin{pmatrix}
    g_{\mu\nu} - B_{\mu\rho}g^{\rho\lambda}B_{\lambda\nu} & B_{\mu\rho}g^{\rho\nu} \\
    -g^{\mu\rho}B_{\rho\nu} & g^{\mu\nu}
    \end{pmatrix}.
\end{equation}
This combines the metric and $B$-field into a single $O(d,d)$-covariant tensor. Thus, the case with $B$-field may be found by applying the methodology of section \ref{metricdef} to the doubled target space sigma-model
\begin{equation}\label{doubledS}
    S = \frac{1}{4}\int\limits_{\Sigma} \cH_{IJ}d{\mathbbold{X}^I}\wedge \ast d{\mathbbold{X}^J}+\frac{1}{2}\int\limits_{\Sigma}\Omega_{IJ}d{\mathbbold{X}^I}\wedge d{\mathbbold{X}^J},
\end{equation}
where $\Omega_{IJ}$ is a constant antisymmetric tensor that will play no further role\footnote{See \cite{Hull:2004in} and \cite{Hull:2009sg} for a more general discussion of such terms}. The relations \eqref{SDC_1} can be derived from the self-duality constraints\footnote{Where $\ast dz=-dz, \ast d\bar{z}=\bar{d}z$.} \cite{Hull:2009sg, Hull:2004in}
\begin{equation}
    d\mathbbold{X}^I = L^{IJ}(\cH_{JK}\ast d \mathbbold{X}^K).
\end{equation}
A deformation of the target space $\cH_{IJ}\rightarrow \cH_{IJ}+\delta\cH_{IJ}$ is given by the marginal operator
\begin{equation}\label{doubledO}
    {\cal O} = \frac{1}{2}\int\limits_{\Sigma} \delta \cH_{IJ}\partial {\mathbbold{X}^I}\bar{\partial}{\mathbbold{X}^J}.
\end{equation}
Note that $L_{IJ}$ does not change. Using the undoubled OPEs, it is straightforward to verify that 
\begin{equation}\label{doubledOPE}
    \partial {\mathbbold{X}^I}(z)\partial {\mathbbold{X}^J}(w) \sim -\frac{1}{2}\frac{\cH^{IJ}-L^{IJ}}{(z-w)^2}.
\end{equation}
As in the previous section, we can write the doubled metric in terms of doubled vielbeins as
\begin{equation}
    \cH_{IJ} = \tensor{\cV}{_I^A}\delta_{AB}\tensor{\cV}{^B_J},
\end{equation}
where $\delta_{AB}$ is the doubled frame metric (which does \textit{not} raise/lower frame indices, just as $\cH$ does not raise/lower spacetime indices) and
\begin{equation}
    \tensor{\cV}{_I^A} = \begin{pmatrix}
    e^T & Be^{-1} \\ 0 & e^{-1},
    \end{pmatrix}, \qquad
    \tensor{\cV}{^A_I} = \begin{pmatrix}
    e & 0\\ -e^{-T}B & e^{-T}
    \end{pmatrix}.
\end{equation}
We want $\delta\partial {\mathbbold{X}_I} = \delta (\tensor{\cV}{_I^A}\partial {\mathbbold{X}_A})$. Given some deformation of the doubled metric $\delta\cH_{IJ}$, we have the relations
\begin{equation}
    \delta \cH^{IJ} = -\cH^{IK}\delta\cH_{KL}\cH^{LJ} = L^{IK}\delta \cH_{KL}L^{LJ},
\end{equation}
and we can combine these two to obtain the often more useful identity $\cH_{IJ}\delta \cH^{JK} = - \delta \cH_{IJ}\cH^{JK}$. Note also that we can safely define such objects as $\tensor{\delta\cH}{_I^J} = \delta\cH_{IK}L^{KJ}$, since $\delta L=0$. We also have
\begin{equation}\label{VL}
    \delta \tensor{\cV}{_I^A} = \frac{1}{2}\delta \cH_{IJ}\tensor{\cV}{^J_B}\delta^{BA} - \tensor{\cV}{_I^B}\tensor{\Lambda}{_B^A},
\end{equation}
analogous to the undoubled case. Thus, the $\partial {\mathbbold{X}_I}$ transformation is largely the same as in the undoubled case, except for the fact that there are two metrics, each playing a different role. We are interested in the OPE of ${\cal O}$ given by (\ref{doubledO}) and $\partial \mathbbold{X}_I(w)$. Evaluating the OPE using (\ref{doubledOPE}) and comparing coefficients as we did for the circle deformation gives the doubled mode deformation as
\begin{equation}\label{doubleOalpha}
    \delta_{\cal O}(\cH_{IJ}\mathbbold{a}_n^J) = \frac{1}{4}\Big(\delta\cH_{IJ} - \tensor{\cH}{^K_I}\delta\cH_{KJ}\Big)\,\epsilon^{2n}\Bar{\mathbbold{a}}_{-n}^J,
\end{equation}
where we have introduced the mode expansion
\begin{equation}
    \partial \mathbbold{X}_I(z) = \frac{i}{\sqrt{2}}\sum_n\cH_{IJ}\,\mathbbold{a}_n^J\,z^{-n-1},
\end{equation}
and $\mathbbold{a}_n^I = (\alpha_n^\mu, \tilde{\alpha}_{n\mu})$ are the doubled oscillator modes. Using the above identities, we can rewrite \eqref{doubleOalpha} as
\begin{equation}
    \delta_{\cal O}(\cH_{IJ}\mathbbold{a}_n^J) = \frac{1}{2}\delta\cH_{IJ}\epsilon^{2n}\Bar{\mathbbold{a}}_{-n}^J.
\end{equation}
For the $\hat{\Gamma}$ connection with $|z|=1$, we can then deduce the transformation of $\partial \mathbbold{X}_I$ as
\begin{equation}
    \delta_{\cal O}\partial\mathbbold{X}_I(z)=\frac{1}{2}\frac{\bar{z}}{z}\delta \cH_{IJ}\bar{\partial}{\mathbbold{X}^J}(\bar{z}).
\end{equation}
The $\delta_{E}\partial {\mathbbold{X}_I}$ calculation also has a subtlety
\begin{equation}
    \delta_{E}\partial {\mathbbold{X}_I} = \delta \tensor{\cV}{_I^A}\partial {\mathbbold{X}_A} = \frac{1}{2}\delta \cH_{IJ}\tensor{\cV}{^J_B}\delta^{BA} \partial {\mathbbold{X}_A}.
\end{equation}
It is not the case that $\tensor{\cV}{^J_B}\delta^{BA} \partial {\mathbbold{X}_A} = \partial {\mathbbold{X}^J}$, since $\delta^{BA}$ does not raise/lower indices, since it is not the $O(d,d)$ invariant metric $L$. We have
\begin{align}
    \tensor{\cV}{^J_B}\delta^{BA}\partial {\mathbbold{X}_A} = \begin{pmatrix}
    \tensor{e}{^\nu_b}&0 \\ B_{\nu\rho}\tensor{e}{^\rho_b}&\tensor{e}{_\nu^b}
    \end{pmatrix}
    \begin{pmatrix}
    \delta^{ba}&0 \\ 0& \delta_{ba}
    \end{pmatrix}
    \begin{pmatrix}
    \partial \tilde{X}_a \\ \partial X^a
    \end{pmatrix}= -\partial {\mathbbold{X}^J},
\end{align}
where we have used that $\partial \tilde{X}_\nu = -E^T_{\nu\rho} \partial X^\rho$ and $\partial \tilde{X}_a = -\partial X_a$ (since the background for the frame is just $\delta_{ab}$ and this raises/lowers indices in the undoubled case). Thus, overall, we find that 
\begin{equation}
    \delta \partial {\mathbbold{X}_I}(z) = \frac{1}{2}\delta \cH_{IJ}\left(-\partial {\mathbbold{X}^J}(z) + \frac{\bar{z}}{z}\bar{\partial}{\mathbbold{X}^J}(\bar{z})\right),
\end{equation}
the components of which can be checked to recover the results of \cite{Evans:1995su}.

\subsubsection{The utility of the doubled formalism when $B\neq 0$}\label{SSS: why do we need doubled?}

Once the effects of a metric deformation are understood, the doubled formalism provides an efficient way to generalise to include a $B$-field. In particular, the doubled formalism provides a natural way to incorporate the $B$-field deformation into the $\delta_E$ part of the transformation.
%If we tried to include the $B$-field naively in the undoubled case, we would not get the correct result for the vielbein transformation. What is it about the doubled formalism which allows us to get the correct results? For the transformation of $\partial X_\mu$, if we did the usual $\delta_{\cal O}+\delta_{E}$ transformation, the $\delta_{\cal O}$ transformation would give the correct antiholomorphic contribution, but the $\delta_{E}$ transformation would not give the correct holomorphic contribution. Therefore, we will focus on $\delta_{E}$.
In the undoubled case, we have
\begin{equation}
    \delta_{E} \partial X^\mu = \delta \tensor{e}{^\mu_a}\partial X^a.
\end{equation}
This clearly only gives the metric contribution and there does not seem to be any way to introduce the $B$-field. However, we recall that
\begin{equation}
    \delta \tensor{e}{^\mu_a} = -\frac{1}{2}g^{\mu\nu}\delta g_{\nu\rho}\tensor{e}{^\rho_a} - \tensor{e}{^\mu_b}\tensor{\Lambda}{^b_a},
\end{equation}
where, importantly, $\Lambda_{ab}$ is antisymmetric. If we now write this as
\begin{align}
    \delta \tensor{e}{^\mu_a}\partial X^a =& -\frac{1}{2}g^{\mu\nu}\delta g_{\nu\rho}\tensor{e}{^\rho_a} \partial X^a +\frac{1}{2}g^{\mu\nu}\delta B_{\nu\rho}\partial X^\rho - \frac{1}{2}g^{\mu\nu}\delta B_{\nu\rho}\partial X^\rho - \tensor{e}{^\mu_b}\tensor{\Lambda}{^b_a}\partial X^a\notag\\
    = &-\frac{1}{2}g^{\mu\nu}\delta g_{\nu\rho}\tensor{e}{^\rho_a} \partial X^a +\frac{1}{2}g^{\mu\nu}\delta B_{\nu\rho}\partial X^\rho - \tensor{e}{^\mu_b}\tensor{\Lambda}{^{'b}_a}\partial X^a,\label{PBfield}
\end{align}
where we have absorbed the $- \frac{1}{2}g^{\mu\nu}\delta B_{\nu\rho}\partial X^\rho$ term into the new frame transformation $\Lambda'$, we see that we obtain the correct transformation. It is not obvious why we should do this. However, in the doubled formalism, this is exactly what happens, except it occurs naturally due to the $O(d,d)$ structure. If we compute the first term of (\ref{VL}) in components, we find that 
\begin{equation}
    \frac{1}{2}\delta \cH_{IJ}\tensor{\cV}{^J_B}\delta ^{BA} = \frac{1}{2}\begin{pmatrix}
    \delta g_{\mu\nu}e^{\nu a} - B_{\mu\nu}g^{\nu\rho}\delta B_{\rho\sigma}e^{\sigma a} & \delta B_{\mu\nu}\tensor{e}{^\nu_a} - B_{\mu\nu}g^{\nu\rho}\delta g_{\rho\sigma}\tensor{e}{^\sigma_a} \\ -g^{\mu\nu}\delta B_{\nu\rho} e^{\rho a}& -g^{\mu\nu}\delta g_{\nu\rho} \tensor{e}{^\rho_a}
    \end{pmatrix}.
\end{equation}
Since we want $\delta \partial X^\mu$, it is the second row we are interested in (since $\partial \mathbbold{X}_I = (\partial\tilde{X}_\mu, \partial X^\mu)$). We see that the bottom left entry gives us the additional contribution of 
\begin{equation}
    \frac{1}{2}g^{\mu\nu}\delta B_{\nu\rho} e^{\rho a}\partial X_a,
\end{equation}
using $\partial \tilde{X}_a = -\partial X_a$. This is precisely the contribution that we included arbitrarily in \eqref{PBfield}, but now we see that it arises naturally in the doubled formalism. Thus, it seems as though the $O(d,d)$ structure is precisely the `extra information' we need in order to recover \eqref{dXtorus}. This is somewhat unsatisfactory though, since we intuitively expect that we can derive the correct results without having to resort to the doubled formalism. We will see next that this is indeed possible precisely in the case of the $\hat{\Gamma}$ connection, which is the context in which \eqref{dXtorus} is derived in \cite{Evans:1995su}, and that there is a close connection with universal coordinates. 

\subsubsection{Universal coordinates and doubled geometry}

Let us look at the universality of $X'$ in the $\hat{\Gamma}$ connection (we could also look at $\Pi$, but $X'$ is simpler). We have $X'(z,\Bar{z}) = -\frac{1}{2}g^{-1}(z\partial X(z)-z\bar{\partial}X(\Bar{z}))$. Now, let us assume that 
\begin{equation}
    \delta \tensor{e}{_\mu^a} = \frac{1}{2}\delta g_{\mu\nu}g^{\nu\rho}\tensor{e}{_\rho^a}+\gamma \delta B_{\mu\nu}e^{\nu a}, 
\end{equation}
for some c-number $\gamma$. As we saw earlier, there is the potential to have such a $B$-field term since the $B$-field is antisymmetric. From the doubled treatment above, we saw that the extra $B$-field contribution to the ${\delta_{E}}$ transformation did not seem to naturally come from the vielbein transformation, but did arise naturally in the doubled formalism as a requirement of the $O(d,d)$-covariance of the doubled formalism. If we require that all (target space) scalar operators are killed by ${\delta_{E}}$, this specifies the action of ${\delta_{E}}$ on the operators. Thus, we have
\begin{align}
    \delta_{E}\partial X_\mu = \frac{1}{2}\delta g_{\mu\nu}\partial X^\nu + \gamma \delta B_{\mu\nu}\partial X^\nu,   \qquad  \delta_{E}\bar{\partial} X_\mu = \frac{1}{2}\delta g_{\mu\nu}\bar{\partial} X^\nu + \bar{\gamma} \delta B_{\mu\nu}\bar{\partial} X^\nu,
\end{align}
where $\bar{\gamma}$ is \textit{not} the complex conjugate of $\gamma$. Then, substituting this into $X'$, we have
\begin{align}
    \delta X'^\mu(z,\Bar{z}) &= \frac{1}{2}g^{\mu\rho}g^{\sigma\nu}\delta
     g_{\rho\sigma}(z\partial X_\nu(z)- \Bar{z}\bar{\partial}X_\nu(\Bar{z})) - \frac{1}{2}g^{\mu\nu}\left(\frac{1}{2}\delta g_{\nu\rho}z\partial X^\rho(z)+\gamma\delta B_{\nu\rho}z\partial X^\rho(z) \right.\notag \\
     &\left.- \frac{1}{2}\delta E_{\nu\rho}\Bar{z}\bar{\partial}X^\rho(\Bar{z})
     - \frac{1}{2}\delta g_{\nu\rho}\Bar{z}\bar{\partial}X^\rho(\Bar{z})-\bar{\gamma}\delta B_{\nu\rho}\Bar{z}\bar{\partial}X^\rho(\Bar{z}) + \frac{1}{2}\delta E_{\nu\rho}^Tz\partial X^\rho(z))\right)\notag \\
     &= \frac{1}{4}g^{\mu\nu}\delta B_{\nu\rho}\Big(z\partial X^\rho(z)(1-2\gamma) + \Bar{z}\bar{\partial} X^\rho(\Bar{z})(1+2\bar{\gamma})\Big),
\end{align}
and so we require $\gamma=-\bar{\gamma}=\frac{1}{2}$ for $X'$ to be universal. Thus, we see that the condition of universality is sufficient to give the correct $B$-field contributions to recover \eqref{dXtorus}, \eqref{dbarXtorus}. These are also precisely the contributions that arose naturally in the doubled geometry. This is as we would expect because universality and doubled geometry both preserve the $O(d,d)$ structure. With universality, this comes from the fact that the canonical commutation relations are preserved under parallel transport, and with doubled geometry the $O(d,d)$ structure is explicit in its construction. Given that the $\Hat{\Gamma}$ connection seems to precisely correspond to the existence of universal coordinates, we thus conclude that the $\Hat{\Gamma}$ connection preserves the natural $O(d,d)$-covariance of the embedding fields under parallel transport and gives the same deformation results as the doubled geometry. 

It is instructive to look at the general case where we have not yet chosen a connection, i.e. we do not specify a regime for $\epsilon$ (such as keeping $\epsilon$ fixed and finite, or taking a limit $\epsilon\rightarrow 0$ with some counter-terms). We find that
\begin{equation}
    \delta \alpha_n^\mu =  -\frac{1}{2}g^{\mu\nu}\Big((\delta g_{\nu\rho}-2\gamma\delta B_{\nu\rho})\alpha_n^\rho+\epsilon^{2n}\delta E_{\nu\rho}\bar{\alpha}^\rho_{-n}\Big).
\end{equation}
Now, we would like to compare this to the doubled geometry. In the doubled case, we have doubled oscillator modes $\mathbbold{a}_n^I = (\alpha_n^\mu, \tilde{\alpha}_{n \mu})$, and the self-duality constraints give us 
\begin{equation}
    \tilde{\alpha}_{n\mu} = -E_{\mu\nu}^T\alpha_n^\nu, 
\qquad
    \tilde{\bar{\alpha}}_{n\mu} = E_{\mu\nu}\bar{\alpha}_n^\nu. 
\end{equation}
From our earlier calculations, we have
\begin{equation}
    \delta_{\cal O}\mathbbold{a}_n^I = \frac{1}{2}\epsilon^{2n}\delta \cH{^I_J}\bar{\mathbbold{a}}^J_{-n},
    \qquad
    \delta_{E}\mathbbold{a}_n^I = -\frac{1}{2}\delta \cH{^I_J}\mathbbold{a}_n^J,
\end{equation}
where we have used $\delta \tensor{\cV}{^I_A} \sim -\frac{1}{2}\cH^{IJ}\delta \cH_{JK}\tensor{\cV}{^K_A}$. Thus, by taking the appropriate components and using the self-duality constraints, we get
\begin{equation}
    \delta_{E}\alpha_n^\mu = -\frac{1}{2}g^{\mu\nu}(\delta g_{\nu\rho} - \delta B_{\nu\rho})\alpha_n^\rho.
\end{equation}
Thus, we see that, for agreement with the doubled geometry, we need to set $\gamma=\frac{1}{2}$, and this is independent of connection. Similarly, we would find that $\bar{\gamma}=-\frac{1}{2}$ by looking at $\bar{\alpha}_n$.

Thus, in general we require $\gamma=-\bar{\gamma}=\frac{1}{2}$ to preserve the $O(d,d)$ structure, but it is only for the $\hat{\Gamma}$ connection that these values are required, and in particular it is the existence of universal coordinates which fixes them. As discussed earlier, universal coordinates are special to the $\hat{\Gamma}$ connection and in general we do not have this additional structure. For a general connection, there is some freedom to choose how the $B$-field enters into the transformation, but if the $O(d,d)$ structure is in place then the symmetry between the metric and $B$-field removes this freedom.

\section{Nonlinear Sigma Models and Off-shell Deformations}\label{S: Nonlinear Sigma Models and Off-shell string theory}

So far, we have been discussing genuine CFTs only and the transformations have all been fairly tractable. However, we now look at examples where we no longer have conformal invariance, though the backgrounds we will look at can be used as building blocks for honest string backgrounds\footnote{See \cite{Chaemjumrus:2019ipx} for a recent example.}. The difficulty of working with toy models which are not full string theory solutions is that we can no longer rely on worldsheet conformal invariance and we are forced to consider off-shell correlation functions. 

The approach taken in String Field Theory will be our guide. The worldsheet theory will be taken to be Weyl-invariant, allowing for the local decoupling of worldsheet metric degrees of freedom but at the cost of a loss in diffeomorphism invariance. The ghost sector will not be changed under parallel transport and will not concern us further. Instead of dealing with a complicated worldsheet metric, we imagine a local coordinate system $w_i$ around each puncture. The coordinates around each puncture are related to a reference coordinate system $z$, with respect to which any integration may be done, by functions $z=f_i(w_i)$. It is conventional to choose the location of the punctures as the origin of the local coordinate system, so that $z_i:=f_i(0)$. The details of the metric on the worldsheet $\Sigma$ are then encoded in the set of functions $f_i$.

On-shell correlation functions are independent of the choice of $f_i$ and may be written as the familiar integral over the moduli space of punctured Riemann surfaces ${\cal M}_{g,n}$,
\begin{equation}
    \int\limits_{{\cal M}_{g,n}}\mu_{g,n}\big\langle\Phi_1(z_1)...\Phi_n(z_n)\big\rangle,
\end{equation}
where $\mu_{g,n}$ is the usual measure on ${\cal M}_{g,n}$ built from ghosts and Beltrami differentials.  
More generally, off-shell correlation functions will depend on the choice of $f_i$. One way to address this \cite{Zwiebach:1992ie,Sen:2014pia} is to replace the usual integral over ${\cal M}_{g,n}$ with an integral over ${\cal S}_{g,n}\subset {\cal P}_{g,n}$. Here ${\cal P}_{g,n}$ is the infinite-dimensional bundle with finite-dimensional base ${\cal M}_{g,n}$ and infinite-dimensional fibres describing the possible choices of local coordinate about each puncture. ${\cal S}_{g,n}$ is a section of ${\cal P}_{g,n}$ with the same dimension as ${\cal M}_{g,n}$. If we employ this construction for an on-shell correlation function, the choice of section ${\cal S}_{g,n}$ does not matter and we recover the standard prescription.

In practice, we will only be interested in following the deformation of a small number of insertions (typically one at the point $z=w$ and another, a spectator field, at the origin) and so the details of the maps $z=f_i(w_i)$ and which section of ${\cal P}_{g,n}$ we are working with will not be of immediate concern.

We will only be interested in deformations that preserve $L_0^-=L_0-\bar{L}_0$, the generator of rotations on the worldsheet\footnote{$L_0^-=0$ need only be preserved modulo gauge.}. This is true for CFT cases \cite{Ranganathan:1993vj} and we have shown in Appendix \ref{A: Level matching} that this is true for the constant $H$-flux background we consider here\footnote{Neglecting the effect of terms not invariant under rotations was a property built in to the original formalism of \cite{Sonoda:1991mv,Sonoda:1992hd}. Here it arises naturally from the vanishing of $L^-_0$.}. Even though we shall often discuss the transformation of operators that are not in the kernel of $L_{0}^-$, such as $\partial X^{\mu}$, ultimately one would be interested in operators which would be taken to lie in the kernel of $L_0^-$, as they would have a more direct relevance for physical states.

In these more general cases where we are considering non-CFT deformations, we no longer necessarily have the luxury of universal coordinates. It will prove helpful in this context to be more explicit about the objects at the self-dual point. In particular, we will denote embedding fields of the theory before deformation by $\phi^{\mu}$ (usually a free theory at some point of enhanced symmetry) and those of the deformed theory by the more traditional $X^{\mu}$. We will have in mind a worldsheet $\Sigma$ of genus zero, but we expect our considerations to generalise to higher genus.

\subsubsection*{A note on topology change}

For much of this section, our starting point will be a $T^3$ with one or more circles tuned to the self-dual radius $R=\sqrt{\alpha'}$ and a free worldsheet CFT describing the embedding into this background. The backgrounds of interest will include target spaces with constant curvature or constant $H$-field, described by interacting worldsheet theories. In principle, one would have to contend with a change in the topology of the spacetime (or in the doubled space) when switching on such constant fluxes. We shall sidestep this issue by working in a covering space (such as $T^2\times \R$), where the deformation can be smoothly turned on and then an identification on the coordinates may then be imposed to recover the desired compact background\footnote{This is similar in spirit to the perspective often taken in discussions of the T-fold \cite{Hull:2009sg}.}.

\subsection{$H$-flux deformation} \label{SS: $H$-flux deformation}

As a straightforward example, consider a $T^3$ with a constant $H$-field $H_{\mu\nu\rho}$. This background may be thought of as a $T^2$ bundle over $S^1$ in which the $B$-field in the fibres undergoes a large gauge transformation upon circumnavigating the base. In this case, we choose to work in the cover $T^2\times \R$ where an identification on the base coordinate is taken to be imposed at a later point. The action is
\begin{equation}
S[X]=\frac{1}{2}\int\limits_{\Sigma}{d}^2 z \partial X_{\mu}(z)\bar{\partial}X^{\mu}(\bar{z})+\int\limits_VH+...,
\end{equation}
where $\partial V=\Sigma$ and the ellipsis denotes ghosts and other terms that will not be relevant to our discussion. The classical equation of motion is
\begin{equation}\label{eom_H_flux}
\partial\bar{\partial} X^{\mu}(z,\bar{z})+H^{\mu}{}_{\nu\lambda}\partial X^{\nu}(z,\bar{z})\bar{\partial}X^{\lambda}(z,\bar{z})=0.
\end{equation}
Integrating gives
\begin{equation}\label{classical}
 X^{\mu}(z,\bar{z})=\phi^{\mu}(z,\bar{z})-H^{\mu}{}_{\nu\lambda}\int\limits_{\Sigma}{d}^2 w\,G(z,w)\partial X^{\nu}(w)\bar{\partial}X^{\lambda}(\bar{w}),
\end{equation}
where $\phi^{\mu}(z,\bar{z})\in$ Ker$(\partial\bar{\partial})$, i.e. is a solution on the torus with $H=0$, and $G(z,w)\sim-\ln|z-w|^2$ is the Green's function for $\partial\bar{\partial}$ at genus $0$. In the specific examples below, we will introduce  $\lambda$ such that $H=\lambda\ast1_3$. From this classical consideration the deformation of $\partial \phi_{\mu}(z)$ at first order is
\begin{equation}
\partial \phi_{\mu}(z)\rightarrow \partial X_\mu(z) = \partial \phi_{\mu}(z)+\lambda \epsilon_{\mu\nu\lambda}\int\limits_{\Sigma}\frac{{d}^2 w}{z-w}\partial \phi^{\nu}(w)\bar{\partial}\phi^{\lambda}(\bar{w})+....
\end{equation}
If we put the $\alpha'$ dependence back in, we find that this simply corresponds to $\alpha' \rightarrow \alpha'\lambda^2$. It is significant that $\alpha'$ and $\lambda^2$ appear together\footnote{To see this one must redefine the fields as $X^{\mu}\rightarrow \lambda^{-1}X^{\mu}$.} here and we will comment upon this further in section \ref{SS: Analatycity in lambda}. At higher order in $\lambda$, the full quantum calculation will include higher order contractions not given by classical considerations, but at leading order, where only single contractions contribute, we expect the classical considerations to be reliable. We shall see that this is true.

This constant $H$-flux background was considered in this context in \cite{Mahmood:2020mtq}, where the base coordinate was taken to be independent of the worldsheet coordinates. If one takes the base coordinate $X(z,\bar{z})=x+...$, as given by (\ref{XP}), and only considers the constant piece $x$, the deformation operator may be written in the form
\begin{equation}
    {\cal O}=\frac{1}{3}H_{\mu\nu\rho}x^{\rho}\int_{\Sigma}d^2z\partial \phi^{\mu}(z)\bar{\partial}\phi^{\nu}(\Bar{z})+...,    
\end{equation}
and we are in the adiabatic (free CFT) regime studied in \cite{Mahmood:2020mtq}. Such deformations fall into the class of toroidal deformations considered in section \ref{S: toroidal backgrounds}. We now turn to determining the leading corrections to this adiabatic approximation.

\subsubsection{Deformation at first order}

We refer to the coordinates of the torus as $X^\mu = (X,Y,Z)$. We shall think of this as a $T^2$ with coordinates $(Y,Z)$ fibred over $\R$ with coordinate $X$ and an identification $X\sim X+2\pi$ later imposed. For this background, the important information appears as a large gauge transformation monodromy of the $B$-field in the fibres as $X\rightarrow X+2\pi$.\footnote{Note that, apart from the circle case, we will always use $X^\mu$ to refer to a general coordinate component, so there should not be any confusion in calling the first component $X$.} We usually choose the gauge such that
\begin{equation}
    g = {d} x^2+{d} y^2+{d} z^2, \qquad B = mx {d} y\wedge {d} z,
\end{equation}
where $m\in\Z$ and the deformation operator, given by the pullback of the $B$-field to the worldsheet, is
\begin{equation}\label{HO}
    {\cal O}[\phi^x]=m\int\limits_{\Sigma}{d}^2 z\; \phi^x(z,\bar{z})F^{-}_{yz}(z,\Bar{z}).
\end{equation}
where we have introduced 
\begin{equation}
    F^{\pm}_{\mu\nu}(z,\Bar{z}) \equiv \partial \phi_\mu(z)\Bar{\partial}\phi_\nu(\Bar{z}) \pm \partial \phi_\nu(z)\Bar{\partial}\phi_\mu(\Bar{z}).
\end{equation}

\subsection*{Calculation of $\delta_{E}$}

For the $H$-flux, since we have a non-zero $B$-field, we must evaluate the ${\delta_{E}}$ transformation using one of the approaches described earlier. Since we have already gone through the doubled geometry derivation, it is easiest to just plug in the $H$-flux background into the result we derived, which we recall is 
\begin{equation}
    \delta_{E}\partial \mathbbold{X}_I = -\frac{1}{2}\cH_{IJ}\partial \mathbbold{X}^J.
\end{equation}
Doing so, we find that 
\begin{equation}
    \delta_{E}\partial \phi_x=0,    \qquad  \delta_{E}\partial \phi_y=\frac{1}{2}m\phi^x\partial \phi_z,   \qquad  \delta_{E}\partial \phi_z=-\frac{1}{2}m\phi^x\partial \phi_y.    
\end{equation}
We have the ${\delta_{E}}$ transformation, but if we want to compute the mode transformation we would like to write this in integral form. The reason for this is that, in order to read off the mode transformation $\delta \alpha_n^y$, we would like to have an expression of the form $\sum_{m}w^{-m-1}f_m$, where $f_m$ is some expression in terms of the modes, which can then be read off as the deformation of $\alpha_n^y$. Such expressions are most easily obtained from integral expressions like the ones we have seen already (from expanding $(z-w)^{-1}$ in powers of $z$). We can show (details in Appendix \ref{A:$H$-flux}) that 
\begin{align}
    \delta_{E}\phi_y(w)= \frac{1}{2}m\int\limits_{\Sigma'} \frac{d^2z}{z-w}\bar{\partial}\phi_x(\Bar{z})\partial \phi_z(z) - \frac{1}{2}m\oint\limits_{C_0'}\frac{dz}{z-w}\phi^x(z,\Bar{z})\partial \phi_z(z),
\end{align}
where we define
\begin{equation}
\begin{gathered}
    \int\limits_{\Sigma'} = \lim_{\epsilon\rightarrow 0} \int\limits_{\Sigma^{\epsilon}}, \qquad\int\limits_{C_0'} = \lim_{\epsilon\rightarrow 0} \int\limits_{C_0^\epsilon}
\end{gathered}
\end{equation}
A similar result follows for $\delta_{E}\partial \phi_z$. We can then use this integral representation to compute the mode deformation. The calculation is fairly involved and the details are in Appendix \ref{A:$H$-flux}.

\subsection*{Calculation of $\delta_{\cal O}$}

The deformation operator, that part of the transformation which changes the action, is given by (\ref{HO}). Before proceeding, we need to think about this object makes sense as it stands. If we require $\phi^x$ to be a compact direction, there will be a branch point on the worldsheet whenever the string wraps the $\phi^x$ direction. This is manifest in the periodicity condition
\begin{equation}
    \phi^x(z,\bar{z})\sim \phi^x(e^{2\pi i}z,e^{-2\pi i}\bar{z})=\phi^x(z,\bar{z})+2\pi \omega,
\end{equation}
where $\omega\in\Z$ is the winding number around the $\phi^x$ direction. This leads to an ambiguity ${\cal O}\sim {\cal O}+\Delta {\cal O}$, where
\begin{equation}
    \Delta {\cal O}=2\pi wm\int\limits_{\Sigma}{d}^2 z\; F^{-}_{yz}(z,\Bar{z}).
\end{equation}
There is no problem here though. $\Delta {\cal O}$ generates a large gauge transformation of the $B$-field and is a symmetry of the theory \cite{Evans:1989cs}. Thus, the ambiguity simply reflects the fact that the $B$-field is only locally defined, as the presence of the non-trivial $H$ field strength indicates. We shall see a more general way to deal with such symmetries in section \ref{SS: The General Construction}. For now, we shall assume our definition of the connection is augmented to include an appropriate gauge transformation to account for the branch point. Alternatively (and in practice), we could work in the cover (where $\phi^x$ is not compact), deform our theory and then impose the relevant identification on $\phi^x$ in the new background.

Contraction of ${\cal O}$ with $\partial \phi_y(w)$ gives
\begin{equation}
    \delta_{\cal O}\partial \phi_y(w)=-\frac{1}{2}m\int\limits_{\Sigma'}{d}^2 z\frac{1}{(z-w)^2}\; \phi^x(z,\bar{z})\bar{\partial}\phi_z(\bar{z}),
\end{equation}
which we can more conveniently write as
\begin{equation}
    \delta_{\cal O}\partial \phi_y(w)=-\frac{1}{2}m\int\limits_{\Sigma'}\frac{{d}^2 z}{z-w}\; \partial \phi_x(z)\bar{\partial }\phi_z(\bar{z})-\frac{1}{2}m\oint\limits_{C_0'}\frac{{d} \bar{z}}{z-w}\; \phi^x(z,\Bar{z})\bar{\partial}\phi_z(\bar{z}),
\end{equation}
where we have dropped the $C_w'$ integral since this can be shown to vanish. The first order change in the modes is then given by\footnote{Using
$(z-w)^{-1}=-\sum_{n\geq 0}w^{-n-1}z^n$ when $|z|<|w|$.}
\begin{eqnarray}
\delta_{\cal O} \alpha_n^y&=&\frac{i}{\sqrt{2}}m\int\limits_{\Sigma'}{d}^2z\;z^n\partial \phi_x(z)\bar{\partial}\phi_z(\bar{z})+\frac{i}{\sqrt{2}}m\oint\limits_{C_0'}{d}\bar{z}\;z^n\phi^x(z,\Bar{z})\bar{\partial} \phi_z(\bar{z})+...,
\end{eqnarray}
where the ellipsis denotes divergent terms, which are dealt with according to the choice of connection. As with ${\delta_{E}}$, the details of this mode calculation are given in Appendix \ref{A:$H$-flux}. Similar results also follow for $\delta\partial \phi_x$ and $\delta\partial \phi_z$.

\subsection*{The first order deformation}

Putting the ${\cal O}$ and ${\delta_{E}}$ parts together, gives the first order changes (evaluated on a contour $|w|$=constant)
\begin{eqnarray}
\delta \partial \phi_x(w)&=&\frac{1}{2}m\int\limits_{\Sigma'}\frac{{d}^2 z}{z-w}\; F^{-}_{yz}(z,\Bar{z}),  \label{delta_dX_H_flux} \\
\delta \partial \phi_y(w)&=&\frac{1}{2}m\int\limits_{\Sigma'}\frac{{d}^2 z}{z-w}\; F^{-}_{zx}(z,\Bar{z})-\frac{1}{2}m\oint\limits_{C_0'}\frac{\phi^x(z,\Bar{z}){d} \phi_z(z,\Bar{z})} {z-w},  \label{delta_dY_H_flux} \\
\delta \partial \phi_z(w)&=& \frac{1}{2}m\int\limits_{\Sigma'}\frac{{d}^2 z}{z-w}\; F^{-}_{xy}(z,\Bar{z})+\frac{1}{2}m\oint\limits_{C_0'}\frac{\phi^x(z,\Bar{z}){d} \phi_y(z,\Bar{z})} {z-w}, \label{delta_dZ_H_flux}  
\end{eqnarray}
where we have written ${d} \phi_z(z,\Bar{z})={d} z\partial \phi_z(z)+{d}\bar{z}\bar{\partial}\phi_z(\Bar{z}), {d} \phi_y(z,\Bar{z})={d} z\partial \phi_y(z)+{d}\bar{z}\bar{\partial}\phi_y(\Bar{z})$. In terms of modes, the first order deformation for $\alpha_n^x$ is
\begin{equation}
    \delta \alpha_n^x=-\frac{1}{2}m\int\limits_{\Sigma'}{d}^2z\,z^n\; F^{-}_{yz}(z, \Bar{z}),    
\end{equation}
with the modes $\alpha_n^y$ and $\alpha_n^z$ taking a similar form. With a little work, the integrals can be done and the $\epsilon$ dependence made explicit. This is done in detail in Appendix \ref{A:$H$-flux}, where it is shown that the deformation in the $\partial\phi_y$ modes may be written as
\begin{equation}
    \delta_{\cal O}\alpha^y_n=x{\cal A}_{n}\bar{\alpha}_{-n}^z+\sum_p{\cal B}_{np}\bar{\alpha}_{p}^z\alpha_{n+p}^x + \sum_p{\cal C}_{np}\Bar{\alpha}_p^z\bar{\alpha}^x_{-n-p}, \qquad n\geq 0,
\end{equation}
where ${\cal A}, {\cal B}, {\cal C}$ are $\epsilon$-dependent constants. Only the first term contributes in the adiabatic limit. These expressions give some insight into how turning on the $H$-field deforms the free field algebra of operators.

\subsubsection{Branch points and gauge-invariance}

A few comments are in order. We claim that the second, branch-dependent, terms in \eqref{delta_dY_H_flux} and \eqref{delta_dZ_H_flux} are gauge-dependent.  Note that, in this gauge, $\partial \phi_x$ is in the kernel of ${\delta_{E}}$. This would not necessarily be the case had we chosen to work in a different gauge and is a reflection of the fact that ${\delta_{E}}$ preserves the classical action. Had we chosen to work in the gauge where $B=my{d} z\wedge {d} x$, then $\partial \phi_y$ would have been in the kernel of ${\delta_{E}}$. The difference between these two gauges is the large gauge transformation $B\rightarrow B+\Lambda$, where $\Lambda=\frac{1}{2}m{d}(xy{d} z)$. In terms of the worldsheet, this is generated by
\begin{equation}\label{lambda}
    \Lambda[\phi^x,\phi^y]=\frac{1}{2}m\oint\limits_{\partial \Sigma}\phi^x\phi^y{d} \phi_z.
\end{equation}
Contracting $\Lambda[\phi^x,\phi^y]$ with $\partial \phi_y(w)$ gives the second term in \eqref{delta_dY_H_flux}, suggesting this term really is a gauge artifact. 

There is further evidence for this if we look at the solution obtained from considering the classical equation of motion \eqref{eom_H_flux}, which for $Y$ is
\begin{equation}
    \partial\bar{\partial}Y = -\frac{1}{2}m(\partial X\Bar{\partial}Z - \partial Z\Bar{\partial}X).
\end{equation}
Given a solution $\phi_y$ to the flat equation of motion $\partial\Bar{\partial}\phi_y=0$, this can be solved iteratively. Doing so to first order gives
\begin{equation}
    \delta\partial \phi_y(w) = -\frac{1}{2}m\int \frac{d^2z}{z-w}F^{-}_{xz}(z,\Bar{z}),
\end{equation}
where $\partial Y = \partial \phi_y + \delta \partial \phi_y+O(m^2)$. This can be seen, for example, by noting that $\Bar{\partial}\left(\frac{1}{z-w}\right)=\delta^2(z-w)$. Thus, up to the branch-dependent terms, $(\delta_{\cal O} + \delta_{E})\partial \phi_y$ is of the same form as the classical result, which is what we would expect at first order, and further suggests that the branch-dependent term is a gauge artifact. At higher order, we expect our formalism to give the full quantum corrections and so the results will differ from the classical case.

\subsubsection{Higher order contributions}

In the adiabatic approximation, the deformation of $\partial \phi_\mu$ truncates at first order in $m$, since the $H$-flux deformation $\delta E$ is first order in $m$. However, if we are taking the $X$-dependence in the $B$-field into account, we expect that there will be stringy corrections at all orders in $m$. Therefore, it is of interest to find a way to calculate these corrections. In the CFT case, i.e. when $\cal O$ has no $\phi^x$-dependence, we described a way to systematically generalise the first order procedure to all orders in section \ref{SSS: Higher orders} and in more detail in Appendix \ref{A: Sen}. However, when there is $\phi^x$-dependence in $\cal O$, this procedure would need significant modification. Nevertheless, this is still a starting point to consider the challenges that arise at higher order. For example, at second order, if we are looking at $\delta \partial \phi_y(w)$, we would be interested in the integral
\begin{equation}
  {\cal O}_2{\cal O}_1\partial \phi_y(w)=  \int\limits_{\Sigma_2}d^2z_2\int\limits_{\Sigma_1}d^2z_1 O(z_2,\Bar{z}_2) O(z_1, \Bar{z}_1) \partial \phi_y(w),
\end{equation}
where
\begin{equation}
    {\cal O}_i = m\int\limits_{\Sigma_i} d^2z \phi^x(z,\Bar{z})F^{-}_{yz}(z,\Bar{z}),
\end{equation}
with $\Sigma_i$ given by (\ref{domain}). Naively following the prescription used in the free field case, we would first contract $\partial \phi_y(w)$ with $O(z_1,\Bar{z}_1)$, and then contract the result with $O(z_2,\Bar{z}_2)$, i.e. we would have the sequential contractions 
\begin{alignat}{2}
     &\delta_{{\cal O}_2}\left(\delta_{{\cal O}_1}\Big(\partial \phi_y(w)\Big)\right)&&{}\nonumber\\
     &= -\frac{m^2}{2}\int\limits_{\Sigma_2}d^2z_2\int\limits_{\Sigma_1}\frac{d^2z_1}{(z_1-w)^2} &&{}\wick{\c1{\phi^x}(z_2,\Bar{z}_2)\big( \partial \phi_y(z_2)\Bar{\partial}\c2{\phi_z}(\Bar{z}_2) - \partial \phi_z(z_2)\Bar{\partial}\phi_y(\Bar{z}_2)\big) \c1{\phi^x}(z_1,\Bar{z}_1)\Bar{\partial}\c2{\phi_z}(\Bar{z}_1)},\nonumber\\
     &=\frac{m^2}{4}\int\limits_{\Sigma_2}d^2z_2\int\limits_{\Sigma_1}\frac{d^2z_1}{(z_1-w)^2} &&\bigg( 
     F^-_{yz}(z_2,\Bar{z}_2)\Bar{\partial}\phi_z(\Bar{z}_1)\log|z_1-z_2|^2 \notag \\
     &{} &&\left.+  \frac{\phi^x(z_2,\Bar{z}_2)\partial \phi_y(z_2)\phi^x(z_1,\Bar{z}_1)}{(\Bar{z}_1-\Bar{z}_2)^2} -\frac{1}{2} \frac{\log|z_1-z_2|^2}{(\Bar{z}_1-\Bar{z}_2)^2}\partial \phi_y(z_2)
     \right). \label{O2Hflux}
\end{alignat}
We could then compute this integral to obtain the result. Note that here we have included both the single and double contractions between the $z_1$ and $z_2$ fields, though the double contraction term is most likely a divergent term that we would remove via a regularisation procedure. In Appendix \ref{A: Sen}, we show that, as expected, the deformation resulting from $({\cal O}+\delta_{E})^2$ vanishes for $\partial \phi_\mu$ when we have a constant background $E$. However, in this case, since there is $X$-dependence in $E$, we expect in general that the same cancellations will not occur and that there will be second (and higher) order corrections. As we can see from the above integral, the $\phi^x$-dependence is explicit and makes the calculation significantly more complicated, and will undoubtedly introduce new terms.

If we looked at higher order contractions, what we would see is that, in addition to the terms that we would ordinarily get without the $\phi^x$-dependence, we also get additional terms coming from the $\phi^x$ contractions. The contractions which do not involve $\phi^x$ follow the expected pattern, e.g. at third order, this would be the contraction
\begin{align}
    &\wick{\phi^x(z_3,\Bar{z}_3)\c1{\partial \phi_y}(z_3) \Bar{\partial}\phi_z(\bar{z}_3)\phi^x(z_2,\Bar{z}_2)\c1{\partial \phi_y}(z_2)\phi^x(z_1, \Bar{z}_1)}= -\frac{\Bar{\partial}\phi_z(\bar{z}_3)\phi^x(z_3, \Bar{z}_3)\phi^x(z_2, \Bar{z}_2)\phi^x(z_1, \Bar{z}_1)}{2(z_3-z_2)^2}, 
\end{align}
whereas all other contractions would involve log terms from the `$\phi^x\phi^x$' contractions. Of course, when it comes to doing the integration, the $\phi^x$-dependence will have a large effect here as well, so even terms which would still be there in the constant background case may give extra contributions. 

These calculations make it clear that it is better to make a specific choice of connection, one designed to reduce the complexity of the calculations from the start. A Feynman-diagrammatic approach can be used to streamline the process further. As discussed above, the presence of the branch points is symptomatic of winding modes around the base, and either working in the cover or incorporating an appropriate large gauge transformation in the definition of the connection should deal with such terms. It is unlikely that the calculation can be systematised in the way achieved for the CFT deformations, as ${\cal O}$ can no longer be thought of as a map that preserves the subspace spanned by $\partial \phi$ and $\bar{\partial} \phi$. Instead, the ${\cal O}$ considered here mixes in other operators. It would be interesting to see if there is a relatively simple subspace that ${\cal O}$ preserves. It is also clear that any candidate for a universal coordinate for this deformation would need to be a more general vector within this subspace than that considered in \cite{Evans:1995su}.

\subsection{Nilfold}

As a second example, we consider the nilmanifold. This three-dimensional geometry may be thought of as a $T^2$ bundle over $S^1$ with $SL(2;\Z)$ monodromy, acting as a large diffeomorphism in the fibres. As above, we shall take the base to be $\R$ and then impose an identification in the coordinates. We shall see agreement at first order with the classical result here as well. The Nilfold is T-dual to the $T^3$ with constant $H$-flux discussed above and so provides a natural second example and will return again when we discuss T-duality in this framework in section \ref{S: T-duality}.

The metric is
\begin{equation}\label{nilfold}
    {d} s^2={d} x^2+({d} y-mx{d} z)^2+{d} z^2.
\end{equation}
The Nilfold is usually taken to be compact, with identifications on all coordinates. For clarity, we lift to the cover (the three-dimensional Heisenberg group manifold), perform the deformation, and then impose the appropriate identifications on the coordinates.

The deformation operator is\footnote{Unlike the previous example, there is a quadratic part to ${\cO}$ in which normal ordering of $({\phi^x})^2$ is understood.}
\begin{equation}
    {\cal O}[\phi^x]=-m\int\limits_{\Sigma}  \phi^x F^{+}_{yz}+m^2\int\limits_{\Sigma} ({\phi^x})^2\partial \phi_z\bar{\partial}\phi_z.
\end{equation}
As before, we focus on one of the coordinates. We shall only consider the first order deformation and so neglect the $m^2$ term. The worldsheet equation of motion for $\partial Y$ for the Nilfold is given by
\begin{equation}
    \partial\bar{\partial}Y = \frac{1}{2}m(\partial X\Bar{\partial}Z + \partial Z\Bar{\partial}X) + O(m^2),
\end{equation}
and integrating out the $\bar{\partial}$ gives
\begin{align}
    \delta \partial \phi_y(w) =& -m\phi^x(w,\Bar{w})\partial \phi_z(w) +  \frac{1}{2}m\int \frac{d^2z}{z-w}F^{+}_{xz}(z,\Bar{z})
    = \frac{1}{2}m\int \frac{d^2z}{z-w}F^{-}_{xz}(z,\Bar{z}),
\end{align}
where we have used $\phi^x(w,\Bar{w})\partial \phi_z(w) = \int d^2z \bar{\partial}\left(\frac{1}{z-w}\right)\phi^x(z,\Bar{z})\partial \phi_z(z)$. We will reproduce this up to contour integrals using our formalism. The calculations are qualitatively the same as for the $H$-flux, so we will not give much detail. We have:
\begin{alignat}{2}
    &\delta_{\cal O}\partial \phi_y(w) &&= \frac{1}{2}m \int\limits_{\Sigma'} \frac{d^2z}{z-w}\partial \phi_x(z,\Bar{z})\bar{\partial}\phi_z(z) + \frac{1}{2}\oint\limits_{C_0'} \frac{d\bar{z}}{z-w}\phi^x(z,\Bar{z})\bar{\partial}\phi_z(\bar{z}),\\
    &\delta_{E}\partial \phi_y(w) &&= -\frac{1}{2}m\phi^x(w,\Bar{w})\partial \phi_z(w) \notag\\
    &{}&&= \frac{1}{2}m\oint\limits_{C_0'} \frac{dz}{z-w}\phi^x(z,\Bar{z})\partial \phi_z(z) - \frac{1}{2}m\int\limits_{\Sigma'} \frac{d^2z}{z-w}\partial \phi_z(z)\bar{\partial}\phi_x(\bar{z}),
\end{alignat} 
and so overall we have
\begin{equation}
    (\delta_{\cal O}+\delta_{E})\partial \phi_y(w) = \frac{1}{2}m\int\limits_{\Sigma'} \frac{d^2z}{z-w}F^{-}_{xz}(z,\Bar{z})+ \frac{1}{2}m\oint\limits_{C_0'} \frac{\phi^x(z,\Bar{z})d\phi_z(z,\Bar{z})}{z-w}, 
\end{equation}
which agrees with the classical result up to the branch-dependent terms. The calculation follows similarly for $\partial \phi_x, \partial \phi_z$, and overall we have, to first order in $m$, 
\begin{align}
    &\delta \partial \phi_x(w) = -\frac{1}{2}m\int\limits_{\Sigma'} \frac{d^2z}{z-w}F^{+}_{yz}(z,\Bar{z}), \label{dXNil}\\
    &\delta \partial \phi_y(w) = \frac{1}{2}m\int\limits_{\Sigma'} \frac{d^2z}{z-w}F^{-}_{xz}(z,\Bar{z})+ \frac{1}{2}m\oint\limits_{C_0'} \frac{\phi^x(z,\Bar{z})d\phi_z(z,\Bar{z})}{z-w}, \label{dYNil}\\
    &\delta \partial \phi_z(w) = \frac{1}{2}m\int\limits_{\Sigma'} \frac{d^2z}{z-w}F^{-}_{xy}(z,\Bar{z})+ \frac{1}{2}m\oint\limits_{C_0'} \frac{\phi^x(z,\Bar{z})d\phi_y(z,\Bar{z})}{z-w}.\label{dZNil}
\end{align}
As in the previous example, we view the first terms in the above expressions - those containing $F^{\pm}_{\mu\nu}(z,\bar{z})$ - as the physical deformations of the fields. The gauge ambiguity, represented by the contour integral terms, correspond to target space diffeomorphisms that are T-dual to the gauge transformations generated by (\ref{lambda}).

\subsection{The covariant construction}\label{SS: The General Construction}

A gauge-invariant construction of the $H$-flux deformation (or covariant construction for the nilfold) may be obtained using the background field method\footnote{For example, see \cite{Callan:1989nz}.}, which we describe briefly. The starting point is the Polyakov action\footnote{We reinsert the $\alpha'$ factors in this section for clarity.}
\begin{equation}\label{Poly}
    S_P[X] = \frac{1}{4\pi\alpha'}\int\limits_{\Sigma} d^2\sigma \sqrt{\gamma}\gamma^{ab}\partial_aX^\mu\partial_bX^\nu g_{\mu\nu}(X).
\end{equation}
The first step is to split the worldsheet embedding $X$ into two parts; $X=X_0+\pi$, where $X_0$ obeys the classical equations of motion and $\pi$ can be thought of as a quantum fluctuation as the path integral over $X$ reduces to a path integral over $\pi$. Working with Riemann normal coordinates simplifies the problem and we define new coordinates for the quantum fluctuations, $\eta^\mu$, where 
\begin{equation}
    \pi^\mu = \eta^\mu -\frac{1}{2} \Gamma^\mu_{\nu\rho}(X_0)\eta^\nu\eta^\rho+....
\end{equation}
We consider the background field formulation which preserves the manifest covariance of the theory. Substituting in the new coordinates $X_0, \eta$, gives \cite{Callan:1989nz}
%[SHOULD WE TAKE $X_0$ TO BE CONSTANT? THEN $g_{\mu\nu}(X_0)$ IS CONSTANT AND CAN BE DIAGONALISED TO BE MINKOWSKI (REFERENCE) METRIC, THUS SPLITTING ALL OTHER TERMS OFF AS DEFORMATION TERMS AND THE RIEMANN TENSORS ARE GENUINE COUPLING CONSTANTS...]
\begin{align}
    S_P[X_0+\eta] &= S_P[X_0] + \frac{1}{2\pi \alpha'}\int\limits_{\Sigma} d^2\sigma \sqrt{\gamma}\gamma^{ab}g_{\mu\nu}(X_0)\partial_a X_0^\mu \nabla_b \eta^\nu \notag \\
    &+ \frac{1}{4\pi\alpha'}\int\limits_{\Sigma} d^2\sigma \sqrt{\gamma}\gamma^{ab}\left\{ g_{\mu\nu}(X_0)\nabla_a\eta^\mu\nabla_b\eta^\nu +R_{\mu\lambda\sigma\nu}(X_0)\Big(\partial_aX_0^\mu\partial_bX_0^\nu \eta^\lambda\eta^\sigma \right.\notag \\
    &\left.+ \frac{4}{3}\partial_aX_0^\mu\eta^\lambda\eta^\sigma\nabla_b\eta^\nu +\frac{1}{3}\eta^\lambda\eta^\sigma\nabla_a\eta^\mu\nabla_b\eta^\nu\Big) + ....\right\}.
\end{align}
Note that this is now explicitly gauge-covariant. Note that $S_P[X_0]$ and the terms linear in $\eta$ can be discarded if we choose $X_0$ to obey the classical equations of motion\footnote{If $X_0$ is a classical instanton solution, then $S_P[X_0]$ will still give a finite contribution.}. The kinetic term is awkward in that it involves coupling with the background fields and so a potentially complicated Greens function. One way around this is to work with frame fields $\eta^a=e^a{}_{\mu}(X_0)\eta^{\mu}$. Instead, we expand the metric $g_{\mu\nu}(X_0)$ around a flat reference background
\begin{equation}
    g_{\mu\nu}(X_0)=\eta_{\mu\nu}+h_{\mu\nu}(X_0).
\end{equation}
The covariant deformation operator, with respect to this background, is then
\begin{align}
    {\cal O}_g[X_0,\eta] &= \frac{1}{4\pi\alpha'}\int\limits_{\Sigma} d^2\sigma \sqrt{\gamma}\gamma^{ab}\left\{ h_{\mu\nu}(X_0)\nabla_a\eta^\mu\nabla_b\eta^\nu +R_{\mu\lambda\sigma\nu}(X_0)\Big(\partial_aX_0^\mu\partial_bX_0^\nu \eta^\lambda\eta^\sigma \right.\notag \\
    &\left.+ \frac{4}{3}\partial_aX_0^\mu\eta^\lambda\eta^\sigma\nabla_b\eta^\nu +\frac{1}{3}\eta^\lambda\eta^\sigma\nabla_a\eta^\mu\nabla_b\eta^\nu\Big) + ....\right\}.
\end{align}
To make contact with the discussion in section \ref{S: toroidal backgrounds}, we allow the possibility of a non-trivial $B$-field. Suppose we have an anti-symmetric part of the action given by
\begin{equation}
    S_{AS}[X] = \frac{1}{4\pi\alpha'}\int\limits_{\Sigma} d^2\sigma \epsilon^{ab}\partial_aX^\mu\partial _bX^\nu B_{\mu\nu}(X).
\end{equation}
As with $S_P[X]$, we can substitute in the background field expansion and obtain the action in a covariant form. We will not write it down in full here, but the result is given in \cite{Callan:1989nz}. The terms that are of relevance for us are
\begin{equation}
    \frac{1}{4\pi\alpha'}\int\limits_{\Sigma} d^2\sigma\epsilon^{ab} H_{\mu\nu\rho}(X_0)\partial_aX_0^\mu \nabla_b\eta^\nu\eta^\rho+\frac{1}{12\pi\alpha'}\int\limits_{\Sigma} d^2\sigma\epsilon^{ab}H_{\mu\nu\rho}(X_0)\eta^\mu\nabla_a\eta^\nu\nabla_b\eta^\rho,
\end{equation}
For a flat background with constant $H$, these are the only contributions to ${\cal O}$.

The background fields $g_{\mu\nu}(X_0)$, $H_{\mu\nu\lambda}(X_0)$ and their derivatives play the role of the deformation parameters $\mathfrak{m}_\alpha$. The action is that of an interacting theory with couplings specified by the covariant functions $H_{\mu\nu\rho}(X_0)$, $R_{\mu\nu\lambda\rho}(X_0)$ and their covariant derivatives. A natural construction would be to consider a natural basis of functions $f_I(X_0)$ on the reference spacetime and to then decompose the background metric deformation and $B$-field in terms of this basis, i.e.
\begin{equation}
    h_{\mu\nu}(X_0)=\sum_I c_{\mu\nu}^If_I(X_0).    
\end{equation}
The coefficients $c^I$ might then provide suitable local coordinates on ${\cal M}$ with which to parameterise the deformation. In cases where the initial and final backgrounds have different topology, it is natural to pass to the cover as discussed above\footnote{There are of course well-known cases where one can smoothly change the topology in string theory \cite{Giveon:1993ph,Aspinwall:1994zd,Hull:2006qs}. In such cases, a continuous path, without degenerations in the fibres, is expected to exist between the two topologically distinct backgrounds.}. Given a path in ${\cal M}$ between two backgrounds, the classical solution $X_0$ varies as the action changes as we move along the path. As such, the basis $f_I$ will also change along the path; however, the expression (\ref{parallel}) only requires knowledge of the moduli and their derivatives along the path evaluated at the start of the path (where the theory is free in most cases).

A connection on the space of such backgrounds is given by the variational formula (\ref{ODef}), where the OPEs can in principle be computed in perturbation theory.\footnote{Or, by first finding the beta-functions, the OPE coefficients can be computed using the explicit construction given in \cite{Sonoda:1991mv,Sonoda:1992hd}.} The construction given in \cite{Ranganathan:1993vj} and outlined in section \ref{S: Deformations and connections} then gives the connection associated with deforming the theory by changing the value of the background metric and $B$-field. Taking the reference background as the free theory, we only require knowledge of the OPE of the free theory and we recover the previous construction of section \ref{S: Deformations and connections}, but now in a manifestly covariant form. 

By way of example, in the case of constant $H$-flux $H_{\mu\nu\rho}=\lambda\epsilon_{\mu\nu\rho}$ on a flat background $h_{\mu\nu}(X_0)=0$, the deformation operator is
\begin{equation}\label{Sint}
  {\cal O}_H[X_0,\eta]=  \frac{\lambda}{4\pi\alpha'}\int\limits_{\Sigma} d^2\sigma \epsilon^{ab}\Big(\epsilon_{\mu\nu\rho}\partial_aX_0^{\mu}\partial_b\eta^{\nu}\eta^{\rho}+\frac{1}{3}\epsilon_{\mu\nu\rho}\eta^\mu\partial_a\eta^\nu\partial_b\eta^\rho\Big),
\end{equation}
where $X_0$ is the classical solution at $\lambda\neq 0$\footnote{To relate this to the background field description of the free theory at $\lambda=0$, one could write $X_0$ explicitly in terms of $\lambda$ and the classical solution at $\lambda=0$ using (\ref{classical}).}. This is manifestly gauge invariant. Thus, we see that the choice of branch cut for the $X$ dependence in the $B$-field does not make any difference to the physics, since these choices of branch correspond to $B$-field gauge transformations. The interaction terms then, order by order in $\lambda$, perturbatively describe the deformation of the theory away from the $\lambda=0$ point. Incorporating non-perturbative effects, which will be inaccessible via these techniques, is briefly discussed in the following section.

\subsection{Non-perturbative effects}\label{SS: Analatycity in lambda}

What can go wrong? As alluded to in section \ref{S: Deformations and connections}, there is a close relationship between parallel transport and conventional interaction picture perturbation theory, with the deformation ${\cal O}$ playing a role akin to an interaction Hamiltonian and the parallel transport (\ref{parallel}) akin to a Dyson series. It is well-known that not all physics is accessible via perturbation theory and non-perturbative effects can play an important role. The issue of convergence of the perturbative expansion has an obvious analogue in attempting to parallel transport from one background to another. We can only hope to access that part of the deformed theory that is analytic in the deformation parameter $\lambda$.

In the $H$-flux case, one can show that, up to field redefinitions, $\lambda$ and $\alpha'$ always appear together and so one can use $\alpha'$ as a proxy for $\lambda$. Thus, non-perturbative effects in $\alpha'$ will also be non-perturbative effects in $\lambda$. The parallel transport will be blind to phenomena like worldsheet instantons. This is not always the case and, in section \ref{S: Discussion}, we briefly consider an example where non-perturbative effects in $\alpha'$ are in fact perturbative in $\lambda$.

\section{T-Duality}\label{S: T-duality}

In this final section, we consider how T-duality appears in this construction and how the requirements evident in the Buscher construction \cite{Buscher:1987sk, Buscher:1987qj} emerge in this framework. As demonstrated in \cite{Mahmood:2020mtq}, we can use this method of CFT deformation to do the T-duality at any background by deforming to it from the point of enhanced symmetry. We then use a charge to compute the T-duality by acting on the stress tensor via an automorphism. In some sense, this is a rather trivial process since, provided the automorphism has a well-defined action on the operator algebra of the theory, it will obviously produce a new description of the same underlying physics. Where this is interesting is when the new description is also a conventional string theory, but with a different interpretation of the target space. %As outlined in the introduction, the key point is that an understanding of how T-duality acts at the self-dual point allows us, in principle, to understand how it acts at any other point related to the self-dual point by parallel transport.

The Buscher construction requires the existence of a globally-defined\footnote{It may be that the requirement that the isometry is globally defined may be dropped \cite{Hull:2004in}.}, compact isometry in the target space which preserves all non-trivial field strengths in the background. Do we see such requirements in this formalism and in what capacity? We discuss the existence and compactness of the isometry in turn.

\subsubsection*{Existence of a continuous isometry}

The requirement of an isometry can be seen from the fact that the action of the T-duality charge $Q$ is not well-defined on $X$ itself. In \cite{Mahmood:2020mtq}, we found that 
\begin{equation}
    e^{iQ}X_L(\sigma)e^{-iQ} =-X_L(\sigma)+{\cal C},
\end{equation}
where $Q=\frac{1}{2}\oint\limits d\sigma \cos(2X_L(\sigma))$ and ${\cal C}$ was a constant operator that was dependent on the T-duality charge used.\footnote{The calculations in \cite{Mahmood:2020mtq} were done using a Lorentzian worldsheet metric at a fixed worldsheet time $\tau=0$, hence the dependence on $\sigma$ alone. The general result carries over to the Euclidean case.} Thus, when we have explicit $X$-dependence, we have to deal with the troubling operator ${\cal C}$ and it is not clear how to proceed. In the Buscher procedure, we cannot do anything without an isometry, whereas here the situation is less clear. Though we will not address it in this paper, there is a question of whether we can make sense of the non-isometric case regardless of the aforementioned difficulties.\footnote{It is interesting to note that the transformation is well-defined on those `self-dual' states for which $p=w$. If we set $p=w$, we actually find that the anomalous term ${\cal C}$ vanishes and we get 
$e^{iQ}X_L(\sigma)e^{-iQ} =-X_L(\sigma)$.}

We can appreciate why the non-isometric case is much harder to understand if we consider what is happening from the perspective of theory space. A point on the space ${\cal M}$ is given by a choice of metric and $B$-field. Since an isometry will preserve the sigma model, it is reasonable to identify different points on ${\cal M}$ if they correspond to the same sigma model. Hence, an isometry will preserve the sigma model and keep us at the same point. A more general diffeomorphism, which is not an isometry, takes us to a different point on ${\cal M}$ representing a different sigma model. One would need some non-local (from the sigma model perspective) deformation ${\cal O}$ to relate the original fields with those after the diffeomorphism. This gauge transformation in the target space is thus a non-trivial deformation (or parallel transport) in ${\cal M}$. Therefore, from the worldsheet perspective, there is a significant difference between isometries and non-isometric diffeomorphisms. Understanding this difference better is key to understanding T-duality in the absence of continuous isometries in this framework.

\subsubsection*{Compactness of the isometry}

The Buscher prescription also requires the isometry to be compact. The necessity of this requirement can be illustrated in the one-dimensional case, where we can consider a limit in which the circle `decompactifies'. Recall that we can write $\partial X(R)$ in terms of objects at the self-dual radius as 
\begin{equation}
    \partial X(R)(z) = \frac{1}{2}(1+R^2)\partial X(z)+\frac{1}{2}(1-R^2)\bar{\partial}X(\Bar{z}),
\end{equation}
where $\partial X, \bar{\partial}X$ are at the self-dual radius. We might try to think of the `decompactified' case as taking the limit $R\rightarrow \infty$ or, by T-duality, $R\rightarrow 0$. However, we can see from the above that, if we take this limit, we get
\begin{equation}
    \partial X(0) = \frac{1}{2}(\partial X+\bar{\partial}X) = \Pi,
\end{equation}
and doing the same for $\bar{\partial}X(R)$ gives
$\bar{\partial}X(0) =\Pi = \partial X(0)$, i.e. holomorphic and antiholomorphic derivatives seem to coincide. Another way of saying this is that if we have
\begin{equation}
    \begin{pmatrix}
    \partial X(R)\\ \bar{\partial}X(R)
    \end{pmatrix} = 
    M(R) \begin{pmatrix}
    \partial X \\ \bar{\partial}X
    \end{pmatrix},
\end{equation}
then the matrix $M(R)$ degenerates in the limit $R\rightarrow 0$. This failure of the self-dual basis to extend to this case is unsurprising and suggests a new ingredient would be needed to extend the duality to this unlikely case. One suspects the curvature of the connection would be badly behaved at this point, although we have not checked this.

We now turn our attention to how T-duality fits into our framework and how we can derive T-dual backgrounds using the formalism discussed in this paper.

\subsection{Deriving T-dual backgrounds in the parallel transport formalism}\label{SS: T-duality in O+P}

The transformations we have obtained using the parallel transport formalism are different to what we would expect from the universal coordinate methods of \cite{Evans:1995su}, which give an adiabatic approximation. The application of this formalism to T-duality in trivial torus bundles has been discussed in \cite{Mahmood:2020mtq}. We turn now to study T-duality in the torus bundles discussed above. 

We define the  tensor $T_E$ as
\begin{equation}
T_E = g^{\mu\nu}\partial X_{\mu}(E)\partial X_{\nu}(E),
\end{equation}
where $E = g+B$. For a CFT associated with toroidal backgrounds this is the left-moving component of the stress tensor, but for general sigma models there is no clean split into left- and right-moving sectors. Nonetheless, $T_E$ is a useful object to consider as it is the simplest composite operator that is a target space scalar and so invariant under ${\delta_{E}}$. This invariance under ${\delta_{E}}$ streamlines the analysis somewhat.

By writing the stress tensor in terms of fields defined at the self-dual radius, \cite{Evans:1995su} showed how to use the stress tensor to compute the T-dual of a given background: starting with some background $E$ and writing $T_E$ in terms of the fields at the self-dual point, one can use the action of T-duality at the self-dual point to determine the stress tensor for the dual background, written in terms of the self-dual basis. $T_E$ transforms as  
\begin{equation}
    T_E \rightarrow T_{\tilde{E}}=U T_E U^{-1}
\end{equation}
under the T-duality automorphism $U=e^{iQ}$ given in \cite{Evans:1995su}. Knowledge of how the stress tensor changes under a general enough class of deformations (e.g. marginal deformations) then allows one to read off  $\tilde{E}$ from the dual tensor $T_{\widetilde{E}}$.

For the toroidal backgrounds considered in \cite{Evans:1995su}, this was essentially a novel derivation of the familiar Buscher rules \cite{Buscher:1987sk, Buscher:1987qj}. Much of this structure carries over to the more general constructions considered in the last section. In this section, we sketch how known dualities are realised in the framework presented in this paper and to what extent one may use it to generalise beyond the cases where the Buscher construction is valid.

It is important to stress that it is not true that the deformation operators for T-dual backgrounds ${\cal O}$ and $\widetilde{\cal O}$ are related simply as $e^{iQ}{\cal O}e^{-iQ}$. That this cannot be true is easily seen if we consider the explicit deformation parameters of the $T^3$ with constant $H$-flux, Nilfold and T-fold backgrounds with respect to the $T^3$ CFT\footnote{For the T-fold, we recall that the metric and $B$-field are given by
\begin{equation*}
    ds^2 = {d} x^2 + \frac{1}{1+(mx)^2}({d} y^2+{d} z^2), \qquad B = -\frac{mx}{1+(mx)^2}({d} y\wedge {d} z).
\end{equation*}} :
\begin{align}
    &{\cal O}_H= m\int\limits_{\Sigma} \phi^x F^{-}_{yz}, \\
    &{\cal O}_N= -m\int\limits_{\Sigma} \phi^x F^{+}_{yz} +  m^2\int\limits_{\Sigma}  ({\phi^x})^2\partial \phi_z\bar{\partial}\phi_z, \\
    &{\cal O}_{\text{T-fold}}= -m\int\limits_{\Sigma} \phi^x F^{-}_{yz}+\sum_{n\geq 1}(-1)^n(n+1)(m)^{2n}\int\limits_{\Sigma} ({\phi^x})^{2n}\left( \partial \phi_y\bar{\partial}\phi_y + \partial \phi_z\bar{\partial}\phi_z - m\phi^xF^{-}_{yz} \right).
\end{align}
These are clearly not related simply by a change of sign of the chiral field along the direction the duality is being performed\footnote{Although they are related in this simple way to first order in the parameter $m$.}. The reason is that the dual descriptions of the theories involved different parameterisations of the backgrounds. In general, if we have two dual backgrounds $E = G+\delta E$ and $\tilde{E} = G+\delta \tilde{E}$, where $G$ is the background metric at the self-dual point, the deformation operators are given by
\begin{equation}
    {\cal O}_E = \delta E_{\mu\nu}\int\limits_{\Sigma} \partial \phi^\mu\bar{\partial}\phi^\nu, \qquad {\cal O}_{\tilde{E}} = \delta \tilde{E}_{\mu\nu}\int\limits_{\Sigma} \partial \phi^\mu\bar{\partial}\phi^\nu.
\end{equation}
If we try to T-dualise ${\cal O}_E$ explicitly in the $y$-direction say, this would simply correspond to $\partial \phi_y\rightarrow -\partial \phi_y$. However, this is not sufficient to relate ${\cal O}_E$ to $ {\cal O}_{\tilde{E}}$, since we also need $\delta E\rightarrow\delta \tilde{E}$, which is \textit{not} induced by the simple automorphism on the fields alone. From the worldsheet perspective, $\delta E$ and $\delta \tilde{E}$ are coupling constants for perturbations of two different, but related, theories. In order to relate these deformations, we must understand how the the coupling constants are related.

\subsubsection{Trivial torus bundles}

First, let us briefly verify the constant case, i.e. when $E$ has no coordinate dependence. This is essentially the same as \cite{Evans:1995su}, though we include it for completeness and to illustrate the general idea. Specifically, we want to use the stress tensor to show that the ${\cal O}$ transformation in the dual background can be deduced from the ${\cal O}$ transformation in the original background. As above, suppose we have some background $E = G + \delta E$, where $\delta E$ is not necessarily small (i.e. this is to arbitrary order), as well as the corresponding deformation operator ${\cal O}_E$. From this (together with ${\delta_{E}}$), we can find $\partial X_\mu(E)$ in terms of the self-dual point objects, and we know in this constant case that we simply reproduce the results of \cite{Evans:1995su}. Thus, the stress tensor, $T_E$, can then be T-dualised and we obtain the dual tensor $T_{\tilde{E}}$. We can then read off $\tilde{E}$ and therefore deduce the dual operator ${\cal O}_{\tilde{E}}$, where $\tilde{E} = G+\delta\tilde{E}$. Of course, in this constant background case, we already know that the relation between $E$ and $\widetilde{E}$ is given by the fractional-linear transformation
\begin{equation}\label{Buscher}
    \widetilde{E}=(aE+b)(cE+d)^{-1},
    \qquad 
    U=\left(
    \begin{array}{cc}
      a  & b \\
        c & d
    \end{array}
    \right)\in O(d,d;\Z),
\end{equation}
which leads to a complicated relationship between the deformation parameters in ${\cal O}_E$ and ${\cal O}_{\widetilde{E}}$ of the form 
\begin{equation}\label{BuscherE}
    \delta \widetilde{E}= (a(G+\delta E)+b)(c+d)^{-1}\sum_{n\geq 0}(-1)^n\left((c+d)^{-1}c\delta E\right)^n - G.
\end{equation}
Note that, in general, there is a non-trivial zeroth order term here, but for the cases we are considering (i.e. T-duality), it is always the case that $a+b = c+d = 1$ \cite{Hull:2009sg}, and so we have
\begin{equation}
    \delta \widetilde{E}= a\delta E\sum_{n\geq 0}(-1)^n\left(c\delta E\right)^n .
\end{equation}
The complication caused by the transformation of $\delta E$ reflects the fact that the way in which the data $\mathfrak{m}^\alpha$ parameterise the space of backgrounds depends on the duality frame chosen.

\subsubsection*{Circle example}

It is easiest to see this in a concrete example, so we briefly review how this works for the familiar case of the circle. The metric $G$ at the self-dual radius $\sqrt{G}$ is changed to $g$, with a corresponding change in the field $\phi$ to $X$, so that, using \eqref{dXcircle},
\begin{equation}
    \partial X=\frac{1}{2}\Big(\partial \phi+\bar{\partial}\phi\Big)+\frac{1}{2}gG^{-1}\Big(\partial \phi-\bar{\partial}\phi\Big).    
\end{equation}
As seen in section \ref{S: toroidal backgrounds}, this can be constructed from the deformation operator
\begin{equation}
    {\cal O}=\lambda\int_{\Sigma'}{d}^2 z\,\partial \phi(z)\bar{\partial}\phi(\Bar{z}),\qquad \lambda:=\frac{g-G}{G^2}.    
\end{equation}
The deformed stress tensor $T=g^{-1}\partial X\partial X$ may be written as
\begin{equation}\label{T}
    T=\frac{1}{4}g^{-1}\Big((1+gG^{-1})\partial \phi+(1-gG^{-1})\bar{\partial}\phi\Big)^2.
\end{equation}
Under the T-duality automorphism, $U(\partial \phi,\bar{\partial}\phi)U^{-1}= (-\partial \phi,\bar{\partial}\phi)$, so the dual stress tensor is (note the relative sign change between terms)
\begin{equation}
    \widetilde{T}=\frac{1}{4}g^{-1}\Big((1+gG^{-1})\partial \phi-(1-gG^{-1})\bar{\partial}\phi\Big)^2.
\end{equation}
$\widetilde{T}$ may be written in the form of (\ref{T}), but with $g$ replaced by $\tilde{g}= G^2/g$, thus recovering the standard Buscher rule for $d=1$. The dual theory can also be expressed as a deformation of the self-dual theory by the operator
\begin{equation}
    \widetilde{\cal O}=\widetilde{\lambda}\int_{\Sigma'}{d}^2 z\,\partial \phi(z)\bar{\partial}\phi(\Bar{z}),\qquad \widetilde{\lambda}(\lambda)=-\frac{\lambda}{1+\lambda}.    
\end{equation}
The relationship between the deformation parameters $\lambda$ and $\widetilde{\lambda}$ is an alternative writing of the Buscher rules and is an example of the transformation (\ref{Buscher}).

\subsubsection{Stress tensor deformations and T-duality}

In this section, we consider a more general case where the background has some coordinate dependence. The reason why the stress tensor was so useful for the constant case was because we could write down a general form for $T_E$ for any constant background $E$. When there is coordinate dependence, to do the same using the parallel transport method, we need to know explicitly what the coordinate dependence is.

As always, we start with a reference background, which we take to be a torus, tuned to the self-dual radius with coordinates $\phi^i$ and a single base direction with coordinate $\phi^x$, such that $\phi^{\mu}=(\phi^x,\phi^i)$. We assume that the base is $\R$, but may impose identifications so that it is an $S^1$. The action and stress tensor are
\begin{equation}
    S_0[X]=\int_{\Sigma}\partial \phi^x\bar{\partial}\phi^x+G_{ij}\partial \phi^i\bar{\partial}\phi^j,  \qquad  T_G= \partial \phi^x\partial \phi^x+G_{ij}\partial \phi^i\partial \phi^j.    
\end{equation}
This background is then deformed to a background of interest. The associated deformation operator is
\begin{equation}
    {\cal O}[\phi^x]=\int_{\Sigma}{d}^2z \,\delta E_{ij}(\phi^x)\partial \phi^i\bar{\partial}\phi^j,    
\end{equation}
where we define $\delta E_{ij}:=E_{ij}-G_{ij}$ and we allow this to be a function of $\phi^x$. It would be natural to express $\delta E_{ij}(\phi^x)$ in terms of a basis of functions on the base $f_I(\phi^x)$, so that there is a well-defined decomposition
\begin{equation}\label{deltaE}
\delta E_{ij}(\phi^x)=\sum_n\lambda_{ij}^{(I)}f_I(\phi^x),
\end{equation}
where the $\lambda_{ij}^{(I)}$ give a set of coupling constants for the deformation. Our approach will be to work on the cover of the base (in this case $\R$) and then impose identifications $\phi^x\sim \phi^x+2\pi$ after the deformation. This leads to a natural (although not unique) decomposition of the deformation operators ${\cal O}=\sum_I{\cal O}_I$, where
\begin{equation}
    {\cal O}_I[\phi^x]=\int_{\Sigma}\lambda_{ij}^{(I)}f_I(\phi^x)\partial \phi^i\bar{\partial}\phi^j.
\end{equation}
One then uses this deformation operator to deform the operators of the theory, such as the stress tensor:
\begin{equation}
    T_E=T_G+\delta_{\cal O}(T_G)+\delta_{{\cal O}^2}(T_G)+... .  
\end{equation}
The T-dual stress tensor is given by applying the automorphism\footnote{We assume the duality is performed along one of the fibre directions.} $T_{\widetilde{E}}=e^{iQ}T_Ee^{-iQ}$. This then gives a perturbative description of the dual stress tensor. To leading order, the deformations are related by
\begin{equation}
    T_G+\delta_{\widetilde{{\cal O}}}(T_G)+...=T_G+e^{iQ}\delta_{\cal O}(T_G)e^{-iQ}+...,    
\end{equation}
where we note that $T_G$ is invariant under the action of the automorphism. This is hard to calculate in practice and tends to yield complicated expressions on both sides, as the stress tensor written in terms of the reference background is likely to be a complicated and unfamiliar object. It is possible, in principle, to extract information on the dual background $\widetilde{E}_{ij}$, from which one could construct a dual sigma model. Given a generic enough deformation, one for which both the original and dual theories are particular examples, one can in principle construct the stress tensor of this generic theory in terms of the fields of the reference background. The couplings $\widetilde{\lambda}_{ij}^{(I)}$ of the dual theory can then be read off from $T_{\widetilde{E}}$, and the dual deformation $\delta\widetilde{E}_{ij}$ constructed as with (\ref{deltaE}). The dual couplings will be functions of the couplings of the original theory $\widetilde{\lambda}_{ij}^{(I)}(\lambda)$ and this relationship is, in essence, the Buscher rules relating the two backgrounds. This generalises the torus case (\ref{BuscherE}), a simple example of which is the relationship between the radii of a circle background and its dual.

It is worth stressing that the dual theory may be found in terms of its operators; however, identifying the explicit background for a sigma model construction is more involved. The procedure we have outlined here is the straightforward generalisation of that used in \cite{Evans:1995su} for toroidal target spaces. The simplifying feature there was the existence of universal coordinates with which to calculate.

\subsubsection*{Leading order deformations and duality}

In general, using this approach to deduce T-dual backgrounds is difficult to do computationally, so we will only demonstrate this to first order for a relatively simple case. In particular, we will suppose that we start from a trivial torus bundle as described above and that we deform the background in the fibres to some background $E_{ij}$, where
\begin{equation}
    E_{ij} = G_{ij} + \lambda \phi^x\delta E_{ij} +O(\lambda^2),
\end{equation}
where $\delta E$ is constant, and $\lambda$ is some small parameter. We will deduce a general form for the stress tensor $T_E$. Additionally, we will assume that the dual background $\tilde{E}$ is of the same form, i.e.
\begin{equation}
    \tilde{E}_{ij} = G_{ij} + \lambda \phi^x\delta \tilde{E}_{ij} + O(\lambda^2).
\end{equation}
This is of course true for the Nilfold and $H$-flux, for which we will verify this method explicitly. To compute $T_E$, let us first compute the deformation of $\partial \phi_\mu$. We have
\begin{equation}
    {\cal O}_E = \lambda\delta E_{ij}\int\limits_{\Sigma} \phi^x\partial \phi^i\bar{\partial}\phi^j.
\end{equation}
The calculation of the first order change in $\partial \phi_\mu$ closely follows the method set out in section \ref{S: toroidal backgrounds} and so we will be brief here. Taking the OPE with $\partial \phi_i$ and including the $\delta_{E}$ contribution 
\begin{equation}
    \delta_{E}\partial \phi_i = \frac{\lambda}{2}\delta E_{ij}\phi^x \partial \phi^j,
\end{equation}
which can be written in integral form as similar to (\ref{int_form}), gives
\begin{align}
    \delta\partial \phi_i(w) &= -\frac{\lambda}{2}\delta E_{ij}G^{jk}\int\limits_{\Sigma'} \frac{d^2z}{z-w}F^-_{xk}- \frac{\lambda}{2}\delta E_{ij}\oint\limits_{C_0'} \frac{d\phi^j(z,\Bar{z})}{z-w}\phi^x(z,\Bar{z}),
\end{align}
where we recall that $d\phi^\nu(z,\bar{z}) = \partial \phi^\nu(z) dz + \bar{\partial}\phi^\nu(\bar{z}) d\bar{z}$. For $\partial\phi_x$ there is no $\delta_E$ transformation, so we have
\begin{equation}
    \delta \partial\phi_x(w) = \frac{\lambda}{2}\delta E_{ij}\int\limits_{\Sigma'}\frac{d^2z}{z-w}\partial \phi^i(z)\Bar{\partial}\phi^j(\Bar{z}).
\end{equation}
Thus, we can now substitute this into the stress tensor to get the deformed stress tensor $T = g^{\mu\nu}\partial X_\mu(E)\partial X_\nu(E)$, where $E = g+B$, and if we write $g_{ij} = G_{ij} + \lambda\delta g_{ij}$ then, to first order, $\delta g^{ij} = -G^{ik}\delta g_{kl}G^{lj}$. After a short computation, we find that
\begin{align}
    T_E(w) &=  T_G(w)- \lambda\delta g_{ij}\phi^x(w,\Bar{w})\partial \phi^i(w)\partial \phi^j(\bar{w})+\lambda\partial \phi_x(w)\delta E_{ij} \int\limits_{\Sigma'} \frac{d^2z}{z-w}\partial \phi^i(z)\bar{\partial}\phi^j(\bar{z}) \notag \\
    &+ \lambda\partial \phi^i(w)\left(-\delta E_{ij}G^{jk} \int\limits_{\Sigma'} \frac{d^2z}{z-w}F^-_{xk}(z,\Bar{z})- \delta E_{ij}\oint\limits_{C_0'}\frac{\phi^x(z,\Bar{z})d\phi^j(z,\Bar{z})}{z-w} \right). \label{stressTensorXdependence}
\end{align}
We could also write the last term in integral form if it is more convenient, using \eqref{int_form}. This can now be used to compute the T-dual operator $O_{\tilde{E}}$. We will demonstrate this by doing the $H$-flux/Nilfold example explicitly.

\subsubsection*{Example: $H$-flux/Nilfold T-duality}

Let us look at how we can use the above to compute the T-dual to first order. Starting with the $H$-flux, we have $E_{ij} = G_{ij}+m\phi^x\delta E_{ij}+...$, where $\delta E_{ij}$ and the associated deformation operator are
\begin{equation}
    \delta E_{ij} = \begin{pmatrix}
    0&1\\
    -1&0
    \end{pmatrix},  \qquad  {\cal O}_H = m\int\limits_{\Sigma} \phi^x F^{-}_{yz}.
\end{equation}
Substituting in the results \eqref{delta_dX_H_flux}, \eqref{delta_dY_H_flux}, \eqref{delta_dZ_H_flux}, the stress tensor for the $H$-flux is given by
\begin{equation}
    T_{H}(w) = T_G(w) + m\varepsilon^{\mu\nu\rho}\partial \phi_{\mu}(w) \int\limits_{\Sigma'} \frac{d^2z}{z-w}F^{-}_{\nu\rho}(z,\Bar{z})+\delta_{\Lambda}T_H(w),
 \end{equation}   
where
\begin{equation} 
\delta_{\Lambda}T_H(w)= - m\varepsilon^{ij}\partial \phi_i(w)\oint\limits_{C_0'} \frac{\phi^x(z,\Bar{z})d\phi_j(z,\Bar{z})}{z-w} 
\end{equation}
is a gauge-dependent piece ($\varepsilon^{yz}=-\varepsilon^{zy}=1$). Let us now compute the T-duality transformation for this in the $y$-direction, which we recall is obtained via the transformation $\partial \phi_y\rightarrow -\partial \phi_y$. Doing this and using \eqref{int_form} gives
\begin{align}
    \tilde{T}_H(w) &= T_G(w) - m\partial \phi_x(w) \int\limits_{\Sigma'} \frac{d^2z}{z-w}F^{+}_{yz}(z,\Bar{z}) + m\partial \phi_y(w)\int\limits_{\Sigma'} \frac{d^2z}{z-w}F^{-}_{xz}(z,\Bar{z}) \notag\\\label{T_N}
    &+ m\partial \phi_z(w)\int\limits_{\Sigma'} \frac{d^2z}{z-w}F^{-}_{xy}(z,\Bar{z}) + m\partial \phi_y(w)\oint\limits_{C_0'} \frac{\phi^x(z,\Bar{z})d\phi_z(z,\Bar{z})}{z-w} \notag \\&+ m\partial \phi_z(w)\oint\limits_{C_0'} \frac{\phi^x(z,\Bar{z})d\phi_y(z,\Bar{z})}{z-w} + 2m\phi^x(w,\Bar{w})\partial \phi_y(w)\partial \phi_z(w).
\end{align}
Comparing with the general result \eqref{stressTensorXdependence}, we can read off that this dual background is $\tilde{E}_{ij}= G_{ij} + m\phi^x\delta \tilde{E}_{ij}$, where $\delta \widetilde{E}_{ij}$ and its associated deformation operator are
\begin{equation}
    \delta\tilde{E}_{ij} = \begin{pmatrix}
    0&-1\\
    -1&0
    \end{pmatrix},  \qquad  {\cal O}_{\tilde{E}} = -m\int\limits_{\Sigma} \phi^xF^{+}_{yz} = {\cal O}_N.
\end{equation}
We recognise this as the Nilfold background (\ref{nilfold}) to first order. We can also verify, using \eqref{dXNil}, \eqref{dYNil}, \eqref{dZNil}, that \eqref{T_N} is indeed the stress tensor of the Nilfold. Thus, the known duality is recovered to leading order.

\subsection{Making use of the doubled formalism}

The benefit of the doubled formalism is that T-duality is a symmetry of the sigma model and so many of the complications of the previous section do not arise in this framework. For torus bundles of the kind we have been considering (isometric, non-degenerating), there is an explicit doubled formalism, and the deformations may be understood in terms of a deformation of the doubled metric. As such the deformation operator ${\cal O}$ transforms naturally under $O(d,d;\Z)$ and T-duality may be simply understood. It is rare that we have a concrete doubled formalism\footnote{Identity structure manifolds seem to be the exception.} and so we do not expect to learn anything new, but it is useful to see how a doubled formalism may be put to good use when one is available.

The doubled action is given by\footnote{We explicitly discuss the case where only the fibres of the torus bundle are doubled. The generalisation to cases where all directions are doubled including the associated WZW term in the doubled action is expected to be straightforward.} (\ref{doubledS}) and can be written as $S = S_0+\Delta S_E$, where $S_0$ is the action for the flat doubled torus, and we expect that the associated deformation operators satisfy
\begin{equation}
    e^{iQ}{\cal O}_E(E)e^{-iQ} = {\cal O}_{E'}(E'),
\end{equation}
so the deformation operators of dual backgrounds are indeed T-dual in the doubled formalism. This is easiest to understand by looking at an example, so let us show this explicitly for the Nilfold/$H$-flux case. Starting with the Nilfold, with metric (\ref{nilfold}), we have the deformation
\begin{equation}
\Delta S_N[X]=\frac{1}{2}\int\limits_{\Sigma}\delta{\cal H}_{IJ}(X)\partial \mathbbold{X}^I\bar{\partial}\mathbbold{X}^J,
\qquad
    \delta{\cal H}_{IJ}(X) = \begin{pmatrix}
    0 & -mX & 0 & 0 \\
    -mX & (mX)^2 & 0 & 0 \\
    0 & 0 & (mX)^2 & mX \\
    0 & 0 & mX & 0
    \end{pmatrix}.
\end{equation}
Now, let us do the same for the $H$-flux. In this case, we have the deformation operator
\begin{equation}
\Delta S_H[X]=\frac{1}{2}\int\limits_{\Sigma}\delta{\cal H}'_{IJ}(X)\partial \mathbbold{X}^I\bar{\partial}\mathbbold{X}^J,
\qquad
\delta{\cal H}'_{IJ}(X) = \begin{pmatrix}
    (mX)^2 & 0 & 0 & mX \\
    0 & (mX)^2 & -mX & 0 \\
    0 & -mX & 0 & 0 \\
    mX & 0 & 0 & 0
    \end{pmatrix}.
\end{equation}
Thus, we have the marginal operators 
\begin{align}
    &{\cal O}_N[\phi^x] = \frac{1}{2}\int\limits_{\Sigma}  \left(- m\phi^x(\partial \phi_y\bar{\partial}\phi_z + \partial \phi_z\bar{\partial}\phi_y - \partial \tilde{\phi}_y\bar{\partial}\tilde{\phi}_z - \partial \tilde{\phi}_z\bar{\partial}\tilde{\phi}_y) +(m\phi^x)^2(\partial \tilde{\phi}_y\bar{\partial}\tilde{\phi}_y + \partial \phi_z\bar{\partial}\phi_z)\right), \\
    &{\cal O}_H[\phi^x] = \frac{1}{2}\int\limits_{\Sigma} \left( - m\phi^x(\partial \tilde{\phi}_y\bar{\partial}\phi_z + \partial \phi_z\bar{\partial}\tilde{\phi}_y - \partial \phi_y\bar{\partial}\tilde{\phi}_z - \partial \tilde{\phi}_z\bar{\partial}\phi_y) +(m\phi^x)^2(\partial \phi_y\bar{\partial}\phi_y + \partial \phi_z\bar{\partial}\phi_z)\right),
\end{align}
and we can see that, under the duality transformation $\phi^y\leftrightarrow \widetilde{\phi}^y$, we do indeed have $e^{iQ}{\cal O}_Ne^{-iQ} = {\cal O}_H$, as expected. The utility of the doubled formalism is that it gives a duality-covariant parameterisation of this limited space of backgrounds.

\section{Discussion}\label{S: Discussion}

The goal of this paper was to expand on the ideas of \cite{Sonoda:1991mv, Sonoda:1992hd, Ranganathan:1993vj} and in particular to push the idea of CFT deformations further. One of our main goals was to show how our formalism extends to more general QFTs, and this was demonstrated by looking at the $H$-flux and Nilfold. A crucial aspect of this paper was the use of the ${\delta_{E}}$ transformation, which had to be incorporated to deal with the deformation of operators with spacetime indices, i.e. non-scalar operators on the target space. The stress tensor, which traditionally has been the operator of interest, carries no target space indices and so there is no ${\delta_{E}}$ contribution to its deformation. We have tried to widen the scope of QFT deformations by describing a formalism which can be applied to any operator in a wide class of theories. We built on the methods of \cite{Sen:1990hh} and described a prescription for extending these methods to higher orders. We showed that this prescription works as expected for simple CFTs, and we suspect that it can be extended to non-CFT cases as well. 

%We described connections in general and in particular focused on the $\hat{\Gamma}$ and $c$ connections described in \cite{Ranganathan:1993vj}. We found that the $\hat{\Gamma}$ connection was closely related to the doubled geometry when looking at the general torus deformation. Here, we showed how the doubled formalism naturally recovered the deformation results obtained by assuming universal coordinates, and that this was precisely what we got using the $\hat{\Gamma}$ connection, since this connection implies universal coordinates, as was also shown. The conclusion here was that universal coordinates were equivalent to preserving the $O(d,d)$ structure that is explicit in the doubled geometry.

For marginal deformations of free CFTs, the deformation preserves the subspace spanned by $\partial X$ and $\bar{\partial}X$. For the non-CFT cases that we looked at, we found that the $X$-dependence made a significant difference to the deformation of $\partial \phi_\mu$, mixing in terms of the form $\phi^x\partial \phi_\nu$ and $\phi^x\bar{\partial}\phi_\nu$, as well as the $\bar{\partial}\phi_\nu$ terms seen in trivial torus bundles. More general deformation operators would lead to a more general mixing of the operator basis.
%Normally, this dependence is ignored and most discussions of the $H$-flux and other such `twisted torus' backgrounds resort to the supergravity description, i.e. to leading order in $\alpha'$ \cite{Hull:2009sg}. Here, we have tried to work to all orders in $\alpha'$ and to include all stringy corrections induced by the $X$-dependence of these backgrounds. To first order, we have derived deformations of the $\partial \phi_\mu$ operators which can be thought of as stringy-corrected versions of the deformations that we would get if we had simply worked at the level of the supergravity \cite{Evans:1995su, Mahmood:2020mtq}.
In theory, these ideas can be used to compute the higher order deformations as well, though this would be a computational challenge. Moreover, it seems unlikely that the notion of universal coordinates generalises straightforwardly to more complicated backgrounds.

We also discussed T-duality in the context of these non-CFT backgrounds.
%Here, it was unclear whether the T-duality between the $H$-flux and Nilfold would still hold with the corrections. We found that it did indeed hold to first order, and we described how the T-dual background could be constructed for a general background when described as a deformation of some initial background. Specifically, the key idea was to construct a general form of the stress tensor which could then be used to read off the T-dual background.
This is a direct generalisation of the ideas employed in \cite{Evans:1995su} and has the potential to be applicable to a much wider class of backgrounds. Computationally, using this approach for a completely general background seems like a difficult task beyond first order, though in principle it at least extends the applicability of T-duality beyond the context of the Buscher rules and leads to a different perspective on the duality and its relationship with symmetry enhancement. 

%We also reviewed an approach, known as the background field method, which allows us to express a background in a gauge covariant manner. We showed how this could be used to demonstrate that the choice of branch cut for the $H$-flux background has no bearing on the physics, as we would have hoped. It would be interesting to develop this approach further and to see how the formalism discussed in this paper could be written starting from a general background field expansion. This would essentially be a gauge covariant version of the formalism we have laid out in this paper.

One of the benefits of the approach explored here is that it should be applicable to a very large class of sigma models. Of particular interest would be to explore whether these techniques can be applied to effective worldsheet theories, where specific quantum effects (such as worldsheet instantons) have been incorporated. A specific example is the KK-monopole/NS5-brane duality. Here, worldsheet instantons localise the solutions, breaking the global worldsheet symmetry associated with a target space isometry \cite{Gregory:1997te, Tong:2002rq}. Following the notation of \cite{Chaemjumrus:2019ipx}, the NS5-brane background is given by
\begin{equation}
    ds^2_{10} = V(x^i)ds^2(\R^4) + ds^2(\R^{1,5}),
\end{equation}
where the $x^i$ are coordinates on the transverse space $\R^4$, $ds^2(\R^4)$ and $ds^2(\R^{1,5})$ are the standard flat metrics, the function $V(x^i)$ is a harmonic function and the $H$-flux is given by $H_{ijk} = -\epsilon_{ijkl}\delta^{lm}\partial_m V$.

In the case where $V = V(r)$\footnote{In \cite{Chaemjumrus:2019ipx}, they are mainly interested in the case where $V = V(\tau)$, i.e. $V$ only depends on a single coordinate $\tau$. This is because this case can be interpreted as a $T^3$ with $H$-flux fibred over a line, which is relevant in constructing certain types of hyperkahler manifolds. For this case, the T-duality is relatively easy to study since it reduces to the T-duality chain of the $H$-flux, which is well-known \cite{Hull:2009sg}.}, where $r = |(x^1,x^2,x^3)|$, $V$ is given by
\begin{equation}
    V(r) = \frac{1}{g^2}+\frac{1}{2r}.
\end{equation}
As explained in \cite{Tong:2002rq}, this \textit{smeared} NS5-brane is localised in the $\theta$ direction when worldsheet instantons are taken into account. This results in the replacement $V(r)\rightarrow V(r,\theta)$, where 
\begin{equation}\label{V}
    V(r,\theta) = \frac{1}{g^2}+\frac{1}{2r}\left( 1 + \sum_{k\geq 1}\sum_{\pm} e^{-kr\pm ik\theta} \right).
\end{equation}
In \cite{Mahmood:2020mtq}, it was found that the T-duality automorphism did not have a well-defined action on the worldsheet coordinate $X$, but that it \textit{did} have a well-defined action on exponentials of the form $e^{inX_L}$, for integers $n$. Therefore, it may be that the T-dual of the $\textit{localised}$ NS5-brane can be computed explicitly using the T-duality methods of \cite{Mahmood:2020mtq}\footnote{We note that the localising instantons are not arising from non-perturbative effects in $\lambda$. Reinserting the $\alpha'$ dependence \cite{Harvey:2005ab}: $ g = R/\sqrt{\alpha'}$, $r = Rr'/\alpha'$,  $\theta = R\theta'$ and $ds^2 = -ds'^2/\alpha'$, giving 
\begin{equation}
    V = \frac{\alpha'}{R^2}\left(1 + \frac{R}{2r'}\right).
\end{equation}
To view the NS5-brane as a deformation of the flat background, we introduce a parameter $\lambda$ and so rescale $R\rightarrow \lambda R$, which gives
\begin{equation}
    V \rightarrow \frac{\alpha'}{(\lambda R)^2}\left(1 + \frac{\lambda R}{2r'}\right).
\end{equation}
In these coordinates, the instanton corrections become $e^{-\frac{kr'R}{\alpha'}}$ and we see that, with $\lambda$ included, the instanton corrections go like $e^{-\frac{kr'R\lambda}{\alpha'}}$, i.e. they are analytic in $\lambda$.}.

Specifically, one could start with an \emph{effective} action, based on the instanton-corrected potential (\ref{V}), imagine tuning $\theta$ to be at the self-dual radius\footnote{There is a tension here. The effective action describes the large volume supergravity limit. At the self-dual radius, we would need to include other $\alpha'$ corrections to the supergravity. As such this discussion is only illustrative.} and then perform a T-duality automorphism on this effective sigma model to recover the KK-monopole background. The analysis of \cite{Tong:2002rq} predicts
\begin{equation}
    V(r,\theta) = \frac{1}{g^2}+\frac{1}{2r}\left( 1 + \sum_{k\geq 1}\sum_{\pm}m(k) e^{-kr\pm ik\theta} \right),
\end{equation}
where the $m(k)$ are unknown constants. In principle, the results of \cite{Mahmood:2020mtq} predict a relationship between the $m(k)$ and their counterparts $\tilde{m}(k)$ in the KK-monopole solution.

It would be interesting to apply the construction developed in this paper to these localised backgrounds. Usually, one performs the duality with the smeared backgrounds and then localises by incorporating the instanton corrections, but is it possible to incorporate the instanton effects directly? The idea would be to apply the construction above to the quantum effective worldsheet action in which these instanton effects have already been incorporated. 

Aside from this particular example, it would be interesting to see how the results of this paper can be extended to fermions and a supersymmetric context. It would also be interesting to see if there is a way the methods we have developed can be extended to higher orders, particularly for the non-CFT cases, or if there are any non-perturbative results that could be derived. 

Finally, we note that there is also a close connection with the constructions discussed here and the linear sigma model approach used to prove mirror symmetry \cite{Hori:2000kt,Hori:2003ic}. There, the starting point is an `off-shell' model, in that it does not describe a genuine string background. The model then flows to a string background under renormalization, constrained by the superpotential. It would be interesting to explore this connection further.

%Finally, something which we have not discussed here is the path integral approach. Our analysis has been entirely from the perspective of the operator formalism. This is largely because of the convenience of working with OPEs. However, it would be intriguing to see how this works in the path integral formalism and if there is any additional insight that can be gained. 

\begin{center}
    \textbf{Acknowledgments}

This work has been partially supported by STFC consolidated grant ST/T000694/1.

RR is grateful for the continuing support of the Avery-Tsui Foundation.
\end{center}

\appendix

\section{$H$-flux Mode Transformation}\label{A:$H$-flux}

Given everything that we have done, we should be in a position to compute the deformation to the $H$-flux background, at least to first order. We will go through the calculation of $\delta\alpha_n^y$ in detail.

\subsection{$\delta_{\cal O}$ calculation}

Taking the OPE of $\partial \phi_y(w)$ with the deformation operator and integrating by parts gives 
\begin{equation}\label{$H$-flux_O}
    -\frac{1}{2}m\left(\oint\limits_{C'} d\bar{z} \frac{\phi^x(z,\Bar{z})\bar{\partial}\phi_z(\bar{z})}{z-w} + \int\limits_{\Sigma'} d^2z \frac{\partial \phi_x(z) \bar{\partial}\phi_z(\Bar{z})}{z-w}\right),
\end{equation}
where $C' = C_0'\cup C_w'$. Let us deal with the contour integrals first. Expanding in terms of modes, we have:
\begin{equation}
    \frac{im}{2\sqrt{2}}\sum_{\substack{k\geq 0\\ p}}\bar{\alpha}_p^zw^{-k-1}\oint\limits_{C'} d\bar{z}\,\phi^x(z,\bar{z})
    z^k\bar{z}^{-p-1},\label{$H$-flux_O_contour}
\end{equation}
where
\begin{equation}
    \phi^x(z,\bar{z})=
    x - \frac{i}{\sqrt{2}}(\alpha_0^x\log z  +\bar{\alpha}_0^x\log \bar{z}) + \frac{i}{\sqrt{2}}\sum_{n\neq 0}\frac{1}{n}(\alpha_n^xz^{-n} + \bar{\alpha}_n^x\bar{z}^{-n}).
\end{equation}
Let us compute the kinds of integrals that appear here. The most involved ones are those with logs, which generally looks like
\begin{equation}
    \oint\limits_{C'} d\bar{z}z^k \bar{z}^{-p-1}\log z,
\end{equation}
and if we are on a contour where $|z|=r$ then this simplifies to 
\begin{equation}
    -r^{-2p}\oint\limits_{C'} dz z^{k+p-1}\log z.
\end{equation}
Let us first deal with the circle around $z=0$. We will take the contour to have radius $\epsilon$ and take the limit $\epsilon\rightarrow 0$ at the end. There is a branch point at $z=0$, but, as discussed in section \ref{S: Nonlinear Sigma Models and Off-shell string theory}, the choice of branch simply amounts to a choice of gauge. Therefore, we will still consider a circle around $z=0$, bearing in mind that there is a gauge-dependent piece. Setting $z=\epsilon e^{i\theta}$, the integral becomes 
\begin{equation}
    -\frac{\epsilon^{k-p}}{2\pi}\int\limits_0^{2\pi} d\theta e^{i(k+p)\theta}(\log \epsilon + i\theta) = \left\{ \begin{matrix}
        \frac{\epsilon^{k-p}}{2\pi i(k+p)^2}, \qquad k+p\neq 0, \\
        -\epsilon^{2k}(\log \epsilon + i\pi), \qquad k+p=0.
    \end{matrix} \right.
\end{equation}
In the above integral, the terms which come from the $i\theta$ part of the integral are the branch-dependent part, and if we neglect these terms we get
\begin{equation}\label{log}
    -\frac{\epsilon^{k-p}}{2\pi}\int\limits_0^{2\pi} d\theta e^{i(k+p)\theta}(\log \epsilon + i\theta) = -\delta_{k+p, 0}\epsilon^{2k}\log \epsilon  + ...,
\end{equation}
where $...$ represents the branch-dependent contributions.
We also need to check whether there is any contribution from the $z=w$ boundary. There is no branch cut here, so we can just use a circular contour with radius $\epsilon\rightarrow 0$, so we have
\begin{equation}
    \frac{\epsilon}{2\pi}\int\limits_0^{2\pi} d\theta e^{i\theta}w^k\bar{w}^{-p-1}\left(1+\frac{\epsilon}{w}e^{i\theta}\right)^k\left(1+\frac{\epsilon}{\bar{w}}e^{-i\theta}\right)^{-p-1}\log(w+\epsilon e^{i\theta}),
\end{equation}
which we can see vanishes in the limit $\epsilon \rightarrow 0$, so there is no contribution here. 
Similarly, 
\begin{equation}\label{log_bar}
    \oint\limits_{C'} d\bar{z}z^k\bar{z}^{-p-1}\log \bar{z} = -\delta_{k+p,0}\epsilon^{2k}\log \epsilon  - ...,
\end{equation}
where the minus before the $...$ indicates that the branch-dependent contributions are minus those of \eqref{log}.

The other type of integral which we are interested in is
\begin{equation}
    \oint\limits_{C'} d\bar{z} z^{k-n}\bar{z}^{-p-1}.
\end{equation}
This is fairly straightforward and we simply need to compute the contributions from contours around $z=0,w$ as usual. We will state the results. From $z=0$, we get $-\epsilon^{k-n-p}\delta_{k-n+p,0}$. From $z=w$, we once again find that there is no contribution when we take the radius to zero. 

Thus, we now have all of the integrals we need to compute \eqref{$H$-flux_O_contour}. Substituting everything in and neglecting branch-dependent contributions, we have\footnote{Where the $n=0$ case is understood as the limit
$\lim_{n\rightarrow 0}\frac{\epsilon^{k-n-p}}{n} = -\epsilon^{k-p}\log\epsilon$.}
\begin{align}\label{contour1}
    -\frac{im}{2\sqrt{2}}\sum_{k\geq 0}\sum_p w^{-k-1}\bar{\alpha}_p^z\lim_{\epsilon\rightarrow 0}\Bigg[&
    x\epsilon^{k-p}\delta_{k+p,0}+\frac{i}{\sqrt{2}}\sum_{n}\frac{1}{n}\epsilon^{k-n-p}\left(\alpha_n^x\delta_{k-n+p,0}+\bar{\alpha}_n^x\delta_{k+n+p,0} \right)
    \Bigg].
\end{align}

\subsubsection*{A note on evaluating contour integrals}

The above calculation is perfectly valid, but it is not the only way of evaluating the contour integral. To illustrate the idea, we will look at the much simpler case of the circle CFT of radius $R$. Here, as we saw earlier, if we wish to deform $\partial X(R)$ to the circle of radius $R+\delta R$, the $\delta_{\cal O}$ part of the deformation involves taking the OPE with the marginal operator, and we end up with an integral proportional to
\begin{equation}
    \oint\limits_{C_0^\epsilon} \frac{dz}{z^2(w-z)}\bar{\partial}X(\bar{z}), 
\end{equation}
as well as an integral around $w$ which is unimportant here. Now, the usual way of evaluating this would be to expand $\bar{\partial}X(\bar{z})$ and $(w-z)^{-1}$ in powers of $z$, using the fact that $|z|=\epsilon$, and then compute the resulting integrals. Doing this, we get
\begin{equation}\label{contour_z_less_w}
    \oint\limits_{C_0^\epsilon} \frac{dz}{z^2(w-z)}\bar{\partial}X(\bar{z}) = -\frac{i}{\sqrt{2}}\sum_{m\geq 0}\bar{\alpha}_{-m}\epsilon^{-2(-m+1)}w^{-m-1},
\end{equation}
and, together with the ${\delta_{E}}$ transformation, we would compare coefficients to get the mode transformations for $\alpha_n$, $n\geq 0$, and then argue that this must extend to all $n$ to preserve commutation relations. However, alternatively, we could treat the contour as a contour around the singularity at \textit{infinity}. This then changes the expansion of $(w-z)^{-1}$, since 
\begin{equation}
    \frac{1}{w-z} = \sum_{m\geq 0}w^{-m-1}z^m, \qquad |z|<|w|,
\end{equation}
\begin{equation}
    \frac{1}{w-z} = -\sum_{m < 0}w^{-m-1}z^m, \qquad |z|>|w|.
\end{equation}
Thus, if we set $y = 1/z$ and use
\begin{equation}
    \frac{dz}{w-z} = -\frac{dy}{y(wy-1)}, 
\end{equation}
we instead find that 
\begin{equation}
    \oint\limits_{C_0^\epsilon} \frac{dz}{z^2(w-z)}\bar{\partial}X(\bar{z}) = -\frac{i}{\sqrt{2}}\sum_{m\geq 0, n}\bar{\alpha}_n\epsilon^{-2(n+1)}w^m\oint\limits_{|y|=1/\epsilon} dy y^{-n+m}, 
\end{equation}
which, after evaluating and relabelling, gives
\begin{equation}
    -\frac{i}{\sqrt{2}}\sum_{m< 0}\bar{\alpha}_{-m}\epsilon^{-2(-m+1)}w^{-m-1},
\end{equation}
i.e. the same as \eqref{contour_z_less_w}, but with $m<0$. We can now compare coefficients as we would normally do and obtain the mode transformation for $\alpha_n$, $n<0$, and clearly this agrees with what we would get if we did it the `canonical' way, as we would hope. For the circle case, apart from giving a way to directly compute the mode expansions for \textit{all} modes (instead of only half and then inferring the other half), there is no particular benefit here. However, for the $H$-flux case, it makes the prescription much clearer since we have 2d integrals where the region of integration includes both $|z|>|w|$ and $|z|<|w|$. We come to this calculation now.

\subsubsection*{The integral over the worldsheet}

Let us evaluate the second term in \eqref{$H$-flux_O}. First, we look at the region $\epsilon<|z|<|w|$, where we have
\begin{align}
    &-\frac{1}{2}m\int\limits_{\substack{\Sigma'\\|z|<|w|}} \frac{d^2z}{z-w} \partial \phi_x(z)\bar{\partial}\phi_z(\Bar{z})=-\frac{m}{4}\sum_{\substack{n,p\\k\geq 0}}\alpha_n^x\bar{\alpha}_p^zw^{-k-1}\int\limits_{\substack{\Sigma'\\|z|<|w|}} d^2z z^{k-n-1}\bar{z}^{-p-1}.
\end{align}
This is straightforward once we set $z=re^{i\theta}$, so we will simply state the results. We get
\begin{equation}\label{2dint}
    \frac{m}{8}\sum_{\substack{p\\k\geq 0}} \frac{\alpha_{k+p}^x \bar{\alpha}_p^z}{p}w^{-k-1}\left(|w|^{-2p}-\epsilon^{-2p}\right),
\end{equation}
where the $p=0$ term is understood in terms of a limit in a similar way to \eqref{contour1}.
For $|z|>|w|$, to deal with the singularity at infinity, we will regularise the integral by integrating over the region $|w|<|z|<1/\epsilon$. The calculation is very similar and we get
\begin{equation}
    \frac{m}{8}\sum_{\substack{p\\k< 0}} \frac{\alpha_{k+p}^x \bar{\alpha}_p^z}{p}w^{-k-1}\left(|w|^{-2p}-\epsilon^{2p}\right).
\end{equation}
Thus, up to divergences in the $\epsilon\rightarrow 0$ limit, we get the same result in both regions $|z|<|w|$ and $|z|>|w|$, except for the range of the summation variable $k$. As we mentioned when discussing the circle case above, we want this to be the case so that our results are the same whichever region we look at and whichever method we use. 

Now, we would like to combine \eqref{contour1} and \eqref{2dint} to obtain $\delta_{\cal O}\alpha_n^y$. As discussed above, we have different prescriptions depending on whether $n\geq 0$ or $n<0$. For $n\geq 0$, which is the case we usually focus on, we choose the contour integral representation where the contour is $|z|=\epsilon$, and we take the part of the 2d integral where $|z|<|w|$. To read off the deformation of $\alpha^y_n$, as usual we must take the coefficient of $w^{-n-1}$ in \eqref{contour1}, \eqref{2dint}. However, note that the coefficient in \eqref{2dint} has $w$-dependence. To deal with this, we note that, when we are extracting the $w^{-n-1}$ coefficient, formally what we are doing is multiplying by $w^n$ and doing the contour integral $\oint_{|w|=\eta} dw$, for some constant $\eta$. When the coefficient is independent of $w$, this amounts to simply reading off the coefficient and the result is independent of $\eta$. However, in the case of \eqref{2dint}, we find that extracting the $w^{-n-1}$ coefficient amounts to setting $|w|=\eta$. Thus, we obtain
\begin{equation}
    \delta_{\cal O}\alpha^y_n=x{\cal A}_{n}\bar{\alpha}_{-n}^z+\sum_p{\cal B}_{np}\bar{\alpha}_{p}^z\alpha_{n+p}^x + \sum_p{\cal C}_{np}\Bar{\alpha}_p^z\bar{\alpha}^x_{-n-p}, \qquad n\geq 0,
\end{equation}
where
$$
    {\cal A}_{n} = -\frac{im}{2\sqrt{2}}\lim_{\epsilon\rightarrow 0}\epsilon^{2n}, \qquad {\cal B}_{np} = \frac{m}{4}\lim_{\epsilon\rightarrow 0}\left( \frac{\epsilon^{-2p}}{n+p} + \frac{\eta^{-2p} - \epsilon^{-2p}}{2p} \right), \qquad
{\cal C}_{np} = -\frac{m}{4}\lim_{\epsilon\rightarrow 0}\frac{\epsilon^{2n}}{n+p}.
$$
Given that the above is for $n\geq 0$, we can take the limits for ${\cal A}_n$ and ${\cal C}_{np}$ since these have no divergences and, extending the result to $n<0$, we get
\begin{align}
    &\delta_{\cal O}\alpha_n^y = \sum_p{\cal B}_{np}\bar{\alpha}_{p}^z\alpha_{n+p}^x, \qquad n\neq 0,\\
    &\delta_{\cal O}\alpha_0^y = -\frac{imx}{2\sqrt{2}}\Bar{\alpha}^z_{0}+ \sum_p{\cal B}_{0p}\bar{\alpha}_{p}^z\alpha_{p}^x - \frac{m}{4}\sum_p \frac{1}{p}\Bar{\alpha}^z_p\Bar{\alpha}^x_{-p}.
\end{align}
If we wanted to compute the $n<0$ case directly, we could use the corresponding integral results for $|z|>|w|$. As discussed above, both methods should give the same results. 

Some comments are in order:
\begin{itemize}
    \item We have left the limits in ${\cal B}_{np}$ unresolved in the above deformation. This is because the way we deal with these limits will depend on the connection we choose, i.e. the way we choose to regularise divergences. Recall that, in the circle case, we found from \eqref{c_conn} that there were no divergent terms and we could take the limit $\epsilon \rightarrow 0$ for all terms, which resulted in $\delta_{\cal O}\partial X=0$ for the $c$ and $\Bar{c}$ connections. However, here we find that, due to the $X$-dependence, we now \textit{do} have non-zero terms in the limit $\epsilon\rightarrow 0$.
    \item Note that this result depends on $|w|=\eta$, as opposed to the CFT case. Given that we no longer have conformal symmetry, the distance from the origin of the operator we are deforming does indeed have an effect on the transformation.
    \item Finally, note that, in the adiabatic limit where $X\rightarrow x$, the above result reduces to the expected result for a flat torus, i.e. the ${\cal O}$ deformation is given solely by the ${\cal A}_n$ term.
\end{itemize}

\subsection{${\delta_{E}}$ calculation}
From earlier considerations, we know that the ${\delta_{E}}$ transformation is given by 
\begin{equation}
    \delta_{E} \partial \phi_y= \frac{1}{2}m\phi^x\partial \phi_z.
\end{equation}
In order to extract the transformation of the modes, we can rewrite this as an integral by noting that\footnote{Where we have used
$$
\frac{\partial}{\partial \bar{z}}\left(\frac{1}{z-w}\right)=\delta^2(z-w),
$$
and integrated by parts.}
\begin{align}
    0 = \int\limits_{\Sigma'} d^2z \bar{\partial}\left(\frac{1}{z-w} \right)\phi^x(z,\Bar{z})\partial \phi_z(z)= \oint\limits_{C'} \frac{dz}{z-w}\phi^x(z,\Bar{z})\partial \phi_z(z) - \int\limits_{\Sigma'} \frac{d^2z}{z-w}\bar{\partial}\phi_x(\Bar{z})\partial \phi_z(z).\label{Hflux_P_int}
\end{align}
Now, let us look specifically at the contour integral around the point $w$. If we expand $\phi^x(z,\Bar{z}), \partial \phi_z(z)$ around $w$, this is 
\begin{align}
    \oint\limits_{C_w'} \frac{dz}{z-w}&\left(\phi^x(w, \Bar{w}) + ((z-w)\partial \phi_x(w)+...)+((\bar{z}-\bar{w})\bar{\partial}\phi_x(\Bar{w})+...)\right)\notag \\
    &\times\left(\partial \phi_z(w) + (z-w)\partial ^2\phi_z(w)+... \right),
\end{align}
and on the contour $|z-w|=\epsilon$ we can set $\bar{z}-\bar{w} = \epsilon^2/(z-w)$, which gives infinitely many non-zero terms in the integral above. However, in the limit $\epsilon\rightarrow 0$ only one term survives, so we end up with
\begin{equation}
    \oint\limits_{C_w'} \frac{dz}{z-w}\phi^x(z,\Bar{z})\partial \phi_z(z) = \phi^x(w,\Bar{w})\partial \phi_z(w),
\end{equation}
and so, going back to \eqref{Hflux_P_int}, we find that
\begin{equation}\label{int_form}
    \frac{1}{2}m \phi^x(w,\Bar{w})\partial \phi_z(w) = \frac{1}{2}m\int\limits_{\Sigma'} \frac{d^2z}{z-w}\bar{\partial}\phi_x(\Bar{z})\partial \phi_z(z) - \frac{1}{2}m\oint\limits_{C_0'}\frac{dz}{z-w}\phi^x(z,\Bar{z})\partial \phi_z(z).
\end{equation}
Now we simply have to compute these integrals. The second integral was already dealt with in the ${\cal O}$ case. The first is very similar to what we had in the ${\cal O}$ case, but not quite the same. However, the details are the same in essence, so we will simply state the final result. We find:
\begin{equation}\label{delta_E_hflux_mode}
    \delta_{E}\alpha^y_n=\frac{imx}{2\sqrt{2}}\alpha_{n}^z+\sum_p{\cal D}_{np}\alpha_{p}^z\bar{\alpha}_{p-n}^x +\sum_p{\cal E}_{np}\alpha_p^z\alpha^x_{n-p},
\end{equation}
where
\begin{align}
    &{\cal D}_{np} = -\frac{m}{4}\lim_{\epsilon\rightarrow 0}\left( \frac{ \eta^{-2(p-n)} + \epsilon^{-2(p-n)} }{2(p-n)}\right), \qquad {\cal E}_{np} = -\frac{m}{4}\frac{1}{n-p},
\end{align}
where we have also used the same notation as above for cases where the denominator seems to vanish, i.e. these correspond to log terms.\footnote{For the terms where there is no power of $\epsilon$ in the numerator, we do as follows. Take $\frac{1}{n-p}$ as an example. Look at $\frac{\epsilon^{n-p-k}}{n-p}$ in the limit $n-p\rightarrow0$ and then set $k=n-p$.} Note that the first and last terms in \eqref{delta_E_hflux_mode} are independent of $\epsilon$, which is reassuring given the results we have derived previously for the flat torus case. Once again, the adiabatic limit, i.e. the term proportional to $x$, gives us what we would expect.

\section{Higher Order Deformations}\label{A: Sen}

\subsection{Multiple ${\cal O}$ insertions}

Our prescription for deforming an operator involved the insertion of a deformation operator ${\cal O}=\int\limits O$, which was then integrated over the worldsheet with discs removed around insertion points. The choice of connection gave a prescription for how these discs are defined. However, if we have multiple insertions of $\cal O$ then more information is needed. Here, we give a prescription that reproduces the expected results.

Suppose we have some operator ${\cal A}(w, \Bar{w})$ which we wish to deform (with target space indices suppressed). As given in section \ref{SSS: Higher orders} for $\partial X$, we define the ${\cal O}^n$ operator insertion as
\begin{equation}
    \int\limits_{\Sigma_n} d^2z_n ... \int\limits_{\Sigma_1} d^2z_1 O(z_n, \Bar{z}_n) ... O(z_1, \Bar{z}_1) {\cal A}(w, \Bar{w}),
\end{equation}
where we recall that the domain of integration $\Sigma_i$ is 
\begin{equation}
    \Sigma_i = \{ z_i \in \mathbb{C} | \quad |z_i|\geq \epsilon, |z_i-w|>0, |z_i-z_j|\geq \epsilon \quad \forall j>i  \}. 
\end{equation}
Once we have this prescription, we can then take OPEs between the various operators and explicitly compute the integral to any order that is desired. As an example, we will look at the $({\cal O}+{\delta_{E}})^2$ calculation for $\partial X_\mu$ for a CFT deformation $g\rightarrow g+\delta g$.

\subsection{$({\cal O}+{\delta_{E}})^2$ for $\partial X_\mu$}

For simplicity, we will suppose that there are no $B$-field deformations involved. We know that, given a deformation $g\rightarrow g+\delta g$, if we have operator insertion 
\begin{equation}
    {\cal O} = \delta g_{\mu\nu}\int\limits_{\Sigma} \partial X^\mu\bar{\partial}X^\nu,
\end{equation}
as well as the ${\delta_{E}}$ operator \eqref{delta_E}, we get the deformation 
\begin{equation}
    \delta \partial X_\mu = \frac{1}{2}\delta g_{\mu\nu}(\partial X^\nu - \bar{\partial}X^\nu),  
\end{equation}
and this is in fact the full transformation. We know this because we already have the full transformation from other methods, such as universal coordinates. If this is the case, it should be that all contributions from higher power insertions of ${\cal O}$ and ${\delta_{E}}$ cancel, e.g. at second order in $\delta g$, we expect
\begin{equation}
    (\delta_{{\cal O}^2}+\delta_{\cal O}\delta_{E}+\delta_{E}\delta_{\cal O}+\delta_{E}^2)\partial X_\mu = 0.
\end{equation}
Similar results should hold for higher orders\footnote{Note that this is only true for $\partial X_\mu$ and is \textit{not} true for $\partial X^\mu$. This is because the $\partial X^\mu$ transformation involves the inverse metric, which induces corrections to all orders in $\delta g$. This is the same reason why the transformation of the modes $\alpha_n^\mu$ involves corrections to all orders, but $g_{\mu\nu}\alpha_n^\nu$ truncates at first order in $\delta g$, as explained in \ref{SSS: Higher orders}}. Let us verify the second order result. 

\subsubsection{${\cal O}^2$}

We will start with ${\cal O}^2$ since this is the most involved calculation. We have
\begin{equation}
    \int\limits_{\Sigma_2}d^2z_2 \int\limits_{\Sigma_1} d^2z_1 O(z_2, \Bar{z}_2)O(z_1, \Bar{z}_1)\partial X_{\mu}(w),
\end{equation}
where $O =\delta g_{\mu\nu} \partial X^\mu\bar{\partial}X^\nu$ and
\begin{align}
    &\Sigma_1 = \{z_1\in \mathbb{C} |\quad |z_1|> \epsilon, |z_1-z_2|> \epsilon, |z_1-w|>0 \}, \\
    &\Sigma_2 = \{z_2\in\mathbb{C} | \quad|z_2| > \epsilon,  |z_2-w|>0 \}.
\end{align}
The second order contribution is given by the following contractions: 
\begin{align}
    \delta g_{\nu\rho}\delta g_{\sigma \tau}&\left( 
    \wick{\int\limits_{\Sigma_2}d^2z_2 \int\limits_{\Sigma_1} d^2z_1 \partial X^\nu(z_2)\bar{\partial}\c2{X}^\rho(\bar{z}_2) \partial \c1{X}^\sigma(z_1)\bar{\partial}\c2{X}^\tau(\bar{z}_1) \partial \c1{X}_\mu(w)}\right.\notag \\
    +&\left.
    \int\limits_{\Sigma_2}d^2z_2 \int\limits_{\Sigma_1} d^2z_1 \wick{\partial \c2{X}^\nu(z_2)\bar{\partial}\c1{X}^\rho(\bar{z}_2) \partial X^\sigma(z_1)\bar{\partial}\c1{X}^\tau(\bar{z}_1) \partial \c2{X}_\mu(w)}\right).\label{O^2 contractions}
\end{align}
Let us look at the first contraction. This is
\begin{equation}
    \frac{1}{4}\delta g_{\nu\rho}\delta g_{\mu \sigma}g^{\rho\sigma}\int\limits_{\Sigma_2}d^2z_2 \int\limits_{\Sigma_1} d^2z_1 \frac{1}{(\bar{z}_2 - \bar{z}_1)^2(z_1-w)^2}\partial X^\nu(z_2).
\end{equation}
For ease of notation, we will ignore the metric factors for now and focus solely on the integral, which is
\begin{align}
    &\int\limits_{\Sigma_2}d^2z_2 \int\limits_{\Sigma_1} d^2z_1 \frac{\partial X^\nu(z_2)}{(\bar{z}_2 - \bar{z}_1)^2(z_1-w)^2} =\int\limits_{\Sigma_2}d^2z_2 \oint\limits_{\partial\Sigma_1} d\bar{z}_1 \frac{\partial X^\nu(z_2)}{(\bar{z}_2-\bar{z}_1)^2(z_1-w)}, \label{O^2}
\end{align}
where we have used that $\frac{\partial}{\partial z_1}\left( \frac{1}{\bar{z}_1 - \bar{z}_2} \right) = 0$, since $\Sigma_1$ excludes a disc around $z_2$. Let us now consider the integral
\begin{equation}
    \sum_i\oint\limits_{\Gamma_i} d\bar{z}_1 \frac{1}{(\bar{z}_1-\bar{z}_2)^2(z_1-w)},
\end{equation}
where $\bigcup_i \Gamma_i = \partial \Sigma_1$. We have boundaries around $z_1 = 0,w, z_2$. Let us call the boundaries $\Gamma_0, \Gamma_w, \Gamma_2$ respectively, and let us consider each of these in turn.

\subsubsection*{$\Gamma_0 = \partial D_0$}

We take $z_1 = \epsilon e^{i\theta}$ and $s: = e^{i\theta}$, which results in the integral 
\begin{equation}
    -\frac{1}{\bar{z}_2^2}\oint\limits_{|s|=1}\frac{ds}{\left(s-\frac{\epsilon}{\bar{z}_2}\right)^2\left(s-\frac{w}{\epsilon} \right)},
\end{equation}
which has a pole at $s = \frac{\epsilon}{\bar{z}_2}$ if we take $|w|,|z_i|>\epsilon$, which we do. Evaluating the residue, we obtain the result
\begin{equation}
    \oint\limits_{|z_1|=\epsilon}d\bar{z}_1 \frac{1}{(\bar{z}_1 - \bar{z}_2)^2(z_1-w)} = -\frac{\epsilon^2}{(w\bar{z}_2 - \epsilon^2)^2}.
\end{equation}

\subsubsection*{$\Gamma_w = \partial D_w$}

Let $z_1 = w+ \epsilon e^{i\theta}$. Then, the contour integral becomes
\begin{equation}
    -i\int\limits_0^{2\pi}d\theta \frac{e^{-2i\theta}}{(\bar{z}_2-\bar{w}-\epsilon e^{-i\theta})^2}
=
    -\frac{1}{(\bar{z}_2-\bar{w})^2}\oint\limits_{|s|=1}\frac{ds}{s\left(s-\frac{\epsilon}{\bar{z}_2-\bar{w}}\right)^2},
\end{equation}
where $s=e^{i\theta}$ has been used in the second equality. The integrand has poles at both $s=0$ and $s=\epsilon(\bar{z}_2-\bar{w})^{-1}$. Doing the residue calculation, we find that we get equal and opposite contributions from each pole, and so we get zero contribution from this boundary.

\subsubsection*{$\Gamma_1 = \partial D_{z_2}$}

Once again we make a substitution by taking $z_1 = z_2 + \epsilon e^{i\theta}, s=e^{i\theta}$, which gives
\begin{equation}
    -\frac{1}{\epsilon^2}\oint\limits_{|s|=1}\frac{ds}{s + (z_2-w)\epsilon^{-1}},
\end{equation}
which has no poles since $|w|, |z_i|>\epsilon$ and $|w-z_i|>\epsilon$, so this boundary gives zero contribution.

Thus, overall we have
\begin{equation}
    \sum_i\oint\limits_{\Gamma_i} d\bar{z}_1 \frac{1}{(\bar{z}_1-\bar{z}_2)^2(z_1-w)} = -\frac{\epsilon^2}{(w\bar{z}_2 - \epsilon^2)^2}.
\end{equation}
Going back to \eqref{O^2}, we now wish to do the $z_2$ integral, i.e. we compute
\begin{equation}
    \int\limits_{\Sigma_2}d^2z_2 \partial X^\nu(z_2) \oint\limits_{\partial \Sigma_1}d\bar{z}_1 \frac{1}{(\bar{z}_1-\bar{z}_2)^2(z_1-w)} = \frac{\epsilon^2}{w}\oint\limits_{\partial \Sigma_2} dz_2 \frac{\partial X^\nu(z_2)}{w\bar{z}_2-\epsilon^2},
\end{equation}
where we have used $-\frac{\epsilon^2}{(w\bar{z}_2 - \epsilon^2)^2} = \frac{\partial}{\partial \bar{z}_2} \frac{\epsilon^2}{w(w\bar{z}_2-\epsilon^2)}$ and integrated out the derivative. Using the mode expansion of $\partial X^\nu(z_2)$ and letting $z_2 = \epsilon e^{i\theta}$, this becomes
\begin{equation}
    \frac{\epsilon}{w\sqrt{2}}\sum_n \alpha_n^\nu \int\limits_0^{2\pi} d\theta \frac{\epsilon^{-n}e^{-in\theta}}{w e^{-i\theta}-\epsilon}
=
    \frac{i}{\sqrt{2}}\sum_n \alpha^\nu_n \epsilon^{-n} \oint\limits_{|s|=1} ds \frac{1}{s^n(s-\frac{w}{\epsilon})}, 
\end{equation}
where we have set $s=e^{i\theta}$ in the second equality. The integrand has a single pole at $s=0$ with residue $-(w/\epsilon)^{-n}$, so overall we have
\begin{equation}
    \int\limits_{\Sigma_2}d^2z_2 \partial X^\nu(z_2) \oint\limits_{\partial \Sigma_1}d\bar{z}_1 \frac{1}{(\bar{z}_1-\bar{z}_2)^2(z_1-w)} = -\frac{i}{\sqrt{2}} \sum_n w^{-n-1} \alpha_n^\nu.
\end{equation}
This completes the calculation of the first contraction in \eqref{O^2 contractions}. The calculation for the second one involves similar integrals and we will give some brief details. The integral is
\begin{equation}
    \frac{1}{4}\delta g_{\mu\nu}\delta g_{\rho \sigma}g^{\nu\sigma}\int\limits_{\Sigma_2}d^2z_2 \int\limits_{\Sigma_1} d^2z_1 \frac{1}{(\bar{z}_2 - \bar{z}_1)^2(z_2-w)^2}\partial X^\rho(z_1)=: \frac{1}{4}\delta g_{\mu\nu}\delta g_{\rho \sigma}g^{\nu\sigma}\,{\cal I}^\rho,
\end{equation}
and so the integral ${\cal I}^\rho$ that we are interested in calculating is
\begin{align}
   {\cal I}^\rho=-\int\limits_{\Sigma_2}d^2z_2 \oint\limits_{\partial\Sigma_1} dz_1 \frac{\partial X^\rho(z_1)}{(\bar{z}_1-\bar{z}_2)(z_2-w)^2} =\frac{i}{\sqrt{2}}\sum_n \alpha_n^\rho \int\limits_{\Sigma_2}\frac{d^2z_2}{(z_2-w)^2}\oint\limits_{\partial \Sigma_1} dz_1 \frac{z_1^{-n-1}}{\bar{z}_1-\bar{z}_2},
\end{align}
where now we must include $\partial X^\rho(z_1)$ from the outset since it has $z_1$ dependence. We are thus interested in the integral
\begin{equation}
    \sum_i\oint\limits_{\Gamma_i}\frac{z_1^{-n-1}}{\bar{z}_1 -\bar{z}_2 }.
\end{equation}
As before, let us look at each boundary in turn. 

\subsubsection*{$\Gamma_0$}

Using $|z_1|=\epsilon$, we have $\Bar{z}_1 - \Bar{z}_2 = \frac{\epsilon^2-z_1\Bar{z}_2}{z_2}$, and substituting this in gives
\begin{align}
    &-\frac{1}{\bar{z}_2}\oint\limits_{|z_1|=\epsilon} dz_1 \frac{z_1^{-n}}{z_1 - \epsilon^2/\Bar{z}_2}= -\frac{1}{\Bar{z}_2}\left[ \frac{1}{(n-1)!} \frac{\partial^{n-1}}{\partial z_1^{n-1}}\left( z_1 - \frac{\epsilon^2}{\Bar{z}_2} \right)^{-1} \right]_{z_1=0} - \frac{1}{\Bar{z}_2}\left( \frac{\epsilon^2}{\Bar{z}_2} \right)^{-n}=0,
\end{align}
so we get zero contribution from this boundary. 

\subsubsection*{$\Gamma_w$}

We set $z_1 = w+\epsilon e^{i\theta}$ as well as $s = e^{i\theta}$, which gives
\begin{equation}
    \frac{\epsilon^{-n}}{\Bar{w} - \Bar{z}_1}\oint\limits_{|s|=1} ds \frac{s}{\left( s+\frac{w}{\epsilon} \right)^{n+1}\left(s+ \frac{\epsilon}{\Bar{w}-\Bar{z}_2} \right)}, 
\end{equation}
which has a pole at $s = -\frac{\epsilon}{\Bar{w} - \Bar{z}_2}$, and evaluating the integral and taking the limit $\epsilon \rightarrow 0$, we find that this also vanishes (recall that our prescription is such that we always take the circle around $w$ to vanish), so once again there is no contribution. 

\subsubsection*{$\Gamma_1$}

We have $|z_1-z_2|=\epsilon$, so the integral simply becomes
\begin{equation}
    \epsilon^2 \oint\limits_{|z_1-z_2|=\epsilon} dz_1 \frac{z_1-z_2}{z_1^{n+1}},
\end{equation}
and setting $z_1 = z_2 + \epsilon s$, where $s=e^{i\theta}$, we get
\begin{equation}
    \epsilon^{-n-1}\oint\limits_{|s|=1}ds\frac{s}{\left( s+\frac{z_2}{\epsilon} \right)^{n+1}} = 0,
\end{equation}
since there are no poles inside the unit circle. Thus, all boundaries give zero, so we conclude that
\begin{equation}
    \int\limits_{\Sigma_1}d^2z_1 \int\limits_{\Sigma_2} d^2z_2 \frac{1}{(\bar{z}_1 - \bar{z}_2)^2(z_2-w)^2}\partial X^\rho(z_1) = 0,
\end{equation}
i.e. the second term in \eqref{O^2 contractions} vanishes, and the only contribution is from the first term. Therefore, overall we have
\begin{equation}
    \delta_{{\cal O}^2}\partial X_\mu = \frac{1}{4}\delta g_{\mu\nu}g^{\nu\rho}\delta g_{\rho\sigma}\partial X^\sigma.
\end{equation}
Note that this is precisely what we would get if we applied $\delta_{\cal O}$ in a naive sequential way, i.e. if we said
\begin{equation}
    \delta_{{\cal O}^2}\partial X_\mu = \frac{1}{2}\delta_{\cal O}\left(\delta g_{\mu\nu}\partial X^\nu\right) = \frac{1}{4}\delta g_{\mu\nu}g^{\nu\rho}\delta g_{\rho\sigma}\partial X^\sigma,
\end{equation}
and so it seems as though we can simply say $\delta_{{\cal O}^n} = (\delta_{\cal O})^n$. We will see shortly that this is indeed how we claim the higher order transformations work.

\subsubsection{${\delta_{E}}^2$}

This is fairly straightforward. The first action gives $\frac{1}{2}\delta g_{\mu\nu}e^\nu_a\partial X^a$, and so overall, after the second action, we get
\begin{equation}
    \delta_{E}^2\partial X_\mu = -\frac{1}{4}\delta g_{\mu\nu}g^{\nu\rho}\delta g_{\rho\sigma}\partial X^\sigma = -\delta_{\cal O}^2\partial X_\mu,
\end{equation}
and so we have $(\delta_{\cal O}^2+ \delta_{E}^2)\partial X_\mu = 0$. 

\subsubsection{${\cal O}\delta_{E} + \delta_{E}{\cal O}$}

This is again straightforward, so we will simply state the results. We have
\begin{equation}
    \delta_{E}\delta_{\cal O}\partial X_\mu = - \delta_{\cal O}\delta_{E}\partial X_\mu = \frac{1}{4}\delta g_{\mu\nu}g^{\nu\rho}\delta g_{\rho\sigma}\bar{\partial} X^\sigma.
\end{equation}
Thus, we find that $(\delta_{\cal O}+\delta_{E})^2\partial X_\mu = 0$, as expected. We expect that $(\delta_{\cal O}+\delta_{E})^n\partial X_\mu =0$ holds for all $n>1$. 

For other operators the story will be different, and each operator must be dealt with on a case-by-case basis. For example, if we are looking at the stress tensor $T$, we should find that there are non-zero contributions only to order $\delta g^2$, and so we expect
\begin{equation}
    (\delta_{\cal O}+\delta_{E})^n T =0, \qquad n\geq 3.
\end{equation}
Of course, we could also just substitute the transformation for $\partial X$ into $T$ instead of deriving it from scratch, and this should give the same result. 

\subsection{An operational approach to higher order $\cal O$ insertions}

The above calculations suggest a way of `operationalising' the $\cal O$ insertions at higher order. What we saw was that, for the $\delta_{{\cal O}^2}\partial X_\mu$ calculation, the only double contraction which gave a contribution was the one which corresponded to the order of integration, i.e. the contraction schematically of the form
\begin{equation}
    \int\limits_{\Sigma_2}\int\limits_{\Sigma_1}\wick{\c2{O_2}\c1{O_1}\c2{\partial}\c1{ X}_\mu}.
\end{equation}
We postulate that this generalises to higher powers, i.e. at order $n$, the only contraction of relevance is the following:
\begin{equation}
    \int\limits_{\Sigma_n}...\int\limits_{\Sigma_1} [O_n[O_{n-1}[...[O_2[O_1,\partial X_\mu]]...]],
\end{equation}
i.e. we contract in the order in which we compute the integrals (we use commutator notation for clarity, as explained in section \ref{S: toroidal backgrounds}). Thus, we would first contract $\partial X_\mu$ with $O_1$, then contract the result with $O_2$, and so on. This provides a way of making the application of multiple $\cal O$ operators more systematic, since this can intuitively be understood as sequentially applying the operator insertions. This allows us to write the deformation to all orders as
\begin{equation}
    \delta \partial X_\mu = \exp(\delta_{\cal O} + \delta_{E})\partial X_\mu.
\end{equation}
Note that we are not giving a mathematical or physical proof that this approach works, we are simply saying that this is a prescription which is intuitive and seems to agree with known results. Also, although we have specifically looked at $\partial X_\mu$ here, we expect this approach to work for any operator.

\section{Level Matching}\label{A: Level matching}

For a variety of reasons, it is important that, in all of our discussions on operator deformations, we still have level matching, or rotational invariance. One reason is that, in \cite{Sonoda:1991mv}, the variational formula that is postulated is averaged over all angular variables.

\subsection{Level matching for the circle}

First we do the circle case to illustrate how it works in a standard CFT context. For a circle of radius $R$ deformed to $R+\delta R$, we have deformation operator 
\begin{equation}
    {\cal O} = \lambda \int\limits_\Sigma \partial X\bar{\partial}X,
\end{equation}
where $\lambda = \delta g/R^2 = (2R\delta R+\delta R^2)/R^2$. To show level matching, we must show that
\begin{equation}
    [{\cal O}, L_0] = [{\cal O}, \bar{L}_0], 
\end{equation}
where
\begin{equation}
    L_0 = \oint\limits_{|z|=1} dz zT(z), \qquad \Bar{L}_0 = \oint\limits_{|z|=1} d\Bar{z} \Bar{z}T(\Bar{z}),
\end{equation}
where both integrals are independent of the radius of the contour. In other words, we must show that $[{\cal O}, L_0]$ is invariant under $\text{holomorphic}\leftrightarrow\text{antiholomorphic}$. We have
\begin{align}
    [{\cal O}, L_0] =& \lambda\int\limits_\Sigma d^2z\oint\limits_{C_z}dw \partial X(z)\Bar{\partial}X(\bar{z}) g^{-1}w\partial X(w)\partial X(w) \notag \\
    \sim& -\lambda \int\limits_\Sigma d^2z\oint\limits_{C_z}\frac{dw}{(z-w)^2}\Bar{\partial}X(\Bar{z})w\partial X(w),
\end{align}
and doing the contour integral and integrating by parts gives
\begin{equation}
    \lambda\oint\limits_{C_0',C_w'}d\Bar{z} z\partial X(z)\Bar{\partial}X(\bar{z}).
\end{equation}
The integral around $C_w$ can easily be seen to vanish when we take the limit $\epsilon \rightarrow 0$, so we are left with 
\begin{equation}
    \lambda\oint\limits_{C_0}d\Bar{z} z\partial X(z)\Bar{\partial}X(z),
\end{equation}
and using $zd\Bar{z} = - \bar{z}dz$ when $|z|$ is constant, we see that this is invariant under $z\leftrightarrow\Bar{z}$, and so we do indeed have $[{\cal O}, L_0^-]=0$, as required.

\subsection{Level matching for the $H$-flux}

Now we come to the more complicated case of the $H$-flux, although we will see that level matching is still preserved. We have
\begin{align}
    &[{\cal O}, L_0]= m\int\limits_\Sigma d^2z \oint\limits_{C_z}dw \phi^x(z,\Bar{z})F^-_{yz}(z,\Bar{z})\Big(\partial \phi_x^2(w) + \partial \phi_y^2(w) + \partial \phi_z^2(w)\Big) \notag \\
    &\sim m\int\limits_\Sigma d^2z\oint\limits_{C_z}dw\frac{w}{z-w}\bigg(
    F^-_{yz}(z,\Bar{z})\partial \phi_x(w)- \frac{\phi^x(z,\Bar{z})}{z-w}(\Bar{\partial}\phi_z(\Bar{z})\partial \phi_y(w) - \Bar{\partial}\phi_y(\Bar{z})\partial \phi_z(w))
    \bigg)
    \notag \\
    &=-m\int\limits_\Sigma d^2z\bigg(
    z\partial \phi_x(z)F^-_{yz}(z,\Bar{z}) + \phi^x(z,\Bar{z})F^-_{yz}(z,\Bar{z})+ z\phi^x(z,\Bar{z})\partial F^-_{yz}(z,\Bar{z})
    \bigg),
\end{align}
where in the last step we have done the $w$ contour integral and used that $\partial\Bar{\partial}\phi^y = \partial \Bar{\partial}\phi^z=0$. The last term in the final equality above can be written as
\begin{align}
    &-m\int\limits_\Sigma d^2z \Big(
    \partial_z(z\phi^x(z,\Bar{z})F^-_{yz}(z,\Bar{z})) - \phi^x(z,\Bar{z})F^-_{yz}(z,\Bar{z}) - z\partial \phi_x(z)F^-_{yz}(z,\Bar{z})
    \Big) \notag \\
    &=m\oint\limits_{C_0',C_w'}d\Bar{z} \phi^x(z,\Bar{z})F^-_{yz}(z,\Bar{z}) + m\int\limits_\Sigma d^2z  \left(\phi^x(z,\Bar{z})+z\partial \phi_x(z)\right)F^-_{yz}(z,\Bar{z}).
\end{align}
By the same argument as for the circle, the contour integral around $w$ vanishes, and so overall we have
\begin{equation}
     [{\cal O}, L_0] = m\oint\limits_{C_0'}d\Bar{z}z\phi^x(z,\Bar{z})\left(\partial \phi_y(z)\Bar{\partial}\phi_z(\Bar{z}) - \partial \phi_z(z)\Bar{\partial}\phi_y(\Bar{z})\right),
\end{equation}
which is indeed invariant under $z\leftrightarrow \Bar{z}$, so we still have $[{\cal O}, L_0^-]=0$ for the $H$-flux.

\end{document}